%% file: main.tex
\documentclass{article}
\usepackage{graphicx} % Required for inserting images
\usepackage{csquotes}
\usepackage{dsfont}
\usepackage{amsmath}
\usepackage{jheppub}
\usepackage{graphbox}
\usepackage[usenames,dvipsnames]{xcolor} 
\usepackage{tikz}	
\usepackage{dirtytalk}
\usepackage{dsfont}
\usepackage{amsthm}
\usepackage{float}
\usepackage{bbm, dsfont}
\usepackage[normalem]{ulem}
\usepackage{slashed}
\usepackage{mathtools}
\usepackage{cancel}
\usepackage{cleveref}
\usepackage{orcidlink}
\usepackage{enumerate}
\usepackage{fontawesome}
\usepackage{adjustbox}
\usepackage{tcolorbox}
\usepackage{colortbl} % For background colors
\usepackage{xcolor}   % For color definitions
\usepackage{bbold}

\definecolor{ourblue}{RGB}{0, 57, 120} % 
\colorlet{ourlightblue}{ourblue!50!white} % Approximateoblue color
\newcommand{\bs}[1]{\boldsymbol{#1}}
\usepackage{hyperref}
\hypersetup{colorlinks,linkcolor={red},citecolor={blue},urlcolor={red}} 
\def\equationautorefname~#1\null{(#1)\null}
\hypersetup{colorlinks=true}

\newcommand{\dd}{\mathrm{d}}

\newcommand{\eps}{\varepsilon}
\newcommand{\cH}{\mathcal{H}}
\newcommand{\st}{\tilde{s}}

\newcommand{\rA}{\mathrm{A}}
\newcommand{\rD}{\mathrm{D}}
\newcommand{\rF}{\mathrm{F}}
\newcommand{\bA}{\bs{A}}
\newcommand{\bB}{\bs{B}}
\newcommand{\bC}{\bs{C}}
\newcommand{\bD}{\bs{D}}
\newcommand{\bI}{\bs{I}}
\newcommand{\bJ}{\bs{J}}
\newcommand{\bK}{\bs{K}}
\newcommand{\bM}{\bs{M}}
\newcommand{\bT}{\bs{T}}
\newcommand{\bS}{\bs{S}}
\newcommand{\bR}{\bs{R}}
\newcommand{\bP}{\bs{P}}
\newcommand{\bg}{\bs{g}}
\newcommand{\bn}{\bs{n}}
\newcommand{\br}{\bs{r}}
\newcommand{\bx}{\bs{x}}
\newcommand{\bz}{\bs{z}}
\newcommand{\bDelta}{\bs{\Delta}}
\newcommand{\bOmega}{\bs{\Omega}}
\newcommand{\bbeta}{\bs{\beta}}
\newcommand{\blambda}{\bs{\lambda}}
\newcommand{\bepsilon}{\bs{\epsilon}}

\newcommand{\bnu}{\bs{\nu}}
\newcommand{\bgamma}{\bs{\gamma}}

\newcommand{\bcA}{\bs{\mathcal{A}}}
\newcommand{\bcB}{\bs{\mathcal{B}}}

\newcommand{\bcAt}{\bs{\tilde{\mathcal{A}}}}
\newcommand{\bcBt}{\bs{\tilde{\mathcal{B}}}}
\newcommand{\bcP}{\bs{\mathcal{P}}}
\newcommand{\bcS}{\bs{\mathcal{S}}}
\newcommand{\bcU}{\bs{\mathcal{U}}}

\newcommand*\RTheta[3]{\Theta\left[\genfrac{}{}{0pt}{1}{#1}{#2};#3\right]}

\theoremstyle{definition}
\newtheorem{example}{Example}

\title{Canonical Differential Equations Beyond Genus One}

\abstract{We discuss for the first time canonical differential equations for hyperelliptic Feynman integrals. We study hyperelliptic Lauricella functions that include in particular the maximal cut of the two-loop non-planar double box, which is known to involve a hyperlliptic curve of genus two. We consider specifically three- and four-parameter Lauricella functions, each associated to a hyperelliptic curve of genus two, and construct their canonical differential equations. Whilst core steps of this construction rely on existing methods — that we  show to  be applicable in the higher-genus case — we use new ideas on the structure of the twisted cohomology intersection matrix associated to the integral family in canonical form  to obtain a better understanding of the appearing new functions. We further observe the appearance of Siegel modular forms in the $\eps$-factorized differential equation matrix, nicely generalizing similar observations from the elliptic case. 
}

\author[a,\orcidlink{0000-0001-5820-3570}]{Claude Duhr,}
\emailAdd{cduhr@uni-bonn.de}

\author[a,\orcidlink{0000-0002-3328-499X}]{Franziska Porkert,}
\emailAdd{fporkert@uni-bonn.de}
\affiliation[a]{Bethe Center for Theoretical Physics, Universität Bonn, D-53115, Germany
}

\author[a,\orcidlink{0009-0007-3246-6357}]{Sven F.\ Stawinski,}
\emailAdd{sstawins@uni-bonn.de}

\allowdisplaybreaks
\begin{document}

\begin{flushright}
BONN-TH-2024-17
\end{flushright}

\maketitle
\normalem
\allowdisplaybreaks
\raggedbottom

\newpage
\section{Introduction}
\label{Sec:Intro}
\input{Sec_Intro}

\section{Review: Feynman Integrals, Maximal Cuts and Differential Equations} 
\label{Sec:Review}
\input{Sec_ReviewFeynman}

\section{Background on Hyperelliptic Curves}
\label{Sec:background}
\input{Sec_Hyperelliptic}

\section{Canonical Forms for Hyperelliptic Maximal Cuts}
\label{Sec_canonicalFormsHyperelliptic}
\input{Intro_Hyperelliptic}
\subsection{Three Parameter Lauricella Function}
\label{Sec:Three}
\input{Sec_Three}
\subsection{Four Parameter Lauricella Function}
\label{Sec:Four}
\input{Sec_Four}

\section{Modularity of the canonical differential equation}
\label{modular}
\input{Sec_Modular}

\section{Outlook and Conclusion} 
\label{Sec:OutConc}
\input{Sec_OutConc}

\section*{Acknowledgments}
We are grateful to Cathrin Semper for collaboration on related projects. We also want to thank Andrew McLeod, Cristoph Nega, Martin Raum, Oliver Schlotterer, Yoann Sohnle, Lorenzo Tancredi and Matthias Wilhelm for inspiring and helpful discussions. FP and SFS would like to thank the Munich Institute for Astro-, Particle and BioPhysics (MIAPbP) for hospitality during the workshop "Special Functions: From Geometry to Fundamental Interactions" where part of this work was carried out. This work is funded by the European Union
(ERC Consolidator Grant LoCoMotive 101043686). Views
and opinions expressed are however those of the author(s)
only and do not necessarily reflect those of the European
Union or the European Research Council. Neither the
European Union nor the granting authority can be held
responsible for them.

\begin{appendix}
\section{Some Background on Intersection Theory}
\label{app:Intersect}
\input{Sec_Intersection}

\section{Genus Two $\Theta$ Constants}
\label{app_genus2thetas}
\input{sec_GenusTwoThetas}
\section{Differentials on the Hyperelliptic Curve}
\label{app_expansions}
\input{ExpansionsDiff}
\section{Constraints on the Form of $a(\blambda)$}
\label{app_newFctConstraints}
\input{Sec_NewFctConstraints}

\section{Details on the Modular Behaviour of the Quasi-Period Matrix}
\label{app_quasiPeriodModular}
\input{Sec_QuasiPeriodModular}
\end{appendix}

%\printbibliography[heading=bibintoc]

\bibliographystyle{JHEP}
\bibliography{References.bib}
\end{document}

%% file: Sec_Intro.tex
Feynman integrals are the universal building blocks of perturbative quantum field theory (QFT) and of central importance to many areas of modern theoretical physics. Despite a lot of progress over the last decades (see \cite{Weinzierl:2022eaz} for a recent review), the computation of Feynman integrals still poses a bottleneck for precision computations in particle and gravitational wave physics. Hence, the development of new methods for their computation will advance the precision physics program, while at the same time allowing to uncover new structures underlying perturbative QFT.

Feynman integrals are generally divergent and need to be regularized, which is typically done by introducing a dimensional regularization parameter $\eps$ \cite{tHooft:1972tcz}. The integrals are then computed as a Laurent series around $\eps=0$. While this is a difficult task in general, many different methods have been developed to tackle this problem (see,~e.g.,~\cite{Smirnov:2012gma,Weinzierl:2022eaz}). Among the most powerful techniques is the method of differential equations \cite{Kotikov:1990kg,Kotikov:1991hm,Kotikov:1991pm}, which trades the integration for the task of solving a system of first order differential equations. If one is able to find a rotation into a distinguished basis, the so-called \enquote{canonical basis}, then the solution of the differential equation is obtained in terms of iterated integrals~\cite{Henn:2013pwa}. The problem of solving a Feynman integral then basically boils down to finding the rotation into the canonical basis (and also an appropriate initial condition).

In the simplest cases the solutions to the canonical differential equation can be expressed in terms of iterated integrals of $\mathrm{dlog}$ forms, which themselves often evaluate to multiple polylogarithms (MPLs) \cite{Goncharov:1998kja,Goncharov:2001iea,Remiddi:1999ew}.\footnote{There are some caveats here, see \cite{Duhr:2020gdd}.} These cases are by now under good control, and there are very general methods allowing one to tackle many such cases \cite{Gituliar:2017vzm,Meyer:2017joq,Lee:2020zfb,Dlapa:2020cwj}. The general idea is to normalize the integrals such that they have unit leading singularities \cite{Cachazo:2008vp}, an important property expected from the elements of the canonical basis \cite{Arkani-Hamed:2010pyv,Henn:2013pwa}. This requires computing the maximal cuts of the integrals, which evaluate to algebraic functions in that case. 

However, it has been known for a long time (cf.,~e.g.,~\cite{Sabry:1962rge}) that there are Feynman integrals which do not admit algebraic maximal cuts, and consequently are not expected to admit a canonical differential equation in terms of $\mathrm{dlog}$ forms. While the maximal cuts of such Feynman integrals cannot be evaluated in terms of algebraic functions, they can be interpreted as periods of an algebraic variety \cite{Bourjaily:2022bwx}. It has indeed been very fruitful to leverage geometric information to solve the associated Feynman integrals. Since the periods of the underlying variety show up in the maximal cut, they are expected to play an important role in the rotation to the canonical basis. However, in these cases much less is known about how to find the canonical basis.

By now various geometries have been uncovered in the context of Feynman integrals, see,~e.g.,~\cite{Bourjaily:2022bwx} for a review. Feynman integrals with algebraic maximal cuts are associated with the geometry of the Riemann sphere, and the simplest non-trivial geometry appearing is the elliptic curve. Elliptic Feynman integrals have received a lot of attention over the last decade, and many of the developed methods and uncovered structures of polylogarithmic Feynman integrals have by now found a suitable generalization for elliptic Feynman integrals \cite{Broedel:2018qkq,Bourjaily:2020hjv,Bourjaily:2021vyj,Kristensson:2021ani,Wilhelm:2022wow}. Also, function spaces generalizing the space of MPLs have been identified, including  elliptic multiple polylogarithms (eMPLs) \cite{Brown:2011wfj,Bloch:2013tra,Adams:2014vja,Broedel:2017kkb,Ablinger:2017bjx} and iterated integrals of modular forms~\cite{ManinModular,Brown:mmv,Adams:2017ejb,Broedel:2021zij}. A canonical basis has been constructed in a variety of examples \cite{Adams:2018yfj,Adams:2018bsn,Broedel:2018rwm,Honemann:2018mrb,Bogner:2019lfa,Duhr:2021fhk,Muller:2022gec,Giroux:2022wav,Dlapa:2022wdu,Gorges:2023zgv,Delto:2023kqv,Giroux:2024yxu,Ahmed:2024tsg,Pogel:2022yat,Duhr:2024bzt,Klemm:2024wtd,Forner:2024ojj}. Whilst there are  no complete algorithms as for polylogarithmic integrals, methods have been developed and applied to multiple non-trivial examples in a somewhat systematic fashion \cite{Dlapa:2022wdu,Pogel:2022vat,Gorges:2023zgv}. In the many known examples of elliptic canonical differential equations,  interesting structures and patterns have been observed. For example, the entries of the canonical differential equation matrix can often be identified as (quasi-) modular forms \cite{Adams:2017ejb,Broedel:2018rwm,Abreu:2019fgk,Broedel:2021zij,Pogel:2022yat,Duhr:2024bzt}, which is not only mathematically very appealing, but also of practical interest, because modular forms are very well studied and can be evaluated numerically very efficiently. 
Over the last few years, also the study of Feynman integrals associated with geometries beyond elliptic curves have received a considerable amount of attention. 
These studies have mostly focused on higher-dimensional generalizations in the form of Calabi-Yau varieties \cite{brownSchnetzK4inPhi4,Bloch:2014qca,Bloch:2016izu,Primo:2017ipr,Bourjaily:2018ycu,Bourjaily:2018yfy,Broedel:2019kmn,Bourjaily:2019hmc,Klemm:2019dbm,Vergu:2020uur,Bonisch:2020qmm,Bonisch:2021yfw,Broedel:2021zij,Pogel:2022yat,Duhr:2022pch,Forum:2022lpz,Pogel:2022ken,Pogel:2022vat,Cao:2023tpx,Doran:2023yzu,Gorges:2023zgv,McLeod:2023doa,Duhr:2023eld,Duhr:2024hjf,Jockers:2024tpc,Duhr:2024bzt,Loebbert:2024fsj,Driesse:2024feo}. More recently, also one-dimensional geometries, in particular Riemann surfaces of higher genus, have received an increasing amount of attention both from the QFT and string theory perspectives~\cite{Huang:2013kh,Georgoudis:2015hca,Doran:2023yzu,Marzucca:2023gto,DHoker:2022xxg,DHoker:2023vax,DHoker:2024ozn,Baune:2024ber,Baune:2024biq}. 

The construction of a canonical basis for a Feynman integral associated with a higher-genus Riemann surface has not been achieved so far. The goal of this paper is to take a first step in this direction. To this end, we will study three- and four-variable Lauricella hypergeometric functions, which resemble the maximal cuts of Feynman integrals associated with Riemann surfaces of genus two, such as the non-planar crossed box integral~\cite{Georgoudis:2015hca,Huang:2013kh,Marzucca:2023gto}. In this case, the Riemann surfaces are always defined by hyperelliptic curves, and thus very direct generalizations of the more familiar elliptic curves. We might therefore expect many of the methods and structures observed for elliptic Feynman integrals to find a  generalization to hyperelliptic geometries. As the Lauricella functions resemble maximal cuts of Feynman integrals, their differential equations resemble the diagonal block of the top sector in the differential equation of a full Feynman integral, which is precisely the block directly associated to the geometry. Understanding how to find the canonical rotation for this block is hence one of the key steps in finding the rotation for the full family of Feynman integrals.
The goal of this paper is to explicitly construct such a rotation for the Lauricella functions in question, generalizing the method of \cite{Gorges:2023zgv} to hyperelliptic curves. As already observed there, this method may require the introduction of some {new functions}, which are not expressible in terms of the periods of the geometry, but are rather integrals over the (quasi-)periods. While it is in general a difficult task to determine which of these functions are independent or might be expressible directly in terms of the periods, we show how recent results on the structure of the {twisted cohomology intersection matrix}\footnote{Twisted cohomology is a mathematical framework for multivalued differential forms. See appendix \ref{app:Intersect} for a brief introduction or \cite{aomotoBook} for more details. } in the canonical basis can help us to understand these functions. 

While the differential equation matrices we construct are certainly $\eps$-factorized, it is a priori not clear that they are also canonical.
We therefore provide further evidence for the three-variable case by studying the modularity properties of the differential equation. We show that the entries of the differential equation matrix can be identified as \emph{Siegel modular forms}, a higher-genus generalization of the more familiar (elliptic) modular forms, as well as certain new functions defined as integrals over Siegel (quasi-)modular forms. This nicely generalizes similar observations made for elliptic canonical differential equations.

The rest of this paper is structured as follows: In section \ref{Sec:Review} we will review the necessary background on Feynman integrals, in particular the method of differential equations and the notion of maximal cuts. We will showcase the method of \cite{Gorges:2023zgv} by applying it to Gauß' hypergeometric ${}_2F_1$ function with a certain choice of parameters, a prototype for the maximal cut of an elliptic Feynman integral. In section \ref{Sec:background}, we will generalize Gauß' hypergeometric function to general Lauricella functions, and we will then review the necessary mathematical basics on the geometry of hyperelliptic curves needed in the remainder of the paper. In section \ref{Sec_canonicalFormsHyperelliptic} we  construct the $\eps$-factorized differential equations for the three- and four-variable Lauricella functions associated with a hyperelliptic curve of genus two, generalizing the method of \cite{Gorges:2023zgv}. In doing this we will in particular show how the recent results of \cite{Duhr:2024xsy} on the structure of the intersection matrix in the canonical basis can be used in this process. In section \ref{modular}, we review the necessary basics of Siegel modular forms and then show how these objects can be identified in the differential equation matrix. We will end the paper by drawing our conclusions and discussing open questions and future directions in section \ref{Sec:OutConc}. Some additional information and technical details, including a brief review of intersection theory as well as some more details on differentials on the hyperelliptic curve and their expressions in $\Theta$ functions, are deferred to various appendices.  We have also attached an ancillary file with the arXiv submission of this paper, where various matrices, that are too big to be displayed, are explicitly given.

%% file: Sec_ReviewFeynman.tex
Since the main motivation for this paper is the computation of Feynman integrals with the method of (canonical) differential equations, we briefly review these concepts in section \ref{SubSec:FeynmanIntegrals} and then introduce maximal cuts in section \ref{SubSec:maxcut}, on which we will focus in the rest of the paper. We will end the section in \ref{SubSec:Elliptic} with a review of the method of \cite{Gorges:2023zgv} for finding the canonical differential equation by applying it to the example of a Gauß' hypergeometric ${}_2F_1$ function with a certain choice of parameters, that is a prototype for the maximal cut of an elliptic Feynman integral. This example will furthermore serve as a stepping stone to the hyperelliptic cases which we will study in section \ref{Sec_canonicalFormsHyperelliptic}.

\subsection{Feynman Integrals and Their Differential Equations}
\label{SubSec:FeynmanIntegrals}
We consider families of Feynman integrals of the  form
\begin{equation}
    I_{\bs{\nu}}^D(\{p_i\cdot p_j\},\{ m_i^2 \})=e^{L\gamma_{\mathrm{E}}\eps}\int\left(\prod_{j=1}^L\frac{\dd^D\ell_j}{i\pi^{D/2}}\right)\frac{1}{\prod_{j=1}^{m}(q_j^2-m_j^2)^{\nu_j}} \,,
\end{equation}
where $\bs{\nu}=(\nu_1,\dots,\nu_m)$ is a vector of integers, $L$ is the number of loops, $\gamma_{\mathrm{E}}=-\Gamma'(1)$ is the Euler-Mascheroni constant and $D$ is the spacetime dimension. The $q_j$ entering the propagators are integer linear combinations of the loop momenta $\ell_k$ and the external momenta $p_k$, and we denote the internal (squared) masses by $m_j^2$. These integrals are typically divergent and need to be regularized. To this end, we use dimensional regularization \cite{tHooft:1972tcz}, i.e., we will analytically continue the space-time dimension $D=d-2\eps$ away from some integer value $d$, and we consider the Feynman integrals as Laurent expansions around $\eps=0$. 

The Feynman integrals in a family can be organized in different \emph{sectors}, defined by the set of positive $\nu_i$, i.e., the set of \enquote{active} propagators. This yields a partial ordering on the Feynman integrals labeled by the vector of integers $\bnu$ by considering $\bnu\geq \bnu'$, if $\nu_i\geq\nu_i'$ for all $i=1,\dots ,m$. The sector with all $\nu_i>0$, is referred to as the \emph{top sector}. 

A given family of Feynman integrals possesses the structure of a finite-dimensional vector space \cite{Smirnov:2010hn,Bitoun:2017nre}. A basis of this vector space is called a set of \emph{master integrals},
\begin{equation}
    \bs{I}(\bs{x},\eps)=(I_1(\bs{x},\eps),\dots,I_N(\bs{x},\eps))^T \,.
\end{equation}
Every member of the family can be expressed as a linear combination of master integrals. We assume that the master integrals have been rescaled in such a way that they only depend on dimensionless ratios of Lorentz invariants $\bs{x}=(x_1,\dots,x_r)$. There are various methods to achieve the decomposition into master integrals, based on integration-by-part (IBP) reduction~\cite{Tkachov:1981wb,Chetyrkin:1981qh,Laporta:2000dsw} or methods from twisted cohomology~\cite{Mastrolia:2018uzb,Frellesvig:2019uqt,Brunello:2023rpq}.

An important consequence of the vector space structure of Feynman integrals is that they satisfy first-order differential equations \cite{Kotikov:1990kg,Kotikov:1991hm,Kotikov:1991pm}. Explicitly, we can take the total derivative of the vector of master integrals $\bs{I}(\bx,\eps)$ with respect to the kinematical variables $\bx$ and re-express the result in the basis. We obtain in this way a differential equation of the form
\begin{equation}
\label{eq:FeynmanIntegralDEQ}
    \dd\bs{I}(\bs{x},\eps)=\bs{B}(\bs{x},\eps)\bs{I}(\bs{x},\eps) \,,
\end{equation}
where $\bs{B}(\bs{x},\eps)$ is a matrix of one-forms with rational dependence on $\bs{x}$ and $\eps$. If we organize the master integrals according to the partial order on the sectors, the differential equation matrix is block-triangular.

There is significant freedom in the choice of master integrals, and we can rotate the basis into a new basis $\bs{I}'(\bs{x},\eps)$ via
\begin{equation}
    \bs{I}'(\bs{x},\eps)=\bs{R}(\bs{x},\eps)\bs{I}(\bs{x},\eps) \,.
\end{equation}
The new master integrals then satisfy the differential equation
\begin{equation}
    \dd\bs{I}'(\bs{x},\eps)=\bs{B}'(\bs{x},\eps)\bs{I}'(\bs{x},\eps) \,,
\end{equation}
where the new differential equation matrix is related to the old one through a gauge transformation,
\begin{equation}
    \bs{B}'(\bs{x},\eps)=\bs{R}(\bs{x},\eps)\bs{B}(\bs{x},\eps)\bs{R}^{-1}(\bs{x},\eps)+\dd \bs{R}(\bs{x},\eps)\bs{R}^{-1}(\bs{x},\eps) \,.
\end{equation}
It is conjectured \cite{Henn:2013pwa} that there always exists a rotation into a basis $\bs{J}(\bs{x},\eps)$, such that the master integrals satisfy an $\eps$\emph{-factorized} differential equation
\begin{equation}
    \dd\bs{J}(\bs{x},\eps)=\eps\bs{A}(\bs{x})\bs{J}(\bs{x},\eps) \,.
\end{equation}
While there are examples of families of Feynman integrals that admit different bases satisfying $\eps$-factorized differential equations (see e.g., \cite{Frellesvig:2023iwr}), it is believed that there is a distinguished one, the so-called \emph{canonical basis}, which satisfies additional properties. For example, the one-forms appearing in the differential equation matrix of the canonical basis are expected to have at most logarithmic singularities \cite{Henn:2013pwa}. 
In the simplest cases the canonical differential equation matrix only contains $\mathrm{dlog}$ forms. These cases are under relatively good control. There are well-understood methods and public codes to find the rotation to the canonical form in many such cases \cite{Gituliar:2017vzm,Meyer:2017joq,Lee:2020zfb,Dlapa:2020cwj}. Roughly speaking, the idea to find the canonical form is to find a set of master integrals with unit leading singularities~\cite{Cachazo:2008vp}, as these integrals are expected to evaluate to pure functions \cite{Arkani-Hamed:2010pyv}. 

Leading singularities are closely connected to \emph{maximal cuts}. They can typically be computed using residues and often evaluate to rational or algebraic functions.
However, there are also many cases where the maximal cuts cannot all be evaluated in terms of algebraic functions. The concepts of leading singularity and pure functions are then less clear (see, however, \cite{Broedel:2018qkq,Bourjaily:2020hjv,Bourjaily:2021vyj}). The maximal cuts (in the $\eps\rightarrow 0$ limit) are then typically identified with the periods of an algebraic variety, which allows one to leverage geometric approaches to study those integrals. Feynman integrals which admit purely algebraic maximal cuts/leading singularities are associated with the  geometry of the Riemann sphere, otherwise they are associated to a non-trivial geometry admitting transcendental periods. The simplest non-trivial geometry appearing in the maximal cut of a Feynman integral is the elliptic curve, but also higher-dimensional generalizations in the form of Calabi-Yau varieties \cite{brownSchnetzK4inPhi4,Bloch:2014qca,Bloch:2016izu,Primo:2017ipr,Bourjaily:2018ycu,Bourjaily:2018yfy,Broedel:2019kmn,Bourjaily:2019hmc,Klemm:2019dbm,Vergu:2020uur,Bonisch:2020qmm,Bonisch:2021yfw,Broedel:2021zij,Pogel:2022yat,Duhr:2022pch,Forum:2022lpz,Pogel:2022ken,Pogel:2022vat,Cao:2023tpx,Doran:2023yzu,Gorges:2023zgv,McLeod:2023doa,Duhr:2023eld,Duhr:2024hjf,Jockers:2024tpc,Duhr:2024bzt,Loebbert:2024fsj}, and other Riemann surfaces of higher genus \cite{Huang:2013kh,Georgoudis:2015hca,Doran:2023yzu,Marzucca:2023gto} have been uncovered, see \cite{Bourjaily:2022bwx} for a recent review.

In the case of non-trivial geometries, our understanding of $\eps$-factorized or canonical differential equations is still very incomplete. Nonetheless, by now many examples are known where an $\eps$-factorized basis has been found even for Feynman integrals associated with non-trivial geometries
\cite{Adams:2018yfj,Adams:2018bsn,Broedel:2018rwm,Honemann:2018mrb,Bogner:2019lfa,Duhr:2021fhk,Muller:2022gec,Giroux:2022wav,Dlapa:2022wdu,Gorges:2023zgv,Delto:2023kqv,Giroux:2024yxu,Ahmed:2024tsg,Pogel:2022yat,Pogel:2022ken,Pogel:2022vat,Duhr:2024bzt,Klemm:2024wtd,Driesse:2024feo,Forner:2024ojj}, and first methods have been developed that allow one to find this basis in a somewhat algorithmic way \cite{Dlapa:2022wdu,Pogel:2022vat,Gorges:2023zgv}. In particular, the method of \cite{Gorges:2023zgv}, builds on earlier observations of \cite{Broedel:2018qkq} and is very similar in spirit to the leading singularity methods for Feynman integrals associated with the Riemann sphere. The appropriate generalization of the concept of leading singularity is the so-called \emph{semi-simple part} of the period matrix of the associated geometry. We will review this method in more detail in section \ref{SubSec:Elliptic}. Note that this method seems to always provide the particular $\varepsilon$-factorized form considered to be canonical (e.g., it is seen to always have simple poles). For all examples that we have considered, this criterion is fulfilled for the bases obtained with the method of \cite{Gorges:2023zgv}.

The goal of this paper is to understand how to extend the method of \cite{Gorges:2023zgv}, developed to find the canonical basis for Feynman integrals associated with elliptic curves (and also applied for Calabi-Yau geometries \cite{Gorges:2023zgv,Duhr:2024bzt,Klemm:2024wtd,Driesse:2024feo,Forner:2024ojj}), to certain algebraic curves of higher genus, in particular \emph{hyperelliptic curves}. For simplicity we will focus on the diagonal block of the differential equation matrix corresponding to the top sector, which is associated to the maximal cuts of the Feynman integral, that we will review in the next section.

\subsection{Maximal Cuts of Feynman Integrals} 
\label{SubSec:maxcut}

It can be shown that the homogeneous solution in a give sector of the differential equation~\eqref{eq:FeynmanIntegralDEQ} is given by the maximal cuts of the master integrals in that sector~\cite{Primo:2016ebd,Primo:2017ipr,Frellesvig_2017,Bosma:2017hrk}.
The maximal cuts can in turn be obtained by putting all propagators on shell. This amounts to taking residues or, said differently, to choosing a contour that encircles all poles coming from the propagators. Such a contour is not necessarily unique, and there may be $N$ independent such contours.  

A useful tool to compute maximal cuts is the so-called \emph{Baikov representation} of a Feynman integral,
\begin{align}
\label{eq:Baikovreo}
    I_{\bs{\nu}}(\{p_i\cdot p_j\}, \{m_i^2\}) = e^{L\varepsilon \gamma_E} \frac{\left[\det \mathcal{G}(p_1,\dots, p_E)\right]^{\frac{E+1-D}{2}}}{\pi^{\frac{1}{2}(n_{\text{isp}}-L)}\left[\det {C}\right]\prod_{j=1}^L \Gamma\left(\frac{D-E+1-j}{2}\right)}\, \int_{\mathcal{C}} \dd^{n_{\text{isp}}}z \left[\mathcal{B}(\bs{z})\right]^{\frac{D-L-E-1}{2}} \prod_{s=1}^{n_{\text{isp}}} z_s^{-\nu_s}  \, . 
\end{align}
The integration variables $\bs{z}=(z_1,\dots, z_{n_{\text{isp}}})$ are given by  propagators $z_i=q_i^2-m_i^2$, and $n_{\text{isp}}$ is the number of linearly independent scalar products involving the loop momenta. The number of propagators may not be sufficient to express all irreducible scalar products. In that case, one adds propagators and sets the powers $\nu_i=0$ for them. The Gram determinant and the Baikov polynomial are defined as 
\begin{align}
\label{eq:Gramdet}
    \det \mathcal{G}(q_1,\dots,q_n)= \det (-q_i\cdot q_j)  \,,\textrm{~~~and~~~} 
    \mathcal{B}(\bs{z}) = \det \mathcal{G}\left(\ell_1,\dots, \ell_L, p_1,\dots, p_E\right)\, . 
\end{align}
The determinant $\det C$ is independent of the integration variables $z_i$ and encodes the Jacobian of the change of variables. The explicit form of the integration contour $\mathcal{C}$ will be irrelevant in the following. For higher loop orders, it is often beneficial to compute the maximal cut in the so-called loop-by-loop approach \cite{Frellesvig_2017}, i.e., to reparametrize one loop momentum at a time. 
The integrand then contains a  product of several Baikov polynomials raised to some non-integer powers $\mu_i$,
\begin{align*}
     I_{\bs{\nu}}(\{p_i\cdot p_j\}, \{m_i^2\})\sim \int_{\mathcal{C}} \dd^{n_{\text{isp}}-\delta} \boldsymbol{z} \, \mathcal{B}_1(\boldsymbol{z})^{\mu_1}\dots \mathcal{B}_m(\boldsymbol{z})^{\mu_{m}}\prod_{s=1}^{m} z_s^{-\nu_s} \, ,
\end{align*}
but with fewer integration variables than in the standard Baikov approach.

The maximal cuts can now be easily computed from the Baikov representation, by taking the residues at all the poles $z_i=0$. This leads to an expression for maximal cuts of the form
\begin{align}
\label{maxcutform}
      \text{MC}\left[ I_{\bs{1}}(\{p_i\cdot p_j\}, \{m_i^2\})\right]\sim 
      \int_{\gamma} \dd^{m'} \boldsymbol{z} \, \mathcal{B}_1(\boldsymbol{z})^{\mu_1}\dots \mathcal{B}_m(\boldsymbol{z})^{\mu_{m}}\Big|_{z_1,\dots, z_m\rightarrow 0} \text{ with } m'=n_{\text{isp}}-\delta-m\, .
\end{align}
In dimensional regularization, the exponents $\mu_i$ are generally non-integer, resulting in a multivalued integrand. Maximal cut integrals as in~\eqref{maxcutform} then allow for an immediate interpretation in the context of twisted cohomology, and we can  apply results from twisted cohomology to study maximal cuts. In \cite{duhr2024twistedriemannbilinearrelations,Duhr:2024xsy} it was discussed how the matrix of maximal cuts can be interpreted as a twisted period matrix, allowing one to find bilinear relations between its entries. Finally, twisted cohomology defines a scalar product on the integrands, the so-called \emph{intersection pairing}, and there are explicit methods for its computation~\cite{Mastrolia:2018uzb,Mizera:2019vvs,Brunello:2024tqf,Chestnov:2022xsy,Chestnov:2022alh,Frellesvig:2019uqt} (see also appendix~\ref{app:Intersect}). In~\cite{Duhr:2024xsy}, it was shown that, as a consequence of self-duality and the bilinear relations, in canonical bases the intersection matrix $\bs{C}$, i.e., the matrix of all intersection pairings between master integrands is expected to be a constant in all external variables, and the $\varepsilon$-dependence factorizes into a scalar function. It is worth noting that this observation aligns with results of \cite{Caron-Huot:2021xqj,Caron-Huot:2021iev,Giroux:2022wav,De:2023xue,Crisanti:2024onv}. 
The fact that the intersection matrix should be constant provides an easy-to-check criterion for being in canonical form. One of the goals of this paper is to show that the constancy of the intersection matrix can also be used constructively.

\subsection{Warm-up: Elliptic Hypergeometric ${}_2F_1$-function}
\label{SubSec:Elliptic}

In this section we show how to find the canonical differential equation for certain classes of Gauß' hypergeometric functions. A canonical differential equation for this family was first obtained in \cite{Broedel:2018rwm}. 
While the result of this section is thus not new, we discuss it in detail for two reasons. First, it allows us to review the algorithm of~\cite{Gorges:2023zgv}, which will serve as a starting point to understand the canonical differential equations for the hyperelliptic cases in the subsequent sections. Second, we will 
illustrate how we can constructively use the fact that the intersection matrix is expected to be constant in a canonical basis to simplify some of the steps of this algorithm. Gauß' hypergeometric is the simplest example to illustrate this novel method.

Gauß' hypergeometric function can be defined by the integral\footnote{Note that this differs from the usual integral representation of the hypergeometric function by a  prefactor.}
\begin{equation}\label{eq:Gauss}
    \mathcal{L}_{\bs{\nu}}(\lambda,\eps)=\int_{\lambda}^{\infty}\dd x\,x^{-\frac{1}{2}+\nu_1+a_1\eps}(x-1)^{-\frac{1}{2}+\nu_2+a_2\eps}(x-\lambda)^{-\frac{1}{2}+\nu_3+a_3\eps} \, .
\end{equation}
Here $\bs{\nu}=(\nu_1,\nu_2,\nu_3)$ is a vector of integers, $a_1,a_2,a_3$ are some rational numbers, $\lambda>1$ is a parameter and $\eps$ is a generic parameter resembling the dimensional regulator. The exponents are chosen in such a way that the integrand for $\eps=0$ is given by a rational function on an elliptic curve $\mathcal{E}$ defined by
\begin{equation}
    y^2=x(x-1)(x-\lambda) \,.
\end{equation}
Gauß' hypergeometric function resembles the maximal cut of an elliptic Feynman integral in dimensional regularization. In the following we will assume some familiarity with elliptic curves, and we refer to \cite{Broedel:2017kkb,Broedel:2018qkq,Weinzierl:2022eaz} for introductions in the physics literature.

To study the differential equations we need some initial choice of master integrals. It is easy to check that the set of $\mathcal{L}_{\bs{\nu}}$ forms a two-dimensional vector space, and we choose as its basis\footnote{This corresponds to choosing $\{1,x\}$ as the basis of the twisted cohomology group $H_{\mathrm{dR}}^1(X,\nabla_\Phi)$ with $X=\mathbb{C}-\{0,1\}$ and twist $\Phi(x)=x^{-\frac{1}{2}+a_1\eps}(x-1)^{-\frac{1}{2}+a_2\eps}(x-\lambda)^{-\frac{1}{2}+a_3\eps}$.}
\begin{equation}
    \bs{I}_0(\lambda,\eps)=(\mathcal{L}_{0,0,0}(\lambda,\eps),\mathcal{L}_{1,0,0}(\lambda,\eps))^T \,,
\end{equation}
Note that, since we are considering differential equations  obtained by differentiating the integrand and performing integration by parts, the integration contour chosen for the master integrals plays no role (as long as all boundary term vanish). Hence, in the following we will typically talk about master \emph{integrands} and when we refer to the master integrals $\bs{I}$, it is understood that the integrands are integrated over some appropriate cycle $\gamma$. 

Our choice of master integrands is motivated by the geometry of elliptic curves, since for $\eps=0$ the two integrands reduce to the two differentials
\begin{equation}
 \varpi_1=\frac{\dd x}{y} \textrm{~~~and~~~} \varpi_2 = \frac{x\dd x}{y}\,.
\end{equation}
These differentials are a canonical choice for a basis of the first cohomology $H^1(\mathcal{E})$ of the elliptic curve $\mathcal{E}$, and they are typically classified into first and second kind, with the first kind differential $\varpi_1$ being holomorphic and the second kind differential $\varpi_2$ being meromorphic without residues. 
The master integrals satisfy the first-order differential equation,
\begin{equation}
\label{eq:deStart2f1}
    \dd \bs{I}_0(\lambda,\eps)=\bs{B}(\lambda,\eps)\bs{I}_0(\lambda,\eps) =\dd\lambda\,\left[\bs{B}^{(0)}(\lambda) + \eps \bs{B}^{(1)}(\lambda)\right]\bs{I}_0(\lambda,\eps)\,,
\end{equation}
with 
\begin{equation}
\label{eq:deMatrix2f1}
    \bs{B}^{(0)}(\lambda)=
    \begin{pmatrix}
    -\frac{1}{2(\lambda-1)} &
    \frac{1}{2\lambda(\lambda-1)} \\
    -\frac{1}{2(\lambda-1)} &
    \frac{1}{2(\lambda-1)}
    \end{pmatrix}\,,\qquad\bs{B}^{(1)}(\lambda)=
    \begin{pmatrix}
    -\frac{a_1-(\lambda-1)a_3}{\lambda(\lambda-1)} &
    \frac{a_1+a_2+a_3}{\lambda(\lambda-1)} \\
    -\frac{a_1}{\lambda-1} &
    \frac{a_1+a_2+a_3}{\lambda-1}
    \end{pmatrix} \,.
\end{equation}
The boundary condition is easily obtained by noting that for $\lambda=1$, the integral can be evaluated in terms of Gamma functions,
\begin{equation}
\mathcal{L}_{\bs{\nu}}(1,\eps) = \frac{\Gamma (\nu_2+\nu_3+(a_2+a_3) \epsilon ) \Gamma \left(\tfrac{1}{2}-\nu_1-\nu_2-\nu_3-(a_1+a_2+a_3) \epsilon\right)}{\Gamma \left(\tfrac{1}{2}-\nu_1-a_1 \epsilon\right)}\,.
\end{equation}

We now show how to apply the algorithm of~\cite{Gorges:2023zgv} to construct a basis $\bs{J}(\lambda,\eps)=\bs{R}(\lambda,\eps)\bs{I}_0(\lambda,\eps)$ such that its differential equation is in canonical form.\footnote{We are only following the steps relevant for the maximal cut, the method of \cite{Gorges:2023zgv} also applies beyond this case.}
Central to this method is the \emph{period matrix} of the elliptic curve, defined as the $2\times 2$ matrix formed by integrating the two basis differentials $\varpi_1,\varpi_2$ over the two canonical cycles $(\gamma_1,\gamma_2)=(a,b)$ of the elliptic curve
\begin{equation}
\label{eq:ellipticperiodmatrix}
    \bcP=\left(\int_{\gamma_j}\varpi_i\right)\equiv
    \begin{pmatrix}
        \omega_1 & \omega_2 \\ \eta_1 & \eta_2
    \end{pmatrix} \,.
\end{equation}
Here the second equality defines the \emph{periods} $\omega_1,\omega_2$ and the \emph{quasi-periods} $\eta_1,\eta_2$ as the integrals of the first and second kind differentials, respectively, over the the two cycles $a,b$. These are (transcendental) functions of the parameter $\lambda$, but we will suppress this dependence in the following for brevity. The two vectors $(\omega_1,\eta_1)$ and $(\omega_2,\eta_2)$ by definition satisfy the differential equation \eqref{eq:deStart2f1} at $\eps=0$, i.e., the period matrix $\bcP$ is the corresponding Wronskian.

The periods and quasi-periods satisfy a quadratic relation, known as the Legendre relation,
\begin{equation}
    \label{eq:legendre}
    \omega_1 \eta_2-\omega_2\eta_1=-8\pi i \,.
\end{equation}
Following \cite{Broedel:2018qkq,Gorges:2023zgv}, we split the period matrix into a \emph{semi-simple} and a \emph{unipotent} part,
\begin{equation}
    \bcP=\bcS \,\bcU=\begin{pmatrix}
        \omega_1 & 0 \\ \eta_1 & \frac{-8\pi i}{\omega_1}
    \end{pmatrix}
    \begin{pmatrix}
        1 & \tau \\ 0 & 1
    \end{pmatrix} \, ,
\end{equation}
where we define the modular parameter $\tau=\omega_2/\omega_1$. While the semi-simple part is only required to be diagonalizable, the defining property of the unipotent part is that it satisfies a first-order differential equation with nilpotent differential equation matrix. Note that the semi-simple part only depends on the periods and quasi-periods defined on the $a$-cycle, with all dependence on the $b$-cycle either removed by the Legendre relation or deferred to the unipotent part in the form of the modular parameter.

The rotation to the canonical form can now be constructed in four simple steps \cite{Gorges:2023zgv}.

\paragraph{Step 1: Derivative Basis.} We will start by rotating from the geometrically motivated basis $\bs{I}_0(\lambda,\eps)$ (which led to the simple Legendre relation \eqref{eq:legendre}) to a derivative basis defined by $\bs{I}_\dd(\lambda,\eps)=(\mathcal{L}_{0,0,0}(\lambda,\eps),\partial_{\lambda} \mathcal{L}_{0,0,0}(\lambda,\eps))$. The derivative basis is related to the initial one by a rotation
\begin{equation}
  \bs{I}_\dd(\lambda,\eps)   =\bs{R}_{\dd}(\lambda,\eps) \bs{I}_0(\lambda,\eps)\,,
\end{equation}
and the rotation matrix can easily be read off from the differential equation matrix in~\eqref{eq:deMatrix2f1},
\begin{equation}
   \boldsymbol{R}_{\dd}(\lambda,\eps)=\begin{pmatrix}
        1 & 0 \\ -\frac{\lambda+2\varepsilon(a_1-(\lambda-1)a_3)}{2\lambda(\lambda-1)} &\frac{1+2\varepsilon(a_1+a_2+a_3)}{2\lambda(\lambda-1)} 
    \end{pmatrix} \,.
\end{equation}
\paragraph{Step 2: Semi-simple Rotation.} The second step is the most crucial one. It consists of rotating with the semi-simple part $\bcS_{\dd}$ of the period matrix, now however in the derivative basis,
\begin{equation}
    \bcP_{\dd}=\begin{pmatrix}
        \omega_1 & \omega_2 \\ \partial_{\lambda}\omega_1 & \partial_{\lambda}\omega_2
    \end{pmatrix} \,.
\end{equation}
This can easily be related to the period matrix $\bcP$ defined above and yields for the semi-simple part
\begin{equation}
    \bcS_{\dd}(\lambda)=   \boldsymbol{R}_{\dd}(\lambda,0)\bcS(\lambda) \,.
\end{equation}
We can now use this to change the basis
\begin{equation}
 \bs{I}_{\mathrm{ss}} (\lambda,\eps)   =\boldsymbol{R}_{\mathrm{ss}}(\lambda)\bs{I}_\dd(\lambda,\eps) \,,
\end{equation}
with $\boldsymbol{R}_{\mathrm{ss}}(\lambda)=\bcS_{\dd}(\lambda)^{-1}$. After this rotation the diagonal elements are already $\eps$-factorized, while the upper right entry only has an $\eps^0$ part and the lower left entry has an $\eps^2$ part. The new differential equation matrix then takes the form 
\begin{align}
    \bs{B}_{{\rm ss}} (\lambda,\eps)
    &= \left(\dd \boldsymbol{R}_{\mathrm{ss}}\right) \boldsymbol{R}_{\mathrm{ss}}^{-1} + \boldsymbol{R}_{\mathrm{ss}}\,  \bs{B}_{\dd}(\lambda,\eps)\, \boldsymbol{R}_{\mathrm{ss}}^{-1}\\
 \nonumber   &=\left(\,
    \begin{array}{@{\hspace{2pt}}c@{\hspace{2pt}} @{\hspace{2pt}}c@{\hspace{2pt}}}
        0 & \cellcolor{ourlightblue}\,   \textcolor{ourblue}{\bullet} \, \,   \\ 
        0 &0   \, \,  
    \end{array}
    \,\right)+\left(\,
    \begin{array}{@{\hspace{2pt}}c@{\hspace{2pt}} @{\hspace{2pt}}c@{\hspace{2pt}}}
        0&0 \\ 
         \cellcolor{ourlightblue}\,   \textcolor{ourblue}{\bullet} \, \,   & \cellcolor{ourlightblue}\,   \textcolor{ourblue}{\bullet}   \, \,  
    \end{array}
    \,\right) \eps+\left(\,
    \begin{array}{@{\hspace{2pt}}c@{\hspace{2pt}} @{\hspace{2pt}}c@{\hspace{2pt}}}
       0  &0 \\ 
        \cellcolor{ourlightblue}\,   \textcolor{ourblue}{\bullet}   \, \,  &0
    \end{array}
    \,\right) \eps^2\, , 
\end{align}
where $\bs{B}_{\dd}(\lambda,\eps)$ is the differential equation matrix obtained after Step 1 and the shaded entries \colorbox{ourlightblue}   {$\textcolor{ourblue}{\bullet}$} indicate non-zero entries that are too large to display explicitly and/or whose precise form is irrelevant for the discussion.

\paragraph{Step 3: Rescaling with $\varepsilon$.} The form of the differential equation matrix $\bs{B}_{{\rm ss}} (\lambda,\eps)$ motivates the third rotation, which is an $\eps$-rescaling,
\begin{equation}
 \bs{I}_{\eps}(\lambda,\eps)   =\boldsymbol{R}_{\eps}(\eps)\bs{I}_{\mathrm{ss}}(\lambda,\eps)\,,
\end{equation}
with
\begin{equation}
    \boldsymbol{R}_{\eps}(\eps)=\begin{pmatrix}
        \eps & 0 \\ 0 & 1
    \end{pmatrix} \,.
\end{equation}
Now the full differential equation is $\eps$-factorized, except for the lower left entry which has an $\eps^0$ term. We find 
\begin{align}
  \nonumber  \bs{B}_{\eps} (\lambda,\eps)
    &= \left(\dd \boldsymbol{R}_{\eps}\right) \boldsymbol{R}_{\eps}^{-1} + \boldsymbol{R}_{\eps}\,  \bs{B}_{\mathrm{ss}}(\lambda,\eps)\, \boldsymbol{R}_{\eps}^{-1}\\
    &=    \bs{B}_{\eps}^{(0)}(\lambda)\,\dd\lambda\,+\bs{B}_{\eps}^{(1)} (\lambda)\, \eps\,\dd\lambda\,\\
  \nonumber  &=\left(\,
    \begin{array}{@{\hspace{2pt}}c@{\hspace{2pt}} @{\hspace{2pt}}c@{\hspace{2pt}}}
        0 &0  \\ 
         \cellcolor{ourlightblue}\,   \textcolor{ourblue}{\bullet} \, \,  &0   \, \,  
    \end{array}
    \,\right)+\left(\,
    \begin{array}{@{\hspace{2pt}}c@{\hspace{2pt}} @{\hspace{2pt}}c@{\hspace{2pt}}}
        0& \cellcolor{ourlightblue}\,   \textcolor{ourblue}{\bullet} \, \, \\ 
         \cellcolor{ourlightblue}\,   \textcolor{ourblue}{\bullet} \, \,   & \cellcolor{ourlightblue}\,   \textcolor{ourblue}{\bullet}   \, \,  
    \end{array}
    \,\right)\,\eps \, ,  
\end{align}
with the matrices explicitly given by
\begin{align}
     \bs{B}_{\eps}^{(0)} (\lambda)&=\begin{pmatrix}
        0 & 0 \\ p(\lambda) & 0 
    \end{pmatrix} \,, \\
      \bs{B}_{\eps}^{(1)} (\lambda)&=\begin{pmatrix}
        0 & \frac{-4\pi i}{\lambda(\lambda-1)\omega_1^2} \\
        \frac{a_3(a_1+a_2+a_3)\omega_1^2}{4\pi i} &
        \frac{(\lambda-1)a_1+\lambda a_2+(2\lambda-1)a_3}{\lambda(\lambda-1)}
    \end{pmatrix} \, ,
\end{align}
where $p(\lambda)$ is a function quadratic in the (quasi-)periods 
\begin{equation}
    p(\lambda)=\frac{1}{8\pi i}\left[\frac{a_2+a_3}{\lambda-1}\omega_1^2-\frac{(\lambda-1)a_1+\lambda a_2+(2\lambda-1)a_3}{\lambda(\lambda-1)}\omega_1\eta_1 \right] \,.
\end{equation}

%%%%%%%%%%%%
\paragraph{Step 4: Integrating out.} We can now  achieve full $\eps$-factorization by performing a final rotation
\begin{equation}
   \bs{J}(\lambda,\eps)= \boldsymbol{R}_{\mathrm{t}}(\lambda)\bs{I}_{\eps}(\lambda,\eps)\,,
\end{equation}
where we make an ansatz for the rotation matrix of the form
\begin{equation}
    \bs{R}_{\mathrm{t}}(\lambda)=\begin{pmatrix}
        1 & 0 \\ t(\lambda) & 1
    \end{pmatrix} \,,
\end{equation}
for some function $t(\lambda)$ to be determined. Indeed, if $t(\lambda)$ satisfies the differential equation
\begin{equation}
\label{eq:tDiffEqF21}
    \dd t(\lambda)+p(\lambda)\dd\lambda=0 \,,
\end{equation}
then the final basis $\bs{J}(\lambda,\eps)$ satisfies an $\eps$-factorized differential equation
\begin{equation}
    \dd\bs{J}(\lambda,\eps)= \eps\bs{A}(\lambda)\bs{J}(\lambda,\eps) \,,
\end{equation}
with the differential equation matrix
\begin{equation}
    \bA(\lambda)=\dd\lambda \begin{pmatrix}
        \frac{4\pi i t(\lambda)}{\lambda(\lambda-1)\omega_1^2} & \frac{-4\pi i}{\lambda(\lambda-1)\omega_1^2} \\
        \frac{2(1-2\lambda)t(\lambda)}{\lambda(\lambda-1)}+\frac{3\omega_1^2}{4\pi i}+\frac{4\pi i t(\lambda)^2}{\lambda(\lambda-1)\omega_1^2} &
        \frac{(4\lambda-2)\omega_1^2-4\pi it(\lambda)}{\lambda(\lambda-1)\omega_1^2}
    \end{pmatrix} \,.
\end{equation}
In the previous equation we have let $a_i=1$ for brevity. We are then left with the task of solving the differential equation for $t(\lambda)$. This is achieved by  integrating along some path,
\begin{equation}
\label{eq:tFormalSolutionF21}
    t_{\lambda_0}(\lambda)=-\int_{\lambda_0}^{\lambda}\dd\lambda'\, p(\lambda') \,,
\end{equation}
with the freedom in the choice of the base-point $\lambda_0$ corresponding to the choice of integration constant of the differential equation. This begs the question if there is a simple closed form for $t_{\lambda_0}(\lambda)$ in terms of periods and quasi-periods. In this particular case one can indeed explicitly solve the integral, showing that $t_{\lambda_0}(\lambda)$ can be expressed as a rational function in $\lambda$ multiplied by $\omega_1^2$ \cite{Broedel:2021zij}. We will now show how $t_{\lambda_0}(\lambda)$ can actually be found by solving only \emph{algebraic equations}, which will be very helpful when we study multivariable examples.

Our key tool is the fact that the intersection matrix should be constant when evaluated in the canonical basis~\cite{Duhr:2024xsy}. For maximal cuts (or hypergeometric functions), on which we are focusing, we even know that in the canonical basis, the intersection matrix takes the form,
\begin{equation}\label{eq:C_J}
    \bs{C}_{\bs{J}}=f(\eps)\bs{\Delta}\,,
\end{equation}
for some rational function $f(\eps)$ and a matrix of rational numbers $\bs{\Delta}$. In practice, we compute the intersection matrix $\bs{C}_{0}$ in the original basis $\bI_0$. In our case, this is a simple algebraic task, (see appendix \ref{app:Intersect}). The intersection matrix in the canonical basis $\bs{J}$, related to the original basis through the rotation $\bs{J}(\lambda,\eps)=\bs{{R}}(\lambda,\eps)\bs{I}_0(\lambda,\eps)$, is then given by
\begin{equation}
    \bC_{\bs{J}}(\lambda,\eps)=\bs{{R}}(\lambda,\eps)\bs{C}_{0}\bs{{R}}^{T}(\lambda,-\eps) \,.
\end{equation}
If we now explicitly insert the rotation matrix constructed above
\begin{equation}
    \bs{{R}}(\lambda,\eps)=\bs{{R}}_{t}(\lambda)\bs{{R}}_{\eps}(\eps)\bs{{R}}_{\mathrm{ss}}(\lambda)\bs{{R}}_{\dd}(\lambda,\eps) \,,
\end{equation}
we indeed find that the intersection matrix $\bs{C}_{\bs{J}}$, which depends on $t_{\lambda_0}(\lambda)$, takes the simple form
\begin{equation}
    \bs{C}_{\bs{J}}=\frac{\eps}{8\pi^2}\begin{pmatrix}
        0 & 1 \\ 1 &
        2t_{\lambda_0}(\lambda)-\frac{((\lambda-1)a_1+\lambda a_2+(2\lambda-1)a_3)\omega_1^2}{4\pi i}
    \end{pmatrix} \,.
\end{equation}
This allows us to read off the function $f(\eps)$ from the entries not depending on the function $t_{\lambda_0}(\lambda)$, yielding
\begin{equation}
    f(\eps)=\frac{\eps}{\pi^2} \,.
\end{equation}
Requiring that the intersection matrix $\bs{C}_{\bs{J}}$ takes the form in~\eqref{eq:C_J} for some constant matrix $\bs{\Delta}$ imposes an algebraic equation for $t_{\lambda_0}(\lambda)$, which can be easily solved by
\begin{equation}
\label{eq:tSolutionF21}
    t_{\lambda_0}(\lambda)=\frac{1}{8\pi i}[(\lambda-1)a_1+\lambda a_2+(2\lambda-1)a_3]\omega_1^2+q \,.
\end{equation}
Here $q$ is an undetermined rational number. Since it appears as an additive constant, it can be interpreted as the integration constant of the differential equation \eqref{eq:tDiffEqF21} or, equivalently, as the choice of base-point $\lambda_0$ in the formal solution \eqref{eq:tFormalSolutionF21}. Hence, we have the freedom to set this number to a convenient value, which we choose to be $q=0$. It is easy to check that the function \eqref{eq:tSolutionF21} indeed solves the differential equation \eqref{eq:tDiffEqF21}, and we have found the $\eps$-factorized differential equation
\begin{equation}
    \dd\bs{J}(\lambda,\eps)=\eps\bs{A}(\lambda)\bs{J}(\lambda,\eps) \,,
\end{equation}
with
\begin{equation}
\label{eq:canonicalDEMatrix2F1}
    \bs{A}(\lambda)=\dd\lambda\begin{pmatrix}
        \frac{(\lambda-1)a_1+\lambda a_2+(2\lambda-1) a_3}{2\lambda(\lambda-1)} & -\frac{4\pi i}{\lambda(\lambda-1)\omega_1^2} \\
        -\frac{((\lambda-1)^2a_1^2+2(\lambda-1)(\lambda a_2-a_3)a_1+(\lambda a_2+a_3)^2)\omega_1^2}{16\pi i\lambda(\lambda-1)} & \frac{(\lambda-1)a_1+\lambda a_2+(2\lambda-1)a_3}{2\lambda(\lambda-1)}
    \end{pmatrix} \,.
\end{equation}
After solving the algebraic equation for $t_{\lambda_0}(\lambda)$ we can in particular find the constant matrix $\bDelta$, which takes the simple form
\begin{equation}
    \bDelta=\frac{1}{8}\bK_2 \,,
\end{equation}
where 
\begin{equation}
    \bK_N=\left(\begin{smallmatrix}
    0 & 0 & \ldots & 0& 1\\
    0 & 0 &  & 1& 0\\
    \vdots &  & {.^{.^{.^{.^.}}}} & & \vdots\\
    \phantom{.}&&&&\\
    0 & 1 & \ldots & 0& 0\\
    1 & 0 &  & 0& 0\\
    \end{smallmatrix}\right) \,,
\end{equation}
is the so-called \emph{exchange matrix}.

Note that the differential equation matrix \eqref{eq:canonicalDEMatrix2F1} only contains the period $\omega_1$, i.e., the quasi-period $\eta_1$ has fully dropped out. This has an important consequence, which is revealed by changing variables from $\lambda$ to the modular parameter $\tau=\omega_2/\omega_1$. First, it is easy to show, using the differential equation for $\omega_1,\omega_2$, that
\begin{equation}
    \dd\tau=-\frac{4\pi i\dd\lambda}{\lambda(1-\lambda)\omega_1^2}\,.
\end{equation}
This is precisely the upper right entry of the differential equation matrix. Furthermore, this gives the Jacobian of the change of variables, and so we can use $\tau$ as the variable in equation \eqref{eq:canonicalDEMatrix2F1}, interpreting $\lambda$ as a function $\lambda(\tau)$, known as the modular lambda function, which can be expressed in terms of (even) Jacobi theta functions \cite{enolskiRichterThomae}
\begin{equation}
    \lambda(\tau)=\frac{\theta_3(\tau)^4}{\theta_2(\tau)^4}\,.
\end{equation}
Also the period $\omega_1$ can be written as a function of the modular parameter in terms of a Jacobi theta function (cf.,~e.g.,~\cite{enolskiRichterThomae,eilersRosenhain})
\begin{equation}
    \omega_1=2\pi \theta_2(\tau)^2 \,.
\end{equation}
Understanding the above differential equation in terms of $\tau$ reveals that the entries are actually modular forms for the congruence subgroup $\Gamma(2)$, with
\begin{equation}
\Gamma(N) = \big\{\gamma\in\textrm{SL}_2(\mathbb{Z}): \gamma = \mathds{1}\!\!\mod N\big\}\,.
\end{equation}
Schematically the differential equation matrix takes the form
\begin{equation}
\label{eq:A(tau)}
    \bs{A}(\tau)=\dd\tau\begin{pmatrix}
        f_2(\tau) & 1 \\
        f_4(\tau) & f_2(\tau)
    \end{pmatrix}\,,
\end{equation}
where $f_2(\tau)$ and $f_4(\tau)$ are modular forms of weights 2 and 4 for $\Gamma(2)$ respectively. This structure can be seen directly from equation \eqref{eq:canonicalDEMatrix2F1}, by realizing that the (squared) period $\omega_1^2$ is a modular form of weight 2 for $\Gamma(2)$.\footnote{The period itself is also a modular form (of weight one), however not for the congruence subgroup $\Gamma(2)$ but for $\Gamma(4)$ \cite{Broedel:2018rwm}.} 

\paragraph{Discussion.} In this section we have derived the canonical form for the differential equation of Gauß' hypergeometric function. While this result is not new, it has allowed us to review the algorithm of~\cite{Gorges:2023zgv}. This algorithm is not only our starting point to understand the higher-genus cases, but it has also allowed us to show how we can use the expected constancy of the intersection matrix in a canonical basis~\cite{Duhr:2024xsy} in a constructive way.

At this point, however, we need to address an important issue. Currently, there is no general consensus for a definition of a canonical form for differential equations associated with geometries beyond the Riemann sphere. Indeed, the factorization of the dimensional regulator $\varepsilon$ alone does not appear to provide a good criterion, cf., e.g., the discussion in~\cite{Frellesvig:2023iwr}. Since one of the goals of this paper is to show that the algorithm of~\cite{Gorges:2023zgv} can be extended beyond genus one, it is important to identify the properties one expects from any purported canonical form. In the following we collect some empirical properties that all commonly accepted canonical differential equations satisfy (in the case where the geometry is a Riemann sphere, these properties are trivially satisfied). We focus on properties that arise in the example of Gauß' hypergeometric function, because it is the natural framework to discuss the generalization to higher genus.
\begin{itemize}
    \item It is $\varepsilon$-factorized.
    \item The intersection matrix is constant.
    \item After changing variables from $\lambda$ to $\tau$, the matrix $\bs{A}(\tau)$ only involves modular forms. In particular, all contributions from quasi-periods and/or the $b$-period have dropped out. It is known that more complicated examples, for example the three-loop equal-mass banana integral~\cite{Pogel:2022yat}, involve new functions that are integrals over modular forms. The latter do not transform like modular forms. 
    \item There is a grading by the weight of the modular forms. In particular, the entry in the upper right corner of $\bs{A}(\tau)$ is simply $\dd\tau$, which has modular weight $-2$, and if we move along the diagonals, the modular weight increases by 2 (cf.~\eqref{eq:A(tau)}). In cases where new functions appear, the modular weight is given by the modular weights of the functions in the integrands, cf.~\cite{Pogel:2022yat}.
    \item The matrix $\bs{A}(\lambda)$ only has simple poles in $\lambda$. 
\end{itemize}
Currently, it is not known if these properties will hold more generally.
In the remainder of this paper, we will show that we can find a basis for hyperelliptic extensions of Gauß' hypergeometric function which still has all the aforementioned properties.

%% file: Sec_Hyperelliptic.tex
After introducing the maximal cut of the non-planar crossed box integral as motivation, we will define in subsection \ref{subsec_Lauricella} a specific set of hypergeometric functions that serve as a prototype for the maximal cuts of hyperelliptic Feynman integrals such as the non-planar crossed box. We then review some basic facts about the geometry of hyperelliptic curves in subsection \ref{subsec_hyper}, which will lay the groundwork for the next sections.

\subsection{(Hyperelliptic) Lauricella Functions}
\label{subsec_Lauricella}

In the previous section we studied Gauß' hypergeometric function, which is a prototype for the maximal cut of an elliptic Feynman integral. Since the main motivation for this paper are hyperelliptic Feynman integrals, we start this section by introducing the maximal cut of the non-planar crossed box as an example of a Feynman integral associated with a hyperelliptic curve.
\begin{example}[The non-planar crossed box]
\label{examplenpcb}
The prime example of a hyperelliptic Feynman integral is the equal-mass non-planar crossed box, whose graph is depicted in figure \ref{fig.npcb}.
\begin{figure}[H]
    \centering
    \includegraphics[align=c, scale=.3]{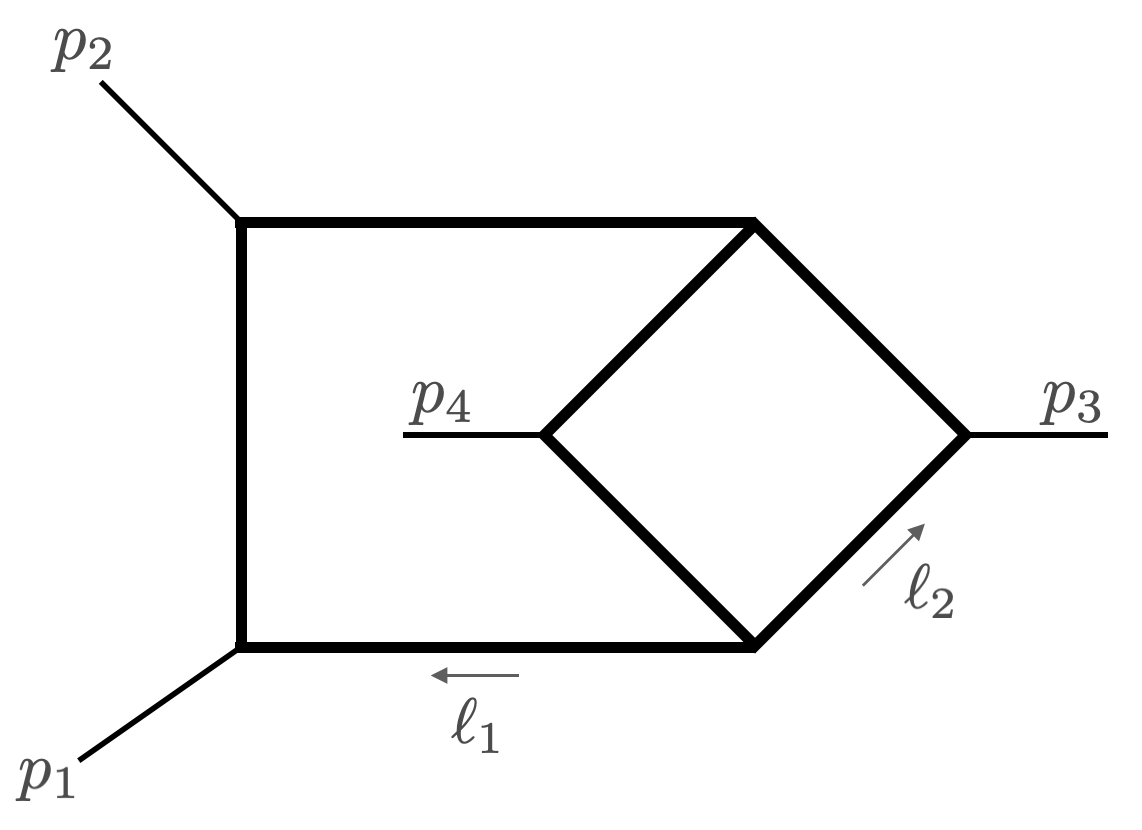}
 \caption{The non-planar crossed box.}
    \label{fig.npcb}
\end{figure}
It is given by the integral 
\begin{align}
\label{npcbintegral}
    I_{\bs{\nu}}^{\rm{npcb}}(s,t,m^2) =e^{2\gamma_E\varepsilon} \int\frac{\dd^D \ell_1}{i\pi^{\frac{D}{2}}}\int\frac{\dd^D\ell_2}{i\pi^{\frac{D}{2}}}\prod_{i=1}^{7} \frac{1}{D_i^{\nu_i}} \,,
\end{align}
with  propagators 
\begin{align}
    D_1&=\ell_1^2-m^2, \quad D_2=(\ell_1-p_1)^2-m^2,\quad D_3=(\ell_1-p_1-p_2)^2-m^2 \,, \nonumber\\
    D_4&=\ell_2^2-m^2,\quad D_5=(\ell_2-p_3)^2-m^2,\quad D_6=(\ell_1+\ell_2)^2-m^2\,, \\
    D_7&=(\ell_1+\ell_2-p_1-p_2-p_3)^2-m^2\, .  \nonumber
\end{align}
We consider all external particles to be massless, i.e., $p_i^2=0$. The integral only depends on two dimensionless variables $s=(p_1+p_2)^2,\,t=(p_1+p_3)^2$, where we set $m^2=1$ without loss of generality. Computed in the loop-by-loop approach, the maximal cuts of the non-planar crossed box take the form 
\begin{align}
\label{npcbmaxcut}
P_{ij}^{\text{npcb}}\sim\int_{\gamma_i} \left[(x-\lambda_1)(x-\lambda_2)\right]^{-\frac{1}{2}} \left[(x-\lambda_3)(x-\lambda_4)(x-\lambda_5)(x-\lambda_6)\right]^{-\frac{1}{2}-\varepsilon} \varphi_j\, ,
\end{align}
where $\varphi_j$ is a single-valued form with poles at most at the $\lambda_i$. The six branch points $\lambda_i$ are 
\begin{align}
\nonumber    \lambda_1 &=-\frac{1}{4}\left(s+\sqrt{s(s-4)}\right), &\lambda_2=-\frac{1}{4} \left(s-\sqrt{s(s-4)}\right) \,,\\
    \lambda_3&=-\frac{1}{4} \left(s+\sqrt{s(s+12)}\right), &\lambda_4= -\frac{1}{4} \left(s-\sqrt{s(s+12)}\right)\,,\\
\nonumber    \lambda_5&=-\frac{s(s+t)+2\sqrt{s^2t+st^2}}{2s},\ &\lambda_6=-\frac{s(s+t)-2\sqrt{s^2t+st^2}}{2s}\, . 
\end{align} 
In the limit $\varepsilon\rightarrow 0$, the integrand of (\ref{npcbmaxcut}) defines a hyperelliptic curve of genus two
\begin{align}
    y^2=(x-\lambda_1)(x-\lambda_2)(x-\lambda_3)(x-\lambda_4)(x-\lambda_5)(x-\lambda_6)\, , 
\end{align}
as discussed in \cite{Huang:2013kh,Marzucca:2023gto}. 
A discussion of  this maximal cut in the context of intersection theory and in particular the bilinear relations it fulfills, can be found in \cite{duhr2024twistedriemannbilinearrelations}.
\end{example}

The maximal cuts of the non-planar crossed box are special instances of Lauricella $F_D$ functions. This motivates the study of these hypergeometric functions as prototypes for the maximal cuts of general hyperelliptic Feynman integrals. More generally, Lauricella $F_D$ functions have the integral representation
\begin{align}
\label{EQU.108} 
F_{D}^{(r)}(\alpha,\bs{\beta},\gamma,\bs{\lambda})= \frac{\Gamma(\gamma)}{\Gamma(\alpha)\Gamma(\gamma-\alpha)}\int_{0}^1x^{\alpha} (1-x)^{\gamma-\alpha}
 \prod_{j=1}^{r} (1-\lambda_{j} x)^{-\beta_j}\frac{ \dd x}{x(1-x)}\, . 
\end{align} 
We consider a particular normalization and parameter choice of these integrals
\begin{align}
\label{lauricellaintegral}
    \mathcal{L}_{\bs{\nu}}(\bs{\lambda})=&\int_{\lambda_n}^{\infty}\dd x\,x^{-\frac{1}{2}+\nu_1+a_1\eps}(x-1)^{-\frac{1}{2}+\nu_2+a_2\eps}(x-\lambda_1)^{-\frac{1}{2}+\nu_3+a_3\eps}\dots(x-\lambda_n)^{-\frac{1}{2}+\nu_{n+2}+a_{n+2}\eps} \\
    =& \frac{\Gamma\left(\frac{1}{2}+\nu_1+\alpha_1\eps\right)\Gamma\left(\nu_2-\nu_1+(\alpha_2-\alpha_1\right)\eps)}{\Gamma\left(\frac{1}{2}+\nu_2+\alpha_2\eps\right)} \notag\\
 \nonumber &  \times F_D^{(n)}\left(\frac{1}{2}+\nu_1+\alpha_1\eps,\frac{1}{2}-\nu_{3}-\alpha_{3}\eps,\dots,\frac{1}{2}-\nu_{n+2}-\alpha_{n+2}\eps,\frac{1}{2}+\nu_2+\alpha_2\eps,\bs{\lambda}\right)\, ,
\end{align}
such that taking $\nu_i=0$ and also setting $\eps\rightarrow 0$, we obtain the polynomial equation
\begin{align}
\label{eq.gencurvefromla}
    y^2= x(x-1)(x-\lambda_1)\dots(x-\lambda_n) \,.
\end{align}
%This equation defines a hyperelliptic curve of genus
%\begin{align}
%    g=\begin{cases}
%      \frac{n}{2}  & \text{ if } n \text{ even} \,,\\
%        \frac{n+1}{2}  & \text{ if } n \text{ odd} \,.
    %\end{cases}
%\end{align}
Interpreting the Lauricella functions as periods of a twisted cohomology group, we can read off the twist
\begin{align}
\label{lauricellatwist}
    \Phi= \,x^{-\frac{1}{2}+a_1\eps}(x-1)^{-\frac{1}{2}+a_2\eps}(x-\lambda_1)^{-\frac{1}{2}+a_3\eps}\dots(x-\lambda_n)^{-\frac{1}{2}+a_{n+2}\eps}. 
\end{align}
Then one can consider the space 
\begin{align}
    X=\mathbb{C}\mathbb{P}- \{0,1,\lambda_1,\dots, \lambda_n,\infty\} \,,
\end{align}
and define the cohomology group $H_{\text{dR}}^1(X,\nabla_\Phi)$, the homology group $H_1(X,\check{\mathcal{L}}_\Phi)$ and their duals. After choosing bases for these spaces we can also associate a twisted period matrix and a cohomology intersection matrix to the respective family of Lauricella functions, as explained in appendix \ref{app:Intersect}. 

\subsection{Geometry of Hyperelliptic Curves}
\label{subsec_hyper}

Motivated by the example of the non-planar crossed box and the Lauricella function appearing in its maximal cut, let us review some basic facts about the geometry of hyperelliptic curves. We will focus on the aspects relevant to the rest of the paper and refer to the literature \cite{farkasKra,Bobenko2011,fayThetaFunctions,tataTheta2} for further details (see also \cite{DHoker:2022xxg,Baune:2024biq} for recent reviews in the physics literature).

A hyperelliptic curve $\Sigma_{\blambda}$ with branch points $\blambda=(\lambda_1,\dots ,\lambda_{n+2})$ can be defined as the set of points $(x,y)$ that satisfy an equation of the form 
\begin{equation}
\label{eq_hypcurve}
    y^2=P_{n+2}(x)=\prod_{i=1}^{n+2}(x-\lambda_i) \,,
\end{equation}
where
\begin{align}
    g=\begin{cases}
      \frac{n}{2}  & \text{ if } n \text{ even} \,,\\
        \frac{n+1}{2}  & \text{ if } n \text{ odd} \,,
    \end{cases}
\end{align}
is the genus of the curve. We refer to a hyperelliptic curve as \emph{even} or \emph{odd} according to whether $n$ is even or odd. Note that a hyperelliptic curve of genus one is just an elliptic curve. As in (\ref{eq.gencurvefromla}) we set $\lambda_1=0$ and $\lambda_2=1$ without loss of generality. Note that the integrals (\ref{lauricellaintegral}) indeed define hyperelliptic curves via the polynomial equation (\ref{eq.gencurvefromla}). Hyperelliptic curves define special instances of Riemann surfaces. We stress that any Riemann surface of genus two is hyperelliptic, and so focussing on hyperelliptic curves is only a restriction for $g\geq 3$.
We will consider both even and odd hyperelliptic curves of genus two in the subsequent sections. Therefore, we introduce both the even and the odd cases side-by-side. Most of the discussion equally applies to both cases.  

\subsubsection{Canonical Cycles for a Hyperelliptic Curve}
\begin{figure}[H]
    \centering
    \includegraphics[align=c, scale=.25]{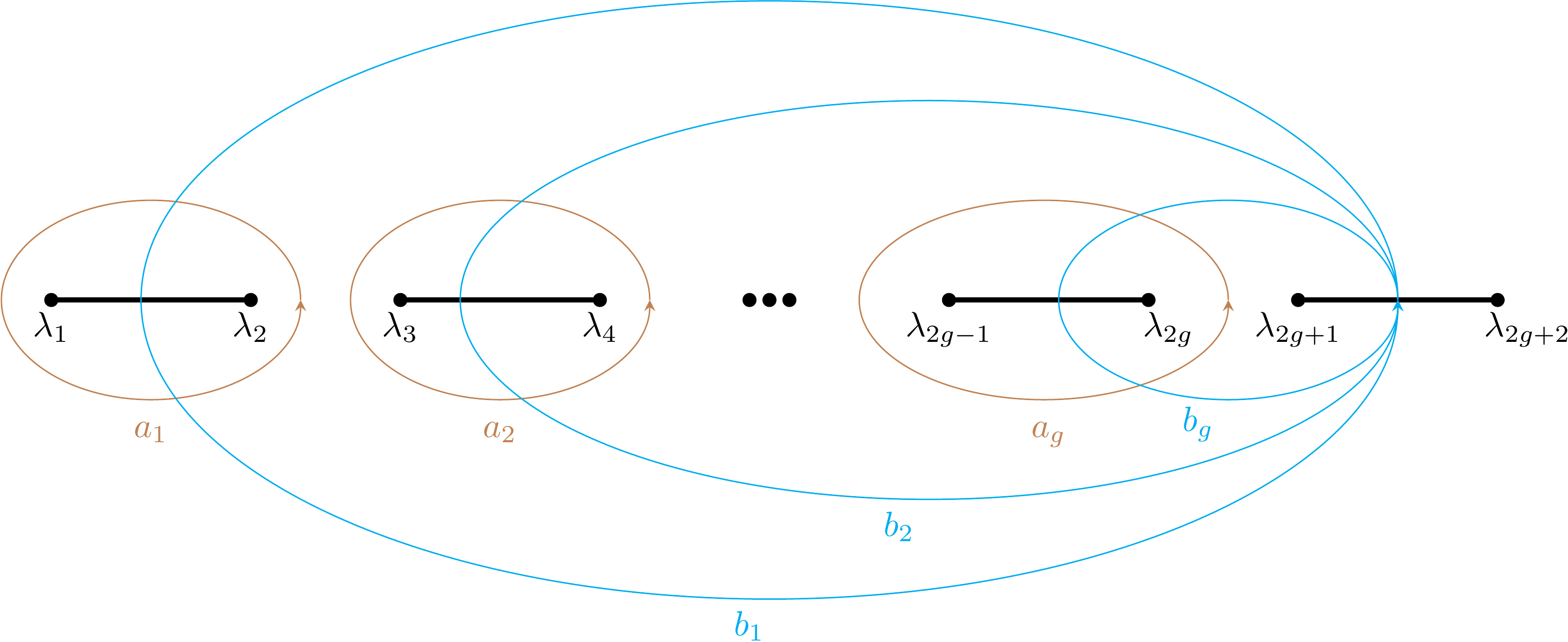}
 \caption{A canonical basis of cycles for an even hyperelliptic curve of genus $g$.}
    \label{fig.fig3}
\end{figure}

The equation defining a hyperelliptic curve describes two Riemann sheets which are glued together along $g+1$ branch cuts, which we choose to be
\begin{align}
\label{eq_branches}  
 [\lambda_1,\lambda_2],~[\lambda_3,\lambda_4],\dots,[\lambda_{2g-1},\lambda_{2g}]\text{ and }\begin{cases}
     [\lambda_{2g+1},\lambda_{2g+2}]\,,&\text{ for } n \text{ even}\,, \\    [\lambda_{2g+1},\infty]\,,&\text{ for } n \text{ odd} \,.
 \end{cases}
\end{align}
The resulting surface is topologically a torus with $g$ handles. The presence of the branch cuts implies the existence of non-trivial cycles on the hyperelliptic curve, and hence a non-trivial first homology group $H_1(\Sigma_{\bs{\lambda}},\mathbb{Z})$. A canonical basis for this homology group of a hyperelliptic curve of genus $g$ can be split into $g$ $a$-cycles, denoted by $a_i$, and $g$ $b$-cycles, denoted by $b_i$. 
Here, canonical means that the topological intersection numbers of the cycles, captured by the intersection pairing $[\bullet |\bullet]$, take the following simple (symplectic) form,
\begin{equation}
    [a_i|a_j]=0,\qquad [b_i|b_j]=0,\qquad [a_i|b_j]=-[b_i|a_j]=\delta_{ij} \,.
\end{equation}
Such a basis of cycles for an even hyperelliptic curve of genus $g$ is shown in figure \ref{fig.fig3}.  The figure looks the same for an odd hyperelliptic curve of genus $g$, with $\lambda_{2g+2}$ replaced by $\infty$. 
Note that for a \emph{punctured} Riemann surface, one needs to add  cycles encircling each of the punctures. Let us next consider the Abelian differentials, i.e., the cohomology group of the hyperelliptic curve.

%%%%%%%%%%%

\subsubsection{Abelian Differentials on the Hyperelliptic Curve}

We can classify all rational differentials (in the coordinates  $x,y$) on the hyperelliptic curve into three kinds of so-called \emph{Abelian differentials}:
\begin{itemize}
\item Abelian differentials of the \textit{first kind} are holomorphic differentials.
\item Abelian differentials of the \textit{second kind} are meromorphic differentials with vanishing residues at the poles.
\item Abelian differentials of the \textit{third kind} are meromorphic differentials with non-vanishing residues. 
\end{itemize}
Since the pole order and the residues of an Abelian differential depend on whether we consider an even or an odd hyperelliptic curve, we need to distinguish these cases for the construction of the Abelian differentials of second and third kinds. 
We will first explain the conceptual points on general grounds and then explicitly show the differences in a genus two example. Details on the expansions of basic differentials can be found in appendix \ref{app_expansions}. 
%%%%%%%%%%%%%%%%%%%%%%%%%%%%%%%%%%%%

\paragraph{Abelian Differentials of the First Kind.} 
The $g$ one-forms 
\begin{equation}
\label{eq_difffirst}
\varpi_i= \frac{x^{i-1}\dd x}{y}\,,\qquad 1\le i\le g \,, 
\end{equation}
are holomorphic and form a basis for the (cohomology classes of) Abelian differentials of the first kind on the hyperelliptic curve. This is independent of whether the curve is even or odd.
\paragraph{Abelian Differentials of the Second Kind.}
For both even and the odd hyperelliptic curves, the basis of (cohomology classes of) Abelian differentials of the second kind is also $g$-dimensional. These differentials generally take the form
\begin{align}
   \frac{\Psi_i(x)}{y}\dd x\,, \text{~~~with~~~} \Psi_i(x) =\sum_{j=1}^{2g}c_j x^j\,,
\end{align}
for some coefficients $c_j$ chosen such that the residue at infinity vanishes. We need to make $g$ linearly independent choices to form a basis, and we will call these differentials $\varpi_{g+1},\dots \varpi_{2g}$. The explicit linear combinations depend on whether the hyperelliptic curve is even or odd, due to the different pole structure of the respective forms (see appendix \ref{app_expansions}). There is considerable freedom in choosing a basis of second kind differentials. Note that for an even hyperelliptic curve the differentials $\tfrac{x^i\dd x}{y}$ for $i\geq g$ have a residue at infinity which we need to make sure to cancel, while this is automatic for an odd curve. Additionally, we can add any linear combination of first kind differentials to any choice of a second kind differential basis element.  We generally choose them in such a way that the generalized Legendre relations take a particularly simple form, see, e.g., \eqref{eq:genus2RiemannBilinearsMatrix}. 
\paragraph{Abelian Differentials of the Third Kind.} For hyperelliptic curves with punctures, we need to add differentials with non-vanishing residues at these punctures to our basis of Abelian differentials. Note that an even hyperelliptic curve  always has a puncture at $\infty$, and so we need to add at least one third kind differential to the basis. We will choose this differential to be
\begin{align}
   \varpi_{2g+1}= \frac{x^g\dd x}{y} \,,
\end{align} 
which indeed has simple poles (with residues of opposite sign) at the infinities on the two Riemann sheets. If there are additional punctures (for both even and the odd hyperelliptic curves), we need to add differentials of the form
\begin{equation}
\varpi^{c}=\frac{\dd x}{y(x-c)} \,,
\end{equation}
to the basis. These have simple poles at $(x,y)=(c,\pm y_c)$ with $y_c=\sqrt{P_g(c)}$ for a fixed choice of branch.

\paragraph{Summary.}
The basis of Abelian differentials for a hyperelliptic curve of genus $g$ with punctures at the points $\{c_1,c_2,\dots\}$ has the following elements: 
\begin{align}\label{alldiff}
   \underbracket[0.5pt]{\underbracket[0.5pt]{ \overbracket[0.5pt]{\varpi_1,\dots,\varpi_g}^{\text{first kind}},  \overbracket[0.5pt]{\varpi_{g+1},\dots, \varpi_{2g}}^{\text{second kind}}}_{\text{odd curve}},\overbracket[0.5pt]{\varpi_{2g+1}}^{\text{third kind}}}_{\text{even curve}},\underbracket[0.5pt]{\overbracket[0.5pt]{\varpi^{c_1},\varpi^{c_2},\dots}^{\text{third kind}}}_{\text{ punctures $c_i$}} \,.
\end{align}
Let us illustrate this in some simple examples
\begin{example}[Abelian Differentials at Genus \textit{One}]
At genus one, for an \emph{odd} curve, we can choose a basis of Abelian differentials made up of one first kind and one second kind differential
\begin{align}
\label{abelianfirstelliptic}
    \frac{\dd x}{y}\,,\quad \frac{x\dd x}{y}\, .
\end{align}
\end{example}
\begin{example}[Abelian Differentials for \textit{Odd} Hyperelliptic Curves of Genus \textit{Two}]\label{exdiff2}
For a genus two hyperelliptic curve we need two first and two second kind differentials. A suitable choice for an odd curve is given by
\begin{equation}
    \frac{\dd x}{y}\,,\quad\frac{x\dd x}{y}\,,\quad \frac{\Psi_1^{2,\mathrm{o}}(x)\dd x}{y}\,,\quad\frac{\Psi_2^{2,\mathrm{o}}(x)\dd x}{y} \,,
\end{equation}
with the polynomials 
\begin{align}
 \label{oddgenus2basis}
\Psi^{2,\text{o}}_1(x)=\left(3x^3-2\tilde{s}_1 x^2+\tilde{s}_2 x\right)\text{~~~and~~~}\Psi_{2}^{2,\text{o}}(x)= x^2\, . 
\end{align}
Here we denote by $\st_k$ the $k^{\textrm{th}}$ elementary symmetric polynomial in the branch points $\lambda_i$ (including $\lambda_1=0,\lambda_2=1$), i.e., $\st_1=\sum_{i=1}^{n+2} \lambda_i, \,\st_2=\sum_{i,j=1, i< j}^{n+2}\lambda_i \lambda_j,$ etc.
\end{example}
\begin{example}[Abelian Differentials for \textit{Even} Hyperelliptic Curves of Genus \textit{Two} ] 
\label{exdiff4}
For an even hyperelliptic curve of genus two, we need an additional third kind differential due to the additional puncture at infinity. We hence have a five-dimensional basis of Abelian differentials  
\begin{equation}
    \frac{\dd x}{y}\,,\quad\frac{x\dd x}{y}\,,\quad\frac{\Psi_1^{2,\mathrm{e}}(x)\dd x}{y}\,,\quad\frac{\Psi_2^{2,\mathrm{e}}(x)\dd x}{y}\,,\quad\frac{x^2\dd x}{y}  \,,
\end{equation}
with a suitable choice for the the polynomials given by
\begin{align}
 \label{evengenus2basis}
\Psi_{1}^{2,\text{e}}(x)= 4\left(x^4-\frac{3}{4}\st_1 x^3+\frac{1}{2}\st_2 x^2-\frac{1}{4}\st_3 x\right) \text{~~~and~~~} \Psi^{2,\text{e}}_2(x)= 2\left(x^3-\frac{1}{2}\st_1 x^2\right) \, . 
 \end{align}
\end{example}
The choices (\ref{oddgenus2basis}) and (\ref{evengenus2basis}) for the second kind differentials will lead to particularly simple quadratic relations between the periods and quasi-periods, see, e.g., \eqref{eq:genus2RiemannBilinearsMatrix}.

%%%%%%%%%%%%%%%%%%%%%%%%%%%%%%%%%%%%
%%%%%%%%%%%%%%%%%%%%%%%%%%%%%%%%%%%%

\subsubsection{Period Matrix and Bilinear Relations} 

After introducing the canonical $a$ and $b$ cycles as a basis for the homology group and the Abelian differentials as a basis for the first cohomology group, we can pair the two to form the period matrix of the hyperelliptic curve.
Integrating the basis elements over the basis cycles, we obtain a period matrix of the form  
\begin{equation}
\label{eq:periodmatrixHyperelliptic}
   \bs{\mathcal{P}} =\left(\int_{\gamma_j} \varpi_i\right)=\begin{pmatrix} \bs{\mathcal{A}} & \bs{\mathcal{B}}& \star \\ \bs{\tilde{\mathcal{A}}} & \bs{\tilde{\mathcal{B}}}& \star  \\
     \star  &  \star  &  \star \end{pmatrix}\, .
\end{equation}
The integrals of the first kind differentials over the $a$ or $b$-cycles are captured by the \emph{$a$- and $b$-period matrices},
\begin{equation}
    \mathcal{A}_{ij}= \int_{a_j} \frac{x^{i-1}\dd x}{y}\,, \qquad \mathcal{B}_{ij}=\int_{b_j} \frac{x^{i-1}\dd x}{y}\,,\qquad i,j=1,\dots ,g \,.
\end{equation}
The integrals of the second kind differentials are captured by the \emph{$a$- and $b$-quasi-period matrices},
\begin{equation}
    \tilde{\mathcal{A}}_{ij}=\int_{a_j}\frac{\dd x\,{\Psi}_i(x)}{y} \,, \qquad 
\tilde{\mathcal{B}}_{ij}=\int_{b_j}\frac{\dd x\,{\Psi}_i(x)}{y}\, ,\qquad i,j=1,\dots ,g \,. 
\end{equation}
The entries of the (quasi-)period matrices can be expressed in terms of Lauricella $D$ functions. The entries represented by $\star$ in \eqref{eq:periodmatrixHyperelliptic} arise from integrals over the additional integration cycles due to punctures. This includes the puncture at $\infty$, that is always present for an even hyperelliptic curve. In the remainder of this section we focus on the upper left blocks consisting of the $a$- and $b$-(quasi)-period matrices. For an odd hyperelliptic curve with no punctures, these blocks make up the full period matrix. 

Note that the only substantial difference between the familiar elliptic case and hyperelliptic curves of higher genus is that the (quasi)-periods are matrices for $g>1$. The same is true for another important quantity, the \emph{(normalized) period matrix}, obtained as a \enquote{quotient} of the $a$- and $b$-period matrices
\begin{equation}
    \bs{\Omega}=\bs{\mathcal{A}}^{-1}\bs{\mathcal{B}}\,.
\end{equation}
The normalized period matrix is  a symmetric $g\times g$ matrix with positive definite imaginary part, generalizing the modular parameter $\tau$ of an elliptic curve to higher genus.

The (quasi-)period matrices defined above fulfill quadratic relations. These follow from the classical Riemann bilinear relations
\begin{equation}
\label{bilrel}\sum_{i=1}^2\left[\int_{a_i}\omega\int_{b_i}\eta- \int_{b_i}\omega\int_{a_i}\eta\right]=\int_{\partial \mathcal{C}}f \eta\, ,
\end{equation}
where $\omega,\eta$ are one-forms and $f$ is a primitive of $\omega$, i.e.,~$\mathrm{d}f=\omega$, see, e.g.,~\cite{2012arXiv1208.0990B} for details. The integral on the right-hand side is over the boundary of the fundamental domain $\mathcal{C}$ (obtained by cutting the hyperelliptic curve along all homology cycles) and can be interpreted as a sum of residues on the hyperelliptic curve. 
Alternatively, one can also derive these relations by taking the $\varepsilon\rightarrow 0$ limit of the twisted Riemann bilinear relations~\cite{duhr2024twistedriemannbilinearrelations} (see appendix~\ref{app:Intersect}).

\begin{example}[Period Matrix and Bilinear Relations for Hyperelliptic Curves of Genus \emph{One}]
\label{exampleellipticlegendreeasy}
For an elliptic curve, the $a$- and $b$-periods are often denoted by 
\begin{align}
    \omega_1= \int_{a} \frac{\dd x}{y}\,,\qquad \omega_2=\int_b \frac{\dd x}{y}\, , 
\end{align}
which is also the convention we used in (\ref{eq:ellipticperiodmatrix}). The normalized period matrix is just a scalar, which is usually referred to as \textit{modular parameter}
\begin{align}
    \tau= \frac{\omega_2}{\omega_1}\, . 
\end{align}
The quasi-periods can be defined as
\begin{align}
    \eta_1= \int_{a} \frac{x\, \dd x}{y}\,, \qquad \eta_2=\int_b \frac{x\, \dd x}{y}\, . 
\end{align}
\end{example}
Note that we could add any multiple of the holomorphic differential to the second kind differential and still obtain a valid basis choice. The choice made here however leads to the {Legendre relation} from~\eqref{eq:legendre}.

\begin{example}[Period Matrix and Bilinear Relations for Hyperelliptic Curves of Genus \emph{Two}]
\label{exampleoddlegendreeasy}
At genus two the $a$- and $b$-periods are $2\times 2$ matrices
\begin{align}
\bs{\mathcal{A}}=\begin{pmatrix}\int_{a_1}\frac{\dd x}{y}&\int_{a_2}\frac{\dd x}{y}\\
\int_{a_1}\frac{x\dd x}{y}& \int_{a_2}\frac{x\, \dd x}{y}\end{pmatrix},\qquad
\bs{\mathcal{B}}
=\begin{pmatrix}\int_{b_1}\frac{\dd x}{y}&\int_{b_2}\frac{\dd x}{y}\\
\int_{b_1}\frac{x \dd x}{y}& \int_{b_2}\frac{x\, \dd x}{y}\end{pmatrix} \,,
\end{align}
and similarly for the quasi-period matrices $\bcAt$ and $\bcBt$. Choosing the bases of Abelian differentials as in the examples above, one can obtain the following simple relations, referred to as \emph{generalized Legendre relations}, from the Riemann bilinear relations \cite{bakerBook} (see also \cite{2012arXiv1208.0990B})
\begin{equation}
\label{eq:genus2RiemannBilinearsMatrix}
    \begin{pmatrix}
        \bcA & \bcB \\ \bcAt & \bcBt
    \end{pmatrix}\begin{pmatrix}
        \bs{0} & \mathds{1} \\ -\mathds{1} & \bs{0}
    \end{pmatrix}
    \begin{pmatrix}
        \bcA & \bcB \\ \bcAt & \bcBt
    \end{pmatrix}^T = 8\pi i
    \begin{pmatrix}
        \bs{0} & \mathds{1} \\ -\mathds{1} & \bs{0}
    \end{pmatrix} \,.
\end{equation}
More explicitly, these provide three independent quadratic relations between $2\times 2$ matrices,
\begin{align}
    \bs{\mathcal{B}}\bs{\mathcal{A}}^T-%
    \bs{\mathcal{A}}\bs{\mathcal{B}}^T&=0 \,, \label{eq:genLegendre1} \\
    \bs{\tilde{\mathcal{B}}}\bs{\tilde{\mathcal{A}}}^T-\bs{\tilde{\mathcal{A}}}\bs{\tilde{\mathcal{B}}}^T&=0 \,, \label{eq:genLegendre2}\\   
    \bs{\tilde{\mathcal{B}}}\bs{\mathcal{A}}^T-\bs{\tilde{\mathcal{A}}}\bs{\mathcal{B}}^T&=8\pi i\mathds{1}   \,. \label{eq:genLegendre3}
\end{align}
By taking the inverse of the above matrix equation \eqref{eq:genus2RiemannBilinearsMatrix} we find two additional relations
\begin{align}
\label{eq:quadraticARelation}
\bs{\mathcal{A}}^T\bs{\tilde{\mathcal{A}}}-\bs{\tilde{\mathcal{A}}}^T\bs{\mathcal{A}}&=0 \,, \\
\nonumber   \bs{\mathcal{B}}^T\bs{\tilde{\mathcal{B}}}-\bs{\tilde{\mathcal{B}}}^T\bs{\mathcal{B}}&=0 \,.
\end{align}
\end{example}

%% file: Intro_Hyperelliptic.tex
In this section, we show that the algorithm originally developed in \cite{Gorges:2023zgv} for finding a canonical form can also be applied for hyperelliptic integral families. We do this by considering Lauricella functions with three and four parameters, which are related to an even and odd hyperelliptic curve of genus two, respectively. The initial differential equations and rotation matrices for both of these examples can be found in the ancillary file.

%% file: Sec_Three.tex
Our starting point is the family of Lauricella functions     $\mathcal{L}_{\bs{\nu}}(\bs{\lambda})$ in  (\ref{lauricellaintegral}) depending on three parameters $\bs{\lambda}=(\lambda_1,\lambda_2,\lambda_3)$. For convenience, we will choose the parameters $\blambda$ to be real and ordered as $\lambda_3>\lambda_2>\lambda_1>1$.  As our initial basis of master integrals we choose
\begin{equation}
\label{eq:masterIntegrandsThree}
    \bs{I}_0=\left(\int_{\gamma}\frac{\dd x}{y}\chi(x),\int_{\gamma}\frac{x \dd x}{y}\chi(x),\int_{\gamma}\frac{ \Psi_1^{2,\text{o}}(x) \dd x}{y}\chi(x),\int_{\gamma}\frac{x^2 \dd x}{y}\chi(x)  \right)^T \,,
\end{equation}
with $\gamma$ some cycle which plays no role in the following (see the discussion in section \ref{SubSec:Elliptic}). Here, we introduced the abbreviations
\begin{equation}
    \chi(x)=x^{a_1\eps}(x-1)^{a_2\eps}(x-\lambda_1)^{a_3\eps}(x-\lambda_2)^{a_4\eps}(x-\lambda_3)^{a_5\eps} \,,
\end{equation}
and the function  $\Psi_1^{2,\text{o}}(x)$ was defined in (\ref{oddgenus2basis}). The differentials in \eqref{eq:masterIntegrandsThree} also define a basis\footnote{Explicitly, this basis is given by $\varpi_1= \dd x,\, \varpi_2=x\, \dd x \, , \varpi_3 = \Psi_1^{2,\text{o}}(x)\dd x\, , \varpi_4 = x^2\dd x$.} of the twisted cohomology group defined by the  twist $\Phi=\chi(x)/y$. We denote the initial intersection matrix obtained from this basis by $\bs{{C}}_0$ (see appendix \ref{app:Intersect} for details).  

The integrand at $\eps=0$ defines a hyperelliptic curve of genus two,
\begin{equation}
    y^2=x(x-1)(x-\lambda_1)(x-\lambda_2)(x-\lambda_3) \,.
\end{equation}
This form of the hyperelliptic curve is sometimes referred to as \emph{Rosenhain normal form}.  The integrands of the master integrals are chosen in such a way that they reduce for  $\eps=0$ to a conveniently normalized set of first and second kind Abelian differentials.  In particular, when integrated over the two $a$ and $b$ cycles, the $2\times 2$ matrices of (quasi-)periods satisfy the simple generalized Legendre relations from section \ref{Sec:background}. We can now derive the differential equation for the chosen master integrals  by applying the total differential and using integration by parts, and we obtain
\begin{equation}
    \dd \bs{I}_0(\bs{\lambda},\eps)=\bs{B}_0(\bs{\lambda},\eps)\bs{I}_0(\bs{\lambda},\eps) \,,
\end{equation}
where the differential equation matrix can be decomposed as
\begin{equation}
    \bs{B}_0(\bs{\lambda},\eps)=\bs{B}_0^{(1)}(\bs{\lambda},\eps)\dd\lambda_1+\bs{B}_0^{(2)}(\bs{\lambda},\eps)\dd\lambda_2+\bs{B}_0^{(3)}(\bs{\lambda},\eps)\dd\lambda_3 \,.
\end{equation}
The matrices $\bs{B}_0^{(i)}(\bs{\lambda},\eps)$ are  rational functions of the parameters $\bs{\lambda}$ and the dimensional regulator $\eps$. They are too large to be displayed here, but they are given in the ancillary file. 

We now construct a rotation that brings the above differential equation into canonical form, generalizing the construction of \cite{Gorges:2023zgv} to the hyperelliptic case.
A key step in constructing the basis is the decomposition of the period matrix, i.e., the Wronskian of the differential equation at $\eps=0$, into a semi-simple and a unipotent part. In the elliptic case, this was achieved through the Legendre relation. Here we can perform the decomposition in a completely analogous fashion using the generalized Legendre relation from~\eqref{eq:genLegendre3}, which is now a relation between $2\times 2$ matrices. We can then perform essentially the same splitting as in the elliptic case
\begin{equation}
    \bs{\mathcal{P}}=\begin{pmatrix}
    \bs{\mathcal{A}} & \bs{\mathcal{B}} \\ \bs{\tilde{\mathcal{A}}} & \bs{\tilde{\mathcal{B}}}\end{pmatrix}
    =
    \begin{pmatrix} \bs{\mathcal{A}} & 0 \\ \bs{\tilde{\mathcal{A}}} & 8\pi i\bs{\mathcal{A}}^{-1T}
    \end{pmatrix}
    \begin{pmatrix}
        \mathds{1} & \bs{\Omega} \\ 0 & \mathds{1}
    \end{pmatrix} 
    \equiv \bcS\,\bcU \, .
\end{equation}
We will now use this splitting to construct the rotation to the canonical basis following precisely the same steps as reviewed in section \ref{SubSec:Elliptic}.

\noindent \paragraph{Step 1:} First, we change from our initial basis, where the second kind differentials are defined to ensure nice generalized  Legendre relations, to a derivative basis, where the second kind differentials are certain first-order derivatives of the first kind differentials. We will choose the derivative basis as follows:
\begin{equation}
 \bs{I}_{\dd}=\left(   \text{\colorbox{Apricot!30}{$ \int_{\gamma}\frac{\dd x}{y}\chi(x)$}},   \text{\colorbox{GreenYellow!30}{$ \int_{\gamma}\frac{x \dd x}{y}\chi(x)$}},\sum_{i=1}^3 \partial_{\lambda_i} \,  \text{\colorbox{Apricot!30}{$\left[ \int_{\gamma}\frac{\dd x}{y}\chi(x)\right]$}},\sum_{i=1}^3\partial_{\lambda_i}\,  \text{\colorbox{GreenYellow!30}{$ \left[\int_{\gamma}\frac{x \dd x}{y}\chi(x)\right]$}} \right)\, .
\end{equation}
This choice is natural, because it preserves the \enquote{block structure} of the differential equation and is fully symmetric in the three parameters $\blambda$.\footnote{We have significant freedom in this choice. The subsequent steps are for example also valid for the choice $\partial_{\lambda_i} \frac{\chi(x)\dd x}{y}$ and $\partial_{\lambda_j} \frac{\chi(x)\dd x}{y}$ with $i\neq j$ as the new differentials of the third and fourth master integrals.} From the original differential equation it is then easy to obtain the concrete rotation matrix $\bs{R}_{\dd}(\blambda,\eps)$ such that
\begin{equation}
\bs{I}_{\dd}(\bs{\lambda},\eps)=\bs{R}_{\dd}(\blambda,\eps)    \bs{I}_0(\bs{\lambda},\eps) \,.
\end{equation}
\paragraph{Step 2:} In the next step, we rotate with the (inverse of the) semi-simple part of the period matrix. We  need to take the semi-simple part of the period matrix in the derivative basis, $\bcP_{\dd}$,
which, exactly like in the elliptic case, can be obtained as 
\begin{equation}
    \bcS_{\dd}(\blambda)=\bs{R}_{\dd}(\blambda,0)\bcS(\blambda) \,.
\end{equation}
We define a new basis by
\begin{equation}
   \bI_{\mathrm{ss}}(\blambda,\eps) =\bR_{\mathrm{ss}}(\blambda) \bI_{\dd}(\blambda,\eps)\,,
\end{equation}
with $\bR_{\mathrm{ss}}(\blambda)=\bcS_{\dd}^{-1}(\blambda)$. Just like in the elliptic case, the differential equation is now already very close to being $\eps$-factorized. Only the upper right block has an order $\eps^0$ term, and only the lower left block has an $\eps^2$ term. In other words, the new differential equation matrix takes the form,
\begin{align}
    \bs{B}_{{\rm ss}} (\lambda,\eps)
    &= \left(\dd \boldsymbol{R}_{\mathrm{ss}}\right) \boldsymbol{R}_{\mathrm{ss}}^{-1} + \boldsymbol{R}_{\mathrm{ss}}\,  \bs{B}_{\dd}(\lambda,\eps)\, \boldsymbol{R}_{\mathrm{ss}}^{-1}\\
    &=\left(\,
    \begin{array}{@{\hspace{2pt}}c@{\hspace{2pt}} @{\hspace{2pt}}c@{\hspace{2pt}}@{\hspace{2pt}}c@{\hspace{2pt}} @{\hspace{2pt}}c@{\hspace{2pt}}}
        0  &0 & \cellcolor{ourlightblue}\,   \textcolor{ourblue}{\bullet} \, \, & \cellcolor{ourlightblue}\,   \textcolor{ourblue}{\bullet} \, \,   \\ 
         0  &0 & \cellcolor{ourlightblue}\,   \textcolor{ourblue}{\bullet} \, \, & \cellcolor{ourlightblue}\,   \textcolor{ourblue}{\bullet} \, \,   \\ 
        0 &0    & 0 &0  \\
        0 &0    & 0 &0 
    \end{array}
    \,\right)+
    %%%%%%%%%% 
    \left(\begin{array}{@{\hspace{3pt}}c@{\hspace{2pt}} @{\hspace{2pt}}c@{\hspace{2pt}}@{\hspace{2pt}}c@{\hspace{2pt}} @{\hspace{2pt}}c@{\hspace{3pt}}}
     \cellcolor{ourlightblue}\,   \textcolor{ourblue}{\bullet} \, \, & \cellcolor{ourlightblue}\,   \textcolor{ourblue}{\bullet} \, \,  &0&0 \\ 
     \cellcolor{ourlightblue}\,   \textcolor{ourblue}{\bullet} \, \, & \cellcolor{ourlightblue}\,   \textcolor{ourblue}{\bullet} \, \,  &0&0 \\ 
     \cellcolor{ourlightblue}\,   \textcolor{ourblue}{\bullet} \, \, & \cellcolor{ourlightblue}\,   \textcolor{ourblue}{\bullet} \, \,  &  \cellcolor{ourlightblue}\,   \textcolor{ourblue}{\bullet} \, \, & \cellcolor{ourlightblue}\,   \textcolor{ourblue}{\bullet} \, \, \\ 
     \cellcolor{ourlightblue}\,   \textcolor{ourblue}{\bullet} \, \, & \cellcolor{ourlightblue}\,   \textcolor{ourblue}{\bullet} \, \,  &  \cellcolor{ourlightblue}\,   \textcolor{ourblue}{\bullet} \, \, & \cellcolor{ourlightblue}\,   \textcolor{ourblue}{\bullet} \, \, 
    \end{array}
    \,\right)
    %%%%%% 
    \, \eps+
    \left(\begin{array}{@{\hspace{2pt}}c@{\hspace{2pt}} @{\hspace{2pt}}c@{\hspace{2pt}}@{\hspace{2pt}}c@{\hspace{2pt}} @{\hspace{2pt}}c@{\hspace{2pt}}} 
     0&0&0&0 \\  0&0&0&0 \\ 
     \cellcolor{ourlightblue}\,   \textcolor{ourblue}{\bullet} \, \, & \cellcolor{ourlightblue}\,   \textcolor{ourblue}{\bullet} \, \, &  0&0 \\ 
     \cellcolor{ourlightblue}\,   \textcolor{ourblue}{\bullet} \, \, & \cellcolor{ourlightblue}\,   \textcolor{ourblue}{\bullet} \, \,  &   0&0 
    \end{array}
    \,\right)\, \eps^2\, , 
\end{align}
where the shaded entries \colorbox{ourlightblue}   {$\textcolor{ourblue}{\bullet}$} indicate non-zero entries of the respective $\eps$-order of the differential equation. Note that we need to use the quadratic relation from eq.~\eqref{eq:quadraticARelation} to see this $\eps$-structure analytically.
\paragraph{Step 3:} Just like in the elliptic case, The structure of the differential equation after the last rotation motivates a simple $\eps$-rescaling,
\begin{equation}
  \bI_{\varepsilon}(\blambda,\eps)  =\bs{R}_{\eps}(\eps) \bI_{\mathrm{ss}}(\blambda,\eps)\,, \qquad \bs{R}_{\eps}(\eps)=\mathrm{diag}(\eps,\eps,1,1)\,.
\end{equation}
The differential equation is now in $\eps$-triangular form. More precisely, it is fully $\eps$-factorized up to the lower left block: 
\begin{align}
    \bs{B}_{{\eps}} (\lambda,\eps)
    &= \left(\dd \boldsymbol{R}_{\eps}\right) \boldsymbol{R}_{\eps}^{-1} + \boldsymbol{R}_{\eps}\,  \bs{B}_{\mathrm{ss}}(\lambda,\eps)\, \boldsymbol{R}_{\eps}^{-1}=  \bs{B}_{{\eps}}^{(0)}(\lambda)+\bs{B}_{{\eps}}^{(1)}(\lambda)\, \eps\\
    &=\left(\,
    \begin{array}{@{\hspace{2pt}}c@{\hspace{2pt}} @{\hspace{2pt}}c@{\hspace{2pt}}@{\hspace{2pt}}c@{\hspace{2pt}} @{\hspace{2pt}}c@{\hspace{2pt}}}
        0  &0 & 0 &0   \\ 
         0  &0 & 0 &0  \\ 
       \cellcolor{ourlightblue}\,   \textcolor{ourblue}{\bullet} \, \, & \cellcolor{ourlightblue}\,   \textcolor{ourblue}{\bullet}    & 0 &0  \\
       \cellcolor{ourlightblue}\,   \textcolor{ourblue}{\bullet} \, \, & \cellcolor{ourlightblue}\,   \textcolor{ourblue}{\bullet}     & 0 &0 
    \end{array}
    \,\right)+
    %%%%%%%%%% 
    \left(\begin{array}{@{\hspace{3pt}}c@{\hspace{2pt}} @{\hspace{2pt}}c@{\hspace{2pt}}@{\hspace{2pt}}c@{\hspace{2pt}} @{\hspace{2pt}}c@{\hspace{3pt}}}
     \cellcolor{ourlightblue}\,   \textcolor{ourblue}{\bullet} \, \, & \cellcolor{ourlightblue}\,   \textcolor{ourblue}{\bullet} \, \,  & \cellcolor{ourlightblue}\,   \textcolor{ourblue}{\bullet} \, \, & \cellcolor{ourlightblue}\,   \textcolor{ourblue}{\bullet}  \\ 
     \cellcolor{ourlightblue}\,   \textcolor{ourblue}{\bullet} \, \, & \cellcolor{ourlightblue}\,   \textcolor{ourblue}{\bullet} \, \,  & \cellcolor{ourlightblue}\,   \textcolor{ourblue}{\bullet} \, \, & \cellcolor{ourlightblue}\,   \textcolor{ourblue}{\bullet}  \\ 
     \cellcolor{ourlightblue}\,   \textcolor{ourblue}{\bullet} \, \, & \cellcolor{ourlightblue}\,   \textcolor{ourblue}{\bullet} \, \,  &  \cellcolor{ourlightblue}\,   \textcolor{ourblue}{\bullet} \, \, & \cellcolor{ourlightblue}\,   \textcolor{ourblue}{\bullet} \, \, \\ 
     \cellcolor{ourlightblue}\,   \textcolor{ourblue}{\bullet} \, \, & \cellcolor{ourlightblue}\,   \textcolor{ourblue}{\bullet} \, \,  &  \cellcolor{ourlightblue}\,   \textcolor{ourblue}{\bullet} \, \, & \cellcolor{ourlightblue}\,   \textcolor{ourblue}{\bullet} \, \, 
    \end{array}
    \,\right)\, \eps\, .
\end{align} 
The entries of the differential equation matrices are: 
\begin{align}
    \bs{B}_{{\eps}}^{(0)}(\lambda)&=\begin{pmatrix}
        0 & 0 \\ \bP(\blambda) & 0
    \end{pmatrix} \,,\\
    \bs{B}_{{\eps}}^{(1)}(\lambda)&= \begin{pmatrix}
       \bcA^{-1}\bbeta_1(\blambda)\bcA & \bcA^{-1}\bbeta_2(\blambda)\bcA^{-1T} \\
       \bcA^T\bbeta_3(\blambda)\bcA & \bcA^T\bbeta_4(\blambda)\bcA^{-1T}
    \end{pmatrix} \,,
\end{align}
where $\bP(\blambda)$ is a $2\times 2$ matrix of differential one-forms, which can schematically be written as 
\begin{equation}
    \bP(\blambda)=\bcA^T\bP_1(\blambda) \bcA+\bcA^T\bP_2(\blambda) \bcAt+\bcAt^T\bP_3(\blambda) \bcA \,,
\end{equation}
with each $\bP_i(\blambda)$ a matrix of {rational} one-forms in the parameter $\blambda$. Their explicit form can be found in the ancillary file. The matrices $\bbeta_i$ are also $2\times 2$ matrices of rational one-forms and read
\begin{align}
 \nonumber   \bbeta_1&=\frac{\dd\lambda_1}{(\lambda_1-\lambda_2)(\lambda_1-\lambda_3)}\begin{pmatrix}
        3\lambda_1-2s_1-1 & 5 \\ -s_2-\lambda_1 & 5\lambda_1
    \end{pmatrix} + (\mathrm{cyclic}) \,, \\
    \bbeta_2&=\frac{4\pi i\,\dd\lambda_1}{d_1}\begin{pmatrix}
        1 & \lambda_1 \\ \lambda_1 & \lambda_1^2
    \end{pmatrix}+ (\mathrm{cyclic}) \,, \\
 \nonumber   \bbeta_3&=\frac{\dd\lambda_1}{4\pi i(\lambda_1-\lambda_2)(\lambda_1-\lambda_3)} \begin{pmatrix}
        \lambda_2\lambda_3(2(s_1-s_2)-5\lambda_1+1) & 5\lambda_2\lambda_3(2\lambda_1-1) \\
        (s_1-\lambda_1-1)(2(s_2-s_1)-1+5\lambda_1) & -5(\lambda_2+\lambda_3-1)(2\lambda_1-1) 
    \end{pmatrix}\\
\nonumber    &+ (\mathrm{cyclic}) \,, \\
\nonumber    \bbeta_4&=\frac{\dd\lambda_1}{d_1} \begin{pmatrix}
        3s_3-2s_2+\lambda_1(s_1(3-\lambda_1)+2\lambda_1(\lambda_1-2)) &
        s_3(2\lambda_1-1) \\
        s_1+s_2-4\lambda_1s_1+5\lambda_1^2+\lambda_1-1 &
        \lambda_1(s_1+s_2-4\lambda_1s_1+5\lambda_1^2+\lambda_1-1)
    \end{pmatrix}+ (\mathrm{cyclic}) \,,
\end{align}
where $s_i$ are the elementary symmetric polynomials in the three parameters $\lambda_1,\lambda_2,\lambda_3$. We defined the abbreviation
\begin{equation}
    d_1=\lambda_1(\lambda_1-1)(\lambda_1-\lambda_2)(\lambda_1-\lambda_3) \,,
\end{equation}
and $\mathrm{(cyclic)}$ refers to the two cyclic images of $(\lambda_1,\lambda_2,\lambda_3)$. For brevity, we only show  the case $a_i=1$. The full matrices can be found in the ancillary file.

\paragraph{Step 4:} We can $\eps$-factorize the differential equation by performing a final rotation of the form
\begin{equation}
  \bJ(\blambda,\eps)  =\bs{R}_{t}(\blambda)\bs{I}_{\varepsilon}(\blambda,\eps) \,,
\end{equation}
with 
\begin{equation}
    \boldsymbol{R}_{t}(\blambda)=\begin{pmatrix}
        \mathbb{1} & 0 \\ \bT(\blambda) & \mathbb{1} 
    \end{pmatrix}\,,
\end{equation}
where $\bT(\blambda)$ is a $2\times 2$ matrix satisfying the first order differential equation
\begin{equation}
    \dd\bT(\blambda)+\bP(\lambda) =0\,.
\end{equation}
This differential equations is now significantly more involved than its counterpart in the elliptic case, and it seems to be a difficult task to solve it in terms of periods and quasi-periods (if possible at all). However, we can make use of the method described in section \ref{SubSec:Elliptic} and require the constancy of intersection matrix in the canonical basis. This imposes constraints on the entries of $\bT(\blambda)$. It will be useful to decompose $\bT(\blambda)$ into a symmetric and antisymmetric part,
\begin{equation}
    \bT(\blambda)=\begin{pmatrix}
        \sigma_1(\blambda) & \sigma_2(\blambda) \\ \sigma_2(\blambda) & \sigma_3(\blambda)
    \end{pmatrix}+
    \begin{pmatrix} 0 & a(\blambda) \\ -a(\blambda) & 0
    \end{pmatrix} \,,
\end{equation}
in terms of some functions $a(\blambda),\sigma_i(\blambda)$ to be determined. We denote by 
\begin{equation}
    \bs{R}(\blambda,\eps)=\bs{{R}}_{t}(\lambda)\bs{{R}}_{\eps}(\eps)\bs{{R}}_{\mathrm{ss}}(\lambda)\bs{{R}}_{\dd}(\lambda,\eps) \,,
\end{equation}
the full transformation and compute the intersection matrix of the transformed bases
\begin{equation}
    \bC_{\bJ}(\blambda,\eps)=\bs{R}^{}(\blambda,\eps)\bC_{0}\bs{R}^{T}(\blambda,-\eps) =-\frac{\eps}{8\pi^2}\begin{pmatrix}
        0 & 0 & 1 & 0 \\ 0 & 0 & 0 & 1 \\ 1 & 0 & v_1 & v_2 \\
        0 & 1 & v_2 & v_3
    \end{pmatrix} \,,
\end{equation}
where $v_1,v_2,v_3$ are some rational combinations of the parameters $\blambda$, the periods and the unknown functions $s_i(\blambda)$. Note that the dependence on the quasi-periods and the function $a(\blambda)$ has dropped out. Since the basis $\bJ(\blambda,\eps)$ is supposed to be canonical, the intersection matrix should take the form~\eqref{eq:C_J} 
for some rational function $f(\eps)$ and matrix of rational numbers $\bDelta$. We can now easily read off the rational function, which is again given by
\begin{equation}
    f(\eps)=\frac{\eps}{\pi^2} \,,
\end{equation}
and then obtain three algebraic constraints for the functions $\sigma_i(\blambda)$, which fully determine them in terms of the periods. We find
\begin{equation}
\label{eq:TsymmPart}
    \begin{pmatrix}
        \sigma_1(\blambda) & \sigma_2(\blambda) \\ \sigma_2(\blambda) & \sigma_3(\blambda)
    \end{pmatrix}=\bcA^T \bS(\blambda) \bcA \,,
\end{equation}
where $\bS(\blambda)$ is a symmetric $2\times 2$ matrix of rational functions, which takes the simple form (again given for $a_i=1$)
\begin{equation}
    \bS(\blambda)=\frac{1}{8\pi i}\begin{pmatrix}
        2(s_2-s_3) & 1-s_1-s_2 \\ 1-s_1-s_2& 4s_1-6
    \end{pmatrix} \,.
\end{equation}
After solving the above constraints, the constant intersection matrix $\bDelta$ takes the simple form
\begin{equation}
    \bDelta=-\frac{1}{8}\begin{pmatrix}
        0 & 0 & 1 & 0 \\ 0 & 0 & 0 & 1 \\ 1 & 0 & 0 & 0 \\ 0 & 1 & 0 & 0
    \end{pmatrix} \,,
\end{equation}
which is proportional to the exchange matrix $\bK_4$ after swapping the third and fourth basis elements. 

We  obtain $a(\blambda)$ by integrating its differential equation along some suitable path
\begin{equation}
    \bT_{\rA}\equiv\begin{pmatrix}
        0 & a(\blambda) \\ -a(\blambda) & 0 
    \end{pmatrix}=-\int_{\blambda_0}^{\blambda}\bP^{\rA} \,,
\end{equation}
for some base point $\blambda_0$, where we defined the anti-symmetric part of $\bP$
\begin{align}
    \bP^{\rA}(\blambda)&=\frac{1}{2}(\bP(\blambda)-\bP^T(\blambda)) \nonumber \\
    &=\frac{1}{2}\left[ \bcA^T(\bP_1(\blambda)-\bP_1^T(\blambda))\bcA +\bcA^T(\bP_2(\blambda)-\bP_3^T(\blambda))\bcAt +\bcAt^T(\bP_3(\blambda)-\bP_2^T(\blambda))\bcA  \right] \nonumber \\
    &\equiv \bcA^T\bP_1^{\rA}(\blambda) \bcA+\bcA^T\bP_2^{\rA}(\blambda) \bcAt+\bcAt^T\bP_3^{\rA}(\blambda) \bcA \,,
\end{align}
where $\bP_i^{\rA}(\blambda)$ are $2\times 2$ matrices of rational one-forms given in the ancillary file. By taking the determinant we find the explicit form of the new function
\begin{equation}
    a(\blambda)=\sqrt{\det\int_{\blambda_0}^{\blambda}\bP^{\rA}} \,.
\end{equation}
So far everything has worked analogously to the elliptic case, with suitable generalizations to higher genus, i.e., with the periods and quasi-periods now being matrices. However, there is a new aspect which was absent at genus 1, namely the fact that the matrix $\bT(\blambda)$ has an antisymmetric part. This antisymmetric part is not determined by the requirement that the intersection matrix becomes constant, and we have not been able to find a closed form solution for its differential equation. Indeed we believe that the function $a(\blambda)$ is not expressible solely in terms of periods and quasi-periods. While we have no proof of this, we provide some evidence in appendix \ref{app_newFctConstraints}, by excluding some general forms that one might have expected the function to take. 
The existence of such \enquote{new functions} has already been noted in more involved elliptic examples as well as in cases of higher-dimensional Calabi-Yau varieties \cite{Gorges:2023zgv,Delto:2023kqv,Klemm:2024wtd,Driesse:2024feo,Forner:2024ojj}. A more solid understanding of this function would be desirable, this is however beyond the scope of this paper.

Finally, the canonical differential equation can now be written as
\begin{equation}
    \dd \bJ(\blambda,\eps)=\eps\bA(\blambda)\dd\bJ(\blambda,\eps) \,,
\end{equation}
 with
\begin{align}
\label{eq:canonicalDELauricella3}
    \bA(\blambda)&=\begin{pmatrix}
        \bcA^{-1}(\bbeta_1-\bbeta_2\bS)\bcA &
        \bcA^{-1}\bbeta_2\bcA^{-1T} \\
        \bcA^{T}(\bbeta_3+\bS\bbeta_1-\bS\bbeta_2\bS-\bbeta_4\bS)\bcA & 
        \bcA^{T}(\bbeta_4+\bS\bbeta_2)\bcA^{-1T} \end{pmatrix} \nonumber\\
        &\quad +\begin{pmatrix}
            -\bcA^{-1}\bbeta_2\bcA^{-1T}\bT_{\rA} & 0 \\
            \bT_{\rA}\bcA^{-1}(\bbeta_1-\bbeta_2\bS)\bcA-\bcA^T(\bS\bbeta_2+\bbeta_4)\bcA^{-1T}\bT_{\rA} &
            \bT_{\rA}\bcA^{-1}\bbeta_2\bcA^{-1T}
        \end{pmatrix} \\
        &\quad +\begin{pmatrix}
            0 & 0 \\
            -\bT_{\rA}\bcA^{-1}\bbeta_2\bcA^{-1T}\bT_{\rA} & 0
        \end{pmatrix} \nonumber\,,
\end{align}
where we decomposed the matrix according to the degree in the new function $\bT_{\rA}$. We will investigate the properties of the functions that enter $\bs{A}(\bs{\lambda})$ in more detail in section~\ref{modular}.

%% file: Sec_Four.tex
We will now construct an $\eps$-factorized differential equation for the Lauricella function with four variables, following the same steps as before. While this still involves a hyperelliptic curve of genus two, the curve is even this time, and hence has an additional puncture at infinity, leading to some new conceptual points which we will highlight. Additionally, note that the maximal cut of the non-planar crossed box of example \ref{examplenpcb} falls into this class this. Our starting point is the family of Lauricella functions  $\mathcal{L}_{\bs{\nu}}(\bs{\lambda})$  as introduced in  (\ref{lauricellaintegral}), now specialized to $n=4$ parameters $\bs{\lambda}=(\lambda_1,\dots, \lambda_4)$, which we take to be real and ordered for definiteness. The integrand defines the multivalued function $    \Phi= \chi(x)/y$ with 
\begin{align}
\nonumber    \chi(x)&=x^{a_1\eps}(x-1)^{a_2\eps}(x-\lambda_1)^{a_3\eps}(x-\lambda_2)^{a_4\eps}(x-\lambda_3)^{a_5\eps}(x-\lambda_4)^{a_6\eps} 
    \,,\\
\label{curveeps0}
    y^2&=x(x-1)(x-\lambda_1)(x-\lambda_2)(x-\lambda_3)(x-\lambda_4) \, .
\end{align} 
The last equation defines an even hyperelliptic curve of genus two. 
As an initial basis for the integral family we  choose
\begin{equation}
\label{I0}
    \bs{I}_0=\left(\int_{\gamma} \chi(x)\frac{\dd x}{y}, \int_{\gamma} \chi(x)\frac{x\dd x}{y},\int_{\gamma} \chi(x)\frac{\Psi_1^{2,\text{e}}(x)\dd x}{y},\int_{\gamma} \chi(x)\frac{\Psi_2^{2,\text{e}}(x)\dd x}{y},\int_{\gamma} \chi(x)\frac{x^2\dd x}{y}
    \right)^T \, .
\end{equation}
with $\Psi_i^{2,\text{e}}(x)$ defined in (\ref{evengenus2basis}). As before, the explicit form of the cycle $\gamma\in H_1(X,\check{\mathcal{L}})$ is irrelevant for our considerations.
We have again chosen the master integrands in \eqref{I0} such that they reduce to suitably chosen Abelian differentials in the limit $\eps\rightarrow 0$. The intersection matrix for the initial basis is denoted by $\bs{{C}}_0$ (see appendix \ref{app:Intersect}).

%%%%%%%%%%%%
In the $\eps\rightarrow 0$ limit  the period matrix and its decomposition into semi-simple and unipotent parts takes the form\footnote{The absence of $\bs{\mathcal{B}}$ and $\bs{\tilde{\mathcal{B}}}$ in the semi-simple part is again due to the application of the bilinear relations \eqref{eq:genus2RiemannBilinearsMatrix}.}
\begin{equation}
\label{periodeven}
    \bs{\mathcal{P}}(\bs{\lambda})=\begin{pmatrix}
    \bs{\mathcal{A}} & \bs{\mathcal{B}}&\star \\ \bs{\tilde{\mathcal{A}}} & \bs{\tilde{\mathcal{B}}}&\star \\ 0 &0 & 1\end{pmatrix}
    =
    \begin{pmatrix} \bs{\mathcal{A}} & 0&0 \\ \bs{\tilde{\mathcal{A}}} & 8\pi i\bs{\mathcal{A}}^{-1T}&0\\0 & 0 & 1
    \end{pmatrix}
    \begin{pmatrix}
        \mathbb{1} & \bs{\Omega} &\star\\ 0 & \mathbb{1}&\star\\ 0 & 0&1 
    \end{pmatrix} 
    \equiv \bcS\,\bcU \,,
\end{equation}
with the  entries $\star$ signifying integrals over the additional cycle around $\infty$ that was not present in the odd case. The differential equation for the initial vector of master integrals is 
\begin{equation}
    \dd \bs{I}_0(\bs{\lambda},\eps)=\bs{B}_0(\bs{\lambda},\eps)\bs{I}_0(\bs{\lambda},\eps) \,,
\end{equation}
with
\begin{equation}
    \bs{B}_0(\bs{\lambda},\eps)=\bs{B}_0^{(1)}(\bs{\lambda},\eps)\dd\lambda_1+\bs{B}_0^{(2)}(\bs{\lambda},\eps)\dd\lambda_2+\bs{B}_0^{(3)}(\bs{\lambda},\eps)\dd\lambda_3 +\bs{B}_0^{(4)}(\bs{\lambda},\eps)\dd\lambda_4\,,
\end{equation} 
and each $\bs{B}_0^{(i)}(\bs{\lambda},\eps)$ is linear in $\varepsilon$.
We will now outline the construction of the rotation into the canonical form, following the same steps as reviewed in section \ref{SubSec:Elliptic} and used in the three-parameter case in  section \ref{Sec:Three}.
\paragraph{Step 1:}
 First we change to the derivative basis, which in this case takes the form,
 \begin{align}
 \label{step1basiseven2}
 \bs{I}_{\dd}=\left(   \text{\colorbox{Apricot!30}{$ \int_{\gamma}\frac{\dd x}{y}\chi(x)$}},   \text{\colorbox{GreenYellow!30}{$ \int_{\gamma}\frac{x \dd x}{y}\chi(x)$}},\sum_{i=1}^4 \partial_{\lambda_i} \,  \text{\colorbox{Apricot!30}{$\left[ \int_{\gamma}\frac{\dd x}{y}\chi(x)\right]$}},\sum_{i=1}^4 \partial_{\lambda_i}\,  \text{\colorbox{GreenYellow!30}{$ \left[\int_{\gamma}\frac{x \dd x}{y}\chi(x)\right]$}},\int_{\gamma}\frac{x^2\dd x}{y}\chi(x)  \right)\, . 
 \end{align}
 We read off the transformation matrix $\bs{R}_d(\bs{\lambda},\eps)$ from the initial differential equation. If we let
\begin{equation}
    \bs{I}_{\dd}(\bs{\lambda},\eps)=\bR_{\dd}(\blambda,\eps)\bs{I}_0(\bs{\lambda},\eps)  \,,
\end{equation}
we see that this vector satisfies a differential equation with the matrix
\begin{align}
    \bs{B}_{\dd}(\bs{\lambda},\varepsilon)= \left(\dd \bR_{\dd}(\bs{\lambda},\varepsilon)\right) \bR_{\dd}(\bs{\lambda},\varepsilon)^{-1} + \bR_{\dd}(\bs{\lambda},\varepsilon)\bs{B}_0(\bs{\lambda},\varepsilon) \bR_{\dd}(\bs{\lambda},\varepsilon)^{-1} \, .
\end{align}
\paragraph{Step 2:}
We rotate with the inverse of the  semi-simple part of the new period matrix at $\eps\rightarrow 0$, 
\begin{equation}
\bI_{\mathrm{ss}}(\blambda,\eps)=\bR_{\mathrm{ss}}(\bs{\lambda})\bI_{\dd}(\blambda,\eps)\, \text{ with }\,  \bR_{\mathrm{ss}}(\bs{\lambda})=  \bcS_{\dd}^{-1}(\blambda)=\left(  \bR_{\dd}(\blambda,0)\bcS(\blambda)\right)^{-1}\, ,
\end{equation}
and this change of basis leads to the new differential equation matrix 
\begin{align}
    \bs{B}_{\mathrm{ss}}(\bs{\lambda},\varepsilon)= \left(\dd \bR_{\mathrm{ss}}(\bs{\lambda})\right) \bR_{\mathrm{ss}}^{-1}(\bs{\lambda}) + \bR_{\mathrm{ss}}(\bs{\lambda})\bs{B}_{\dd}(\bs{\lambda},\varepsilon) \bR_{\mathrm{ss}}(\bs{\lambda})^{-1} \, ,
\end{align}
which takes the form
\begin{align}
\label{eq:fromaftersemisimple}
  \bs{B}_{\mathrm{ss}}(\bs{\lambda},\varepsilon)= \left(\,
    \begin{array}{@{\hspace{2pt}}c@{\hspace{2pt}} @{\hspace{2pt}}c@{\hspace{2pt}}@{\hspace{2pt}}c@{\hspace{2pt}} @{\hspace{2pt}}c@{\hspace{2pt}}@{\hspace{2pt}}c@{\hspace{2pt}}}
        0  &0 & \cellcolor{ourlightblue}\,   \textcolor{ourblue}{\bullet} \, \, & \cellcolor{ourlightblue}\,   \textcolor{ourblue}{\bullet} \, \, &0  \\ 
         0  &0 & \cellcolor{ourlightblue}\,   \textcolor{ourblue}{\bullet} \, \, & \cellcolor{ourlightblue}\,   \textcolor{ourblue}{\bullet} \, \,  &0  \\ 
        0 &0    & 0 &0 &0  \\
        0 &0    & 0 &0 &0  \\
      \cellcolor{ourlightblue}\,   \textcolor{ourblue}{\bullet} \, \, & \cellcolor{ourlightblue}\,   \textcolor{ourblue}{\bullet} \, \,  & \cellcolor{ourlightblue}\,   \textcolor{ourblue}{\bullet} \, \, & \cellcolor{ourlightblue}\,   \textcolor{ourblue}{\bullet} \, \,  &0  \\
    \end{array}
    \,\right)+
    %%%%%%%%%% 
    \left(\begin{array}{@{\hspace{3pt}}c@{\hspace{2pt}} @{\hspace{2pt}}c@{\hspace{2pt}}@{\hspace{2pt}}c@{\hspace{2pt}}@{\hspace{2pt}}c@{\hspace{2pt}} @{\hspace{2pt}}c@{\hspace{3pt}}}
     \cellcolor{ourlightblue}\,   \textcolor{ourblue}{\bullet} \, \, & \cellcolor{ourlightblue}\,   \textcolor{ourblue}{\bullet} \, \,  &0&0 & \cellcolor{ourlightblue}\,   \textcolor{ourblue}{\bullet} \, \,  \\ 
     \cellcolor{ourlightblue}\,   \textcolor{ourblue}{\bullet} \, \, & \cellcolor{ourlightblue}\,   \textcolor{ourblue}{\bullet} \, \,  &0&0 & \cellcolor{ourlightblue}\,   \textcolor{ourblue}{\bullet} \, \, \\ 
     \cellcolor{ourlightblue}\,   \textcolor{ourblue}{\bullet} \, \, & \cellcolor{ourlightblue}\,   \textcolor{ourblue}{\bullet} \, \,  &  \cellcolor{ourlightblue}\,   \textcolor{ourblue}{\bullet} \, \, & \cellcolor{ourlightblue}\,   \textcolor{ourblue}{\bullet} \, \, & \cellcolor{ourlightblue}\,   \textcolor{ourblue}{\bullet} \, \,  \\ 
     \cellcolor{ourlightblue}\,   \textcolor{ourblue}{\bullet} \, \, & \cellcolor{ourlightblue}\,   \textcolor{ourblue}{\bullet} \, \,  &  \cellcolor{ourlightblue}\,   \textcolor{ourblue}{\bullet} \, \, & \cellcolor{ourlightblue}\,   \textcolor{ourblue}{\bullet} \, \, & \cellcolor{ourlightblue}\,   \textcolor{ourblue}{\bullet} \, \, \\
     \cellcolor{ourlightblue}\,   \textcolor{ourblue}{\bullet} \, \, & \cellcolor{ourlightblue}\,   \textcolor{ourblue}{\bullet} \, \,  &  0&0 & \cellcolor{ourlightblue}\,   \textcolor{ourblue}{\bullet} \, \, 
    \end{array}
    \,\right)
    %%%%%% 
    \, \eps+
    \left(\begin{array}{@{\hspace{2pt}}c@{\hspace{2pt}} @{\hspace{2pt}}c@{\hspace{2pt}}@{\hspace{2pt}}c@{\hspace{2pt}}@{\hspace{2pt}}c@{\hspace{2pt}} @{\hspace{2pt}}c@{\hspace{2pt}}} 
     0&0&0&0&0 \\  0&0&0&0 &0\\ 
     \cellcolor{ourlightblue}\,   \textcolor{ourblue}{\bullet} \, \, & \cellcolor{ourlightblue}\,   \textcolor{ourblue}{\bullet} \, \, &  0&0  & \cellcolor{ourlightblue}\,   \textcolor{ourblue}{\bullet} \, \,  \\ 
     \cellcolor{ourlightblue}\,   \textcolor{ourblue}{\bullet} \, \, & \cellcolor{ourlightblue}\,   \textcolor{ourblue}{\bullet} \, \,  &   0& 0&\cellcolor{ourlightblue}\,   \textcolor{ourblue}{\bullet} \, \,  \\
     0&0&0&0&0 
    \end{array}
    \,\right)\, \eps^2\, .
\end{align}
As expected, the upper left $4\times 4$ block is structurally equivalent to the differential equation matrix of the three-parameter case, but we obtain additional $\eps^0$- and $\eps^2$- terms in the fifth row and column, which are related to the additional puncture at $\infty$ that is present in the even hyperelliptic case.
\paragraph{Step 3:}
In order to get rid of the $\varepsilon^2$ terms and bring all $\varepsilon^0$ terms under the diagonal, we perform the simple rotation
\begin{align}\bI_{\eps}(\blambda,\eps)=\bR_{\eps}(\varepsilon)\bI_{\mathrm{ss}}(\blambda,\eps)\,, \text{~~~with~~~}
    \bR_{\eps}(\varepsilon)= \begin{pmatrix}
        1&0&0&0&0\\
        0&1&0&0&0\\
        0&0&0&0&1\\
        0&0&0&\frac{1}{\varepsilon}&0\\
        0&0&\frac{1}{\varepsilon}&0&0
    \end{pmatrix}\, . 
\end{align}
After this rotation, the new differential equation matrix
\begin{align}
    \bs{B}_{\eps}(\bs{\lambda},\varepsilon)= \left(\dd \bR_{\eps}(\varepsilon)\right) \bR_{\eps}^{-1}(\varepsilon) + \bR_{\eps}(\varepsilon)\bs{B}_{\mathrm{ss}}(\bs{\lambda},\varepsilon) \bR_{\eps}(\varepsilon)^{-1} \,,
\end{align}
 takes the form
\begin{align}\label{eq:frombeforeintegratingout}
  \bs{B}_{\eps}(\bs{\lambda},\varepsilon)= \left(\,
    \begin{array}{@{\hspace{2pt}}c@{\hspace{2pt}} @{\hspace{2pt}}c@{\hspace{2pt}}@{\hspace{2pt}}c@{\hspace{2pt}} @{\hspace{2pt}}c@{\hspace{2pt}}@{\hspace{2pt}}c@{\hspace{2pt}}}
        0  &0 &0&0&0  \\ 
        0  &0 &0&0&0  \\ 
      \cellcolor{ourlightblue}\,   \textcolor{ourblue}{\bullet} \, \, & \cellcolor{ourlightblue}\,   \textcolor{ourblue}{\bullet} \, \,  & 0 &0 &0  \\
      \cellcolor{ourlightblue}\,   \textcolor{ourblue}{\bullet} \, \, & \cellcolor{ourlightblue}\,   \textcolor{ourblue}{\bullet} \, \,  & \cellcolor{ourlightblue}\,   \textcolor{ourblue}{\bullet} \, \, & 0 &0  \\
      \cellcolor{ourlightblue}\,   \textcolor{ourblue}{\bullet} \, \, & \cellcolor{ourlightblue}\,   \textcolor{ourblue}{\bullet} \, \,  & \cellcolor{ourlightblue}\,   \textcolor{ourblue}{\bullet} \, \, & 0 &0  \\
    \end{array}
    \,\right)+
    %%%%%%%%%% 
    \left(\begin{array}{@{\hspace{3pt}}c@{\hspace{2pt}} @{\hspace{2pt}}c@{\hspace{2pt}}@{\hspace{2pt}}c@{\hspace{2pt}}@{\hspace{2pt}}c@{\hspace{2pt}} @{\hspace{2pt}}c@{\hspace{3pt}}}
     \cellcolor{ourlightblue}\,   \textcolor{ourblue}{\bullet} \, \, & \cellcolor{ourlightblue}\,   \textcolor{ourblue}{\bullet} \, \,  & \cellcolor{ourlightblue}\,   \textcolor{ourblue}{\bullet} \, \, & \cellcolor{ourlightblue}\,   \textcolor{ourblue}{\bullet} \, \, & \cellcolor{ourlightblue}\,   \textcolor{ourblue}{\bullet} \, \,  \\ 
     \cellcolor{ourlightblue}\,   \textcolor{ourblue}{\bullet} \, \, & \cellcolor{ourlightblue}\,   \textcolor{ourblue}{\bullet} \, \,  & \cellcolor{ourlightblue}\,   \textcolor{ourblue}{\bullet} \, \, & \cellcolor{ourlightblue}\,   \textcolor{ourblue}{\bullet} \, \,  & \cellcolor{ourlightblue}\,   \textcolor{ourblue}{\bullet} \, \, \\ 
     \cellcolor{ourlightblue}\,   \textcolor{ourblue}{\bullet} \, \, & \cellcolor{ourlightblue}\,   \textcolor{ourblue}{\bullet} \, \,  &  \cellcolor{ourlightblue}\,   \textcolor{ourblue}{\bullet} \, \, & \cellcolor{ourlightblue}\,   \textcolor{ourblue}{\bullet} \, \, & \cellcolor{ourlightblue}\,   \textcolor{ourblue}{\bullet} \, \,  \\ 
     \cellcolor{ourlightblue}\,   \textcolor{ourblue}{\bullet} \, \, & \cellcolor{ourlightblue}\,   \textcolor{ourblue}{\bullet} \, \,  &  \cellcolor{ourlightblue}\,   \textcolor{ourblue}{\bullet} \, \, & \cellcolor{ourlightblue}\,   \textcolor{ourblue}{\bullet} \, \, & \cellcolor{ourlightblue}\,   \textcolor{ourblue}{\bullet} \, \, \\
     \cellcolor{ourlightblue}\,   \textcolor{ourblue}{\bullet} \, \, & \cellcolor{ourlightblue}\,   \textcolor{ourblue}{\bullet} \, \,  &  \cellcolor{ourlightblue}\,   \textcolor{ourblue}{\bullet} \, \, & \cellcolor{ourlightblue}\,   \textcolor{ourblue}{\bullet} \, \,  & \cellcolor{ourlightblue}\,   \textcolor{ourblue}{\bullet} \, \, 
    \end{array}
    \,\right)
    %%%%%% 
    \, \eps\, .
\end{align}

\paragraph{Step 4: } We obtain the final rotation by making an ansatz
\begin{align}
\label{u4444}
    \bR_t(\bs{\lambda})= \begin{pmatrix}1&0&0&0&0\\
    0&1&0&0&0\\
    u_{3,1}&u_{3,2}&1&0&0\\
    u_{4,1}&u_{4,2}&u_{4,3}&1&0\\
    u_{5,1}&u_{5,2}&u_{5,3}&0&1\\
    \end{pmatrix} \,,
\end{align}
with the unknown entries $u_{i,j}$ chosen such that the new differential equation matrix
\begin{align}
\eps\bA(\blambda)= \left(\dd \bR_t(\bs{\lambda})\right) \bR_t(\bs{\lambda})^{-1} + \bR_t(\bs{\lambda})\bs{B}_{\eps}(\bs{\lambda},\varepsilon) \bR_t(\bs{\lambda})^{-1} \, ,
\end{align}
$\varepsilon$-factorizes. Note that finding the $u_{i,j}$ from this condition requires solving eight coupled differential equations for the eight undetermined entries. Additionally, we cannot immediately see which  solutions of these eight coupled differential equations will be expressible as rational functions of the branch points $\lambda_i$ and the (quasi-)periods $\bs{\mathcal{A}}$ and $\bs{\tilde{\mathcal{A}}}$ and which ones  will be integrals over these objects that are in fact new functions. This is, however, greatly simplified using the criterion that the intersection matrix is constant in the $\lambda_i$, as already discussed in the previous sections.

 More specifically, we  rotate the initial intersection matrix $\bs{C}_0$ with the full transformation matrix\footnote{As in the three parameter case,  the full transformation matrix is $  \bs{R}(\bs{\lambda},\varepsilon)= \bs{R}_{t}(\bs{\lambda}) \bs{R}_{\eps}(\varepsilon) \bs{R}_{\mathrm{ss}}(\bs{\lambda}) \bs{R}_{\dd}(\bs{\lambda},\varepsilon)$.}  to obtain
\begin{align}
    \bC_{\bJ}= \bs{R}(\bs{\lambda},\varepsilon)\bC_0 \bs{R}(\bs{\lambda},-\varepsilon)^T =\begin{pmatrix}
        0&0&0&0&-\frac{i}{4\pi\varepsilon}\\
        0&0&0&-\frac{i}{4\pi\varepsilon}&0\\
        0&0 & -\frac{1}{\varepsilon \sum_{i=1}^{6} a_i} & v_{3,4}&v_{3,5}\\
        0&-\frac{i}{4\pi\varepsilon}& v_{4,3}&v_{4,4}&v_{4,5}\\
        -\frac{i}{4\pi\varepsilon}&0&v_{5,3}&v_{5,4}&v_{5,5}
    \end{pmatrix}\, , 
\end{align}
where the entries $v_{i,j}$ are rational functions of the branch points $\lambda_i$, the (quasi-)periods $\bs{\mathcal{A}}$ and $\bs{\tilde{\mathcal{A}}}$ and the entries $u_{i,j}$ of (\ref{u4444}).
We again require that 
$\bC_{\bJ}$ should take the form~\eqref{eq:C_J}
for some rational function $f(\eps)$ and matrix of numbers $\bDelta$. From this requirement we find  eight algebraic equations that one can solve for the $u_{i,j}$. However, due to the symmetry of the intersection matrix only five of them are independent. We choose to solve for the entries $\{u_{3,4},u_{3,5},u_{4,2},u_{5,1},u_{5,2}\}$ in terms of the branch points, periods and the three remaining new functions $\{u_{4,1},u_{4,3},u_{5,3}\}$.\footnote{Note that there is a lot of freedom in which functions to solve for. More specifically, we found $24$ combinations for which we can solve. At this point, we could not find any symmetry or preferred choice as in the three-parameter case. } The intersection matrix then takes the particularly simple form
\begin{align}
    \bs{C}_{\bJ} =-\frac{1}{\varepsilon} \begin{pmatrix}
        0&0&0&0&\frac{i}{4\pi}\\
        0&0&0&\frac{i}{4\pi}&0\\
        0&0 & \frac{1}{ \sum_{i=1}^{6} a_i} & 0&0\\
        0&\frac{i}{4\pi}& 0&0&0\\
        \frac{i}{4\pi}&0&0&0&0
    \end{pmatrix}\, .
\end{align}
We cannot exclude that (some of) the remaining entries $u_{4,1},u_{4,3},u_{5,3}$ are expressible in terms of just the periods and branch points.

\subsection{Summary} 

Let us summarize the results of this section. We have analyzed classes of Lauricella functions that for $\varepsilon=0$ compute the periods and quasi-periods of some family of hyperelliptic curves of genus two. We have considered two cases separately, namely the cases of odd and even curves. The latter corresponds to the case where there is an additional puncture at $\infty$.\footnote{Note that there is nothing special about the puncture being at infinity, because we could always send infinity to a point at finite distance using a M\"obius transformation.} 

For both the odd and even cases, we find that we can use the algorithm of~\cite{Gorges:2023zgv} to transform the differential equation into an $\varepsilon$-factorized form. Strikingly, we can follow exactly the same steps as in the elliptic case reviewed in section~\ref{SubSec:Elliptic}, where in the hyperelliptic case we can interpret the $a$ and $b$ periods as matrices. In particular, the Legendre relations, which played an important role in section~\ref{SubSec:Elliptic} to decompose the period matrix into a semi-simple and unipotent part, generalize to matrix relations. While we have focused on the genus-two case, we do not expect any surprises when going to higher genera (at least as long as we consider hyperelliptic curves), because all matrix-valued quantities encountered in the genus-two case have a straightforward extension to higher genera. Moreover, just like in the elliptic case, we find that imposing the constancy of the intersection matrix allows us to considerably simplify the last step of the algorithm from~\cite{Gorges:2023zgv}.

While we observe that all the steps in the  algorithm from~\cite{Gorges:2023zgv} generalize in a straightforward manner to higher genera, there are also some differences. In particular, while imposing the constancy of the intersection matrix allows us to identify most new functions as combinations of periods, we still need to introduce new functions that involve integrals of (quasi-)periods. For the case of even hyperelliptic curves, this does not come as a surprise, because the same feature already arises in the elliptic case. 
The antisymmetric matrix $\bs{T}_A$, however, appears to be a new function already in the odd case, which did not arise for an elliptic curve. Since $\bs{T}_A$ is antisymmetric, it is clear that it must vanish in the elliptic case, but there is no reason why it cannot arise for higher genera.

Finally, we need to address an important question. We have shown so far that the algorithm of~\cite{Gorges:2023zgv} allows us to obtain an $\varepsilon$-factorized differential equation. It is, however, not immediately clear that this is also a good canonical differential equation. In section~\ref{SubSec:Elliptic} we have stated a set of properties which the canonical differential equation in the elliptic case satisfies. Remarkably, we find that our $\varepsilon$-factorized differential equation satisfies these same properties (appropriately extended to higher genera), so that it can be qualified as \enquote{canonical}. Indeed, by construction the intersection matrix is constant, and by inspection we can see that all poles in $\bs{\lambda}$ are simple. The remaining key properties in the elliptic case were related to the modularity properties of the functions that enter the differential equations matrix. In the next section we discuss the modularity properties in the genus two case.

%% file: Sec_Modular.tex
As reviewed in section \ref{SubSec:Elliptic}, the entries of the canonical differential equation matrix of Gauß' hypergeometric function can be interpreted as modular forms. We hence expect the canonical form of the hyperelliptic Lauricella function to involve appropriate higher-genus generalizations. These are the so-called \emph{Siegel modular forms}. After providing a brief review of Siegel modular forms in section \ref{SubSec:Siegel}, we will show in section \ref{SubSec:LauricellaModularity} that this is indeed the case. This provides further strong evidence that the algorithm of \cite{Gorges:2023zgv} has indeed not only produced an $\eps$-factorized differential equation, but it can be considered canonical, also in the hyperelliptic case.

\subsection{Review: Siegel Modular Forms}
\label{SubSec:Siegel}

We will start by reviewing some basic background on Siegel modular forms that is important in the following. For further details, we refer to the classical textbook \cite{freitagSiegel} (in German) or the more recent reviews \cite{vanDerGeerSiegelModularForms,klingenSiegel,andrianovSiegel}.
\paragraph{Basic Definitions.}
Modular forms are holomorphic functions on the upper-half plane $\mathbb{H}$ with a specific transformation behavior under the modular group $\mathrm{SL}(2,\mathbb{Z})$. Siegel modular forms are a generalization to functions on the \emph{Siegel upper half space} 
\begin{equation}
    \cH_g= \{\bOmega\in \mathbb{C}^{g\times g}:\, \bOmega=\bOmega^T \,, \mathrm{Im}\,\bOmega>0 \} \,,
\end{equation}
consisting of complex symmetric $g\times g$ matrices with positive definite imaginary part, which transform in a particular way -- stated precisely in (\ref{eq:TrafoSiegelM}) --  under the \emph{Siegel modular group} $\Gamma_g$, which is  the symplectic group 
\begin{equation}
    \Gamma_g=\mathrm{Sp}(2g,\mathbb{Z})=\left\{\begin{pmatrix} \bA & \bB \\ \bC & \bD\end{pmatrix}\in \mathbb{Z}^{2g\times 2g}:\,
    \begin{pmatrix} \bA & \bB \\ \bC & \bD\end{pmatrix}^T
    \begin{pmatrix} \bs{0} & \mathds{1} \\ -\mathds{1} & \bs{0} \end{pmatrix}
    \begin{pmatrix} \bA & \bB \\ \bC & \bD\end{pmatrix} =
    \begin{pmatrix} \bs{0} & \mathds{1} \\ -\mathds{1} & \bs{0} \end{pmatrix}
    \right\} \,.
\end{equation}
Here $g$ is  some positive integer, but since the normalized period matrix of a hyperelliptic curve (or more general Riemann surface) of genus $g$ naturally lives in $\mathcal{H}_g$, this integer $g$ will later be identified with the genus.

The Siegel modular group acts on the Siegel upper half space by a generalized version of the  Möbius transformation. Explicitly, for $\bs{\gamma}=\begin{pmatrix} \bA & \bB \\ \bC & \bD \end{pmatrix}\in\Gamma_g,\,\bOmega\in\mathcal{H}_g$,
\begin{equation}
    \bs{\gamma} \cdot \bOmega=(\bA\bOmega+\bB)(\bC\bOmega+\bD)^{-1} \,.
\end{equation}
Let $\rho:\,\mathrm{GL}(g,\mathbb{C})\rightarrow \mathrm{GL}(V)$ be a finite-dimensional complex representation with representation space $V$. Then a \emph{Siegel modular form} of weight $\rho$ is a holomorphic map $f:\cH_g\rightarrow V$, such that
\begin{equation}
\label{eq:TrafoSiegelM}
    f(\bgamma\cdot \bOmega)=\rho(\bC\bOmega +\bD)f(\bOmega) \,,
\end{equation}
for all $\bOmega\in\cH_g$ and $\bgamma=\begin{pmatrix} \bA & \bB \\ \bC & \bD\end{pmatrix}\in\Gamma_g$.\footnote{ For $g=1$ there is also an extra holomorphy requirement at $\infty\in\cH_1=\mathbb{H}$, which is unimportant in the following due to Koecher's principle.}

Again,  concepts known from elliptic curves and modular forms have a natural generalization to hyperelliptic curves/Siegel modular forms with various scalar quantities being lifted to $g\times g$ matrices. There is a notable novelty for $g>1$ however: Due to the fact that the \enquote{automorphy factor} $\bC\bOmega+\bD$ appearing in the transformation law above is now a matrix in $\mathrm{GL}(g,\mathbb{C})$, there is some additional freedom of how to choose a representation $\rho$, leading to vector-valued Siegel modular forms.
For $g>1$ there is the possibility of the representation $\rho$ being reducible. In this case we can write the representation $\rho$ as the direct sum of two lower-dimensional representations $\rho=\rho_1\oplus\rho_2$. Then also the vector space of Siegel modular forms can be decomposed into a direct sum
\begin{equation}
    M_\rho(\Gamma_g)=M_{\rho_1}(\Gamma_g)\oplus M_{\rho_2}(\Gamma_g) \,.
\end{equation}
In particular, we can always decompose a Siegel modular form into a sum of Siegel modular forms of irreducible weights.

We will often be interested in functions which do not transform covariantly under the full Siegel modular group, but only under some subgroup. The relevant subgroups are the so-called \emph{congruence subgroups}. The \emph{principle congruence subgroup} of level $N\in\mathbb{Z}_{>0}$ is defined as
\begin{equation}
    \Gamma_g(N)=\left\{\bM\in\Gamma_g:\bM\equiv \mathds{1} \text{ mod } N \right\}\subset \Gamma_g\,.
\end{equation}
A general congruence subgroup of $\Gamma_g$ of level $N$ is then any subgroup that contains the principle congruence subgroup $\Gamma_g(N)$, see the lecture notes \cite{pitaleSiegelLectures} for some examples. One can  define Siegel modular forms with respect to these congruence subgroups in the same way as above, simply restricting the Siegel modular group to the respective congruence subgroup.

\paragraph{Irreducible Representations of $\mathrm{GL}(g,\mathbb{C})$.}
Let us consider some important examples of irreducible representations $\rho$ that will play a role in the following. Besides the trivial representation, $\mathrm{GL}(g,\mathbb{C})$ possesses other representations of dimension one, given by the determinant to some power $k$,
\begin{equation}
    \mathrm{det}^k: \mathrm{GL}(g,\mathbb{C})\rightarrow \mathrm{GL}(1,\mathbb{C}), \,\bM\mapsto (\det\bM)^k \,,
\end{equation}
where the power of the determinant $(\det\bM)^k$ is viewed as a linear map on the
complex numbers acting by multiplication
\begin{equation}
    (\det\bM)^k: \mathbb{C}\rightarrow\mathbb{C}, \, z\mapsto (\det\bM)^k z\,.
\end{equation}
Siegel modular forms with weight $\rho=\mathrm{det}^k$ are referred to as \emph{classical Siegel modular forms} of \emph{weight} $k$ and \emph{degree} $g$.

There are two further irreducible representations of $\mathrm{GL}(g,\mathbb{C})$ that will play an important role in the next section, namely the fundamental representation $\rho_{\mathrm{F}}$ and the dual representation $\rho_{\mathrm{D}}$ with representation space $V=\mathbb{C}^g$. For $\bM\in\mathrm{GL}(g,\mathbb{C})$, we have,
\begin{align}
    &\rho_{\mathrm{F}}(\bM):V\rightarrow V,\, v\mapsto \bM v \,, \\
    &\rho_{\mathrm{D}}(\bM):V\rightarrow V,\, v\mapsto \bM^{-1T}v \,.
\end{align}
Importantly, these are actually related for $g=2$, since for $\bM\in\mathrm{GL}(2,\mathbb{C})$, we have
\begin{equation}\label{eq:eps_def}
    \bepsilon \bM^{-1T} \bepsilon^{-1}=\det(\bM)^{-1}\bM \, \text{ ~~with~~}\, 
    \bepsilon=\begin{pmatrix}
        0 & 1 \\ -1 & 0
    \end{pmatrix} \,,
\end{equation}
which for $g=2$ implies that 
\begin{equation}\label{eq:dualFundamentalIsomorphism}
    \rho_{\mathrm{D}}\simeq \rho_{\mathrm{F}}\otimes \mathrm{det}^{-1}\,.
    \end{equation}

Finally, there is a class of irreducible representations of $\mathrm{GL}(g,\mathbb{C})$ given by the symmetric tensor representation
\begin{equation}
    \mathrm{Sym}^k:\mathrm{GL}(g,\mathbb{C})\rightarrow \mathrm{GL}(\mathrm{Sym}^k V),\,\bM\mapsto \mathrm{Sym}^k(\bM) \,,
\end{equation}
acting on symmetric tensors of rank $k$. Explicitly the representation space can be defined as 
\begin{equation}
    \mathrm{Sym}^k V=V^{\otimes k}/S_k \,,
\end{equation}
with the symmetric group $S_k$ permuting the $k$ copies of the space $V$, which is here given by the representation space of the fundamental representation $V=\mathbb{C}^g$. The representation acts on tensor of rank $k$  as 
\begin{equation}
    \mathrm{Sym}^k(\bM):\mathrm{Sym}^k V\rightarrow \mathrm{Sym}^k V,\, T_{i_1,\dots ,i_k}\mapsto M_{i_1,j_1}\dots M_{i_k,j_k}T_{j_1,\dots ,j_k} \,.
\end{equation}

\paragraph{Riemann $\Theta$ Functions.}
Many examples of Siegel modular forms can be constructed using (Riemann) $\Theta$ functions. 
Here we will just introduce the basic concepts needed for this paper and refer to the textbooks \cite{fayTheta,tataTheta1,tataTheta2,tataTheta3} and the lecture notes \cite{bertolaTheta} for more details.

The basic definition of a \emph{(Riemann) $\Theta$ function} is given by the series representation
\begin{equation}
    \Theta(\bz,\bOmega)=\sum_{\bn\in\mathbb{Z}^g}\exp\left[ i\pi \bn^T\bOmega\bn+2\pi i\bn^T\bz \right] \, , 
\end{equation}
where $\bz=(z_1,\dots,z_g)\in\mathbb{C}^g,\,\bOmega\in\mathcal{H}_g$ and $g$ is a positive integer. We will also need a more general notion of $\Theta$ function, namely the $\Theta$ functions with \emph{(half) characteristics}. Characteristics are pairs of integer vectors $\bepsilon_1,\bepsilon_2\in\mathbb{Z}^g$, typically written in a matrix as $\bepsilon=\left[\genfrac{}{}{0pt}{1}{\bepsilon_1}{\bepsilon_2}\right]$. They are referred to as \emph{odd} or \emph{even} if their scalar product $\bepsilon_1^T\bepsilon_2$ is odd or even. The $\Theta$ functions associated with such a pair of characteristics is then defined by the series
\begin{equation}
    \Theta\left[\bepsilon\right](\bz,\bOmega)=\sum_{\bn\in\mathbb{Z}^g}\exp\left[ i\pi\left(\frac{\bepsilon_1}{2}+\bn \right)^T\bOmega\left(\frac{\bepsilon_1}{2}+\bn \right)+2\pi i\left( \bn+\frac{\bepsilon_1}{2}\right)^T\left( \bz+\frac{\bepsilon_2}{2}\right) \right] \,.
\end{equation}
In general, $\Theta$ functions are quasi-periodic functions of the variable $\bz$
\begin{equation}
    \Theta\left[\bepsilon\right](\bz+\bOmega\blambda_1+\blambda_2,\bOmega)=\exp\left[ 2\pi i\left( \frac{1}{2}(\bepsilon_1^T\blambda_2-\blambda_1^T\bepsilon_2)-\blambda_1^T\bz-\frac{1}{2}\blambda_1^T\bOmega\blambda_1\right)\right]\Theta\left[\bepsilon\right](\bz,\bOmega) \, .
\end{equation}
Hence, by taking appropriate quotients, one can  build well-defined functions on the Jacobian\footnote{This is the natural generalization of the torus description of the elliptic curve to higher genus. However, for $g>1$ the hyperelliptic curve (or general Riemann surface) is only a subvariety of the Jacobian and not equivalent to it as for $g=1$.} $\mathbb{C}^g/(\mathbb{Z}^g\oplus\bOmega\mathbb{Z}^g)$ of a Riemann surface of genus $g$, with normalized period matrix $\bOmega$, out of $\Theta$ functions, which explains their pivotal role in the study of Riemann surfaces. Additionally, $\Theta$ functions are quasi-periodic with respect to their characteristics,
\begin{equation}
    \Theta\left[\genfrac{}{}{0pt}{1}{\bepsilon_1+2\bnu_1}{\bepsilon_2+2\bnu_2}\right](\bz,\bOmega)=\exp\left(i\pi \bepsilon_1^T\bnu_2\right)\Theta\left[\genfrac{}{}{0pt}{1}{\bepsilon_1}{\bepsilon_2}\right](\bz,\bOmega) \,,
\end{equation}
with $\bnu_1,\bnu_2\in\mathbb{Z}^g$, which allows one to restrict to $\bepsilon_1,\bepsilon_2\in\{0,1\}^g$. Thus, there are only $2^{2g}$ independent choices of (half) characteristics. In the following we will only need $\Theta$ functions (and their derivatives with respect to $z_i$) evaluated at $\bz=0$. These functions on $\mathcal{H}_g$ are referred to as \emph{(Riemann) $\Theta$ constants} $\Theta\left[\bepsilon\right](\bOmega)\equiv \Theta\left[\bepsilon\right](\bs{0},\bOmega)$ and \emph{derivative (Riemann) $\Theta$ constants} $\partial_i\Theta\left[\bepsilon\right](\bOmega)\equiv \left .\partial_{z_i}\Theta\left[\bepsilon\right](\bz,\bOmega)\right|_{\bz=0}$. At genus two, there are 16 characteristics, leading to 10 $\Theta$ constants and 6 derivative $\Theta$ constants, which we list for convenience in appendix \ref{app_genus2thetas}, where we also define a brief notation denoting the 10 independent $\Theta$ constants for $g=2$  by $\theta_1,\dots ,\theta_{10}$.

$\Theta$ constants are (classical) Siegel modular forms with weight $\mathrm{det}^{1/2}$ for a congruence subgroup of $\Gamma_g$ of level 2, at least up to a multiplicative complex phase \cite{igusaTheta,igusaForm2}.\footnote{They are referred to as Siegel modular forms with \emph{character}.} This makes them very useful in constructing (genuine) Siegel modular forms for some congruence subgroup or the full Siegel modular group. 

\paragraph{(Quasi-)periods as Siegel (Quasi-)modular Forms.}
Another  example of Siegel modular forms is provided by the $a$-cycle period matrix $\bcA$ of a hyperelliptic curve. To see this, consider a family of hyperelliptic curves dependent on some parameters $\blambda$. If we now perform a closed loop in parameter space, the branch points will also change, but at the end of the loop we should have ended up with the same hyperelliptic curve we started with. This procedure will however in general have changed the integration cycles, i.e., the homology basis, leading to the $a$ and $b$-periods undergoing a linear transformation\footnote{The transposes are due to our convention to label the columns and not the rows of the $\bcA,\bcB$ matrices by the cycles.}
\begin{equation}
    \begin{pmatrix}
        \bcB^T \\ \bcA^T
    \end{pmatrix} \rightarrow \bM \begin{pmatrix}
        \bcB^T \\ \bcA^T
    \end{pmatrix} \,,
\end{equation}
with $\bM$ an integer-valued $2g\times 2g$ matrix, the so-called \emph{monodromy matrix}. Since the monodromy should preserve the symplectic structure of the homology basis, $\bM$ should be an element of $\mathrm{Sp}(2g,\mathbb{Z})$. By performing all possible independent closed loops in parameter space we  generate some subgroup $\Gamma$ of the Siegel modular group $\Gamma_g=\mathrm{Sp}(2g,\mathbb{Z})$, and this subgroup is known to always be a congruence subgroup~\cite{acampo}. The exact congruence subgroup depends on the concrete problem.

The transformation behavior of the periods can now be translated into an action of the congruence subgroup. Explicitly we find, with $\gamma=\begin{pmatrix} \bA & \bB \\ \bC & \bD \end{pmatrix}\in\Gamma$, for the normalized period matrix
\begin{equation}
    \gamma\cdot \bOmega = (\bA\bOmega+\bB)(\bC\bOmega+\bD)^{-1} \,,
\end{equation}
which is the natural action on the Siegel upper half space $\mathcal{H}_g$. Similarly we find the action on the $a$-period matrix 
\begin{equation}
\label{eq:periodModularTransf}
    \gamma\cdot\bcA=\bcA(\bC\bOmega+\bD)^T \,,
\end{equation}
which shows that $\bcA$ is a Siegel modular form with respect to $\Gamma$ with weight $1\otimes \rho_{\mathrm{F}}$, where 1 refers to the trivial representation.

From the above we can also derive the modular transformation behavior of the quasi-period matrix $\bcAt$. Namely, as seen in the previous section, the (quasi-)periods satisfy first-order differential equations. If we choose some first-order differential operator $\partial$, we can express the quasi-period matrix $\bcAt$ in terms of $\bcA$ and its derivative $\partial\bcA$
\begin{equation}
\label{eq:defRMatrices}
    \bcAt=\bR_1\bcA+\bR_2\partial\bcA \,,
\end{equation}
for some $g\times g$ matrices $\bR_i$, which depend on the choice of $\partial$. From this we can now deduce the transformation behavior, with $\gamma\in\Gamma$ as above
\begin{equation}
    \gamma \cdot \bcAt=\bcAt(\bC\bOmega+\bD)^T+\bR_2\bcA\partial\bOmega\bC^T \,.
\end{equation}
Note that while both $\bR_2$ and $\partial\bOmega$ depend on the choice of $\partial$, it is easy to show that the combination $\bR_2\bcA\partial\bOmega$ is independent of this choice and can be computed to be
\begin{equation}
\label{eq:toProveInAppendix}
    \bR_2\bcA\partial\bOmega=8\pi i\bcA^{-1T} \,.
\end{equation}
See appendix \ref{app_quasiPeriodModular} for further details. Hence we find that the quasi-period matrix transforms as 
\begin{equation}
\label{eq:quasiperiodModular}
    \gamma \cdot \bcAt=\bcAt(\bC\bOmega+\bD)^T+8\pi i\bcA^{-1T}\bC^T \,.
\end{equation}
Since $\bcAt$ transforms not like a Siegel modular form but like the derivative of one, we refer to it as a \emph{Siegel quasi modular form}, see \cite{siegelQuasiModular} for a rigorous definition.
\subsection{Modular properties of the Canonical Differential Equation}
\label{SubSec:LauricellaModularity}

As argued in the previous subsection, the matrix of $a$-periods $\bcA$ should transform as a Siegel modular form for an appropriate congruence subgroup $\Gamma\subset \Gamma_2=\mathrm{Sp}(4,\mathbb{Z})$. Since our hyperelliptic curves are in Rosenhain normal form, the relevant congruence subgroup is known to lie in the principal congruence subgroup of level two $\Gamma_2(2)$, see e.g., \cite{rosenhainFormMonodromy}. Numerically we find that $\bcA$ is not a Siegel modular form for $\Gamma_2(2)$, but transforms as a Siegel modular form up to a sign
\begin{equation}\label{eq:A_symp_2}
    \bcA\mapsto -\bcA (\bC\bOmega+\bD)^T \,,\textrm{~~~for~~~}
    \begin{pmatrix}
        \bA & \bB \\ \bC & \bD
    \end{pmatrix}\in \Gamma_2(2) \,,
\end{equation}
which is however sufficient as it will appear quadratically. This is exactly analogous to the elliptic case reviewed in section~\ref{Sec:Review}, where the period of the elliptic curve is not a modular form for $\Gamma(2)$, but its square is.

Consider only the first matrix contributing to $\bA(\blambda)$ in equation \eqref{eq:canonicalDELauricella3}, i.e., the one that contains no factor of the antisymmetric matrix $\bT_{\rA}$ containing the new function $a(\blambda)$. From the transformation behaviour of the periods it directly follows that the four $2\times 2$ blocks all constitute Siegel modular forms of some weights 
\begin{equation}
    \rho_{ij}:\mathrm{GL}(2,\mathbb{C})\rightarrow \mathrm{GL}(V_{ij}),\,\bM\mapsto \rho_{ij}(\bM) \,,
\end{equation}
which are representations of $\mathrm{GL}(2,\mathbb{C})$ acting on some representation spaces. All of them are given by rank-two tensor representations, explicitly given by (for $g,M\in\mathrm{GL}(2,\mathbb{C})$)
\begin{align}
    \nonumber\rho_{11}=\rho_{\mathrm{D}}\otimes\rho_{\mathrm{F}}:\mathrm{GL}(2,\mathbb{C})\rightarrow \mathrm{GL}(V\otimes V),\,\rho_{11}(\bM) \cdot\bT&=\bM^{-1T}\bT\bM^T \,, \\
    \rho_{12}=\rho_{\mathrm{D}}\otimes\rho_{\mathrm{D}}:\mathrm{GL}(2,\mathbb{C})\rightarrow \mathrm{GL}(V\otimes V),\,\rho_{12}(\bM) \cdot\bT&=\bM^{-1T}\bT\bM^{-1} \,, \\
    \nonumber\rho_{21}=\rho_{\mathrm{F}}\otimes\rho_{\mathrm{F}}:\mathrm{GL}(2,\mathbb{C})\rightarrow \mathrm{GL}(V\otimes V),\,\rho_{21}(\bM)\cdot \bT&=\bM \bT\bM^T \,, \label{eq:rho21} \\
    \nonumber\rho_{22}=\rho_{\mathrm{F}}\otimes\rho_{\mathrm{D}}:\mathrm{GL}(2,\mathbb{C})\rightarrow \mathrm{GL}(V\otimes V),\,\rho_{22}(\bM) \cdot\bT&=\bM \bT\bM^{-1} \,,
\end{align}
where $V=\mathbb{C}^2$ is the representation space of the fundamental and dual representations $\rho_{\mathrm{F}}$ and $\rho_{\mathrm{D}}$, as reviewed in the previous subsection.

These representations are reducible, which leads to a decomposition of the Siegel modular forms into Siegel modular forms with irreducible weights. To see this, first recall that the dual representation is related to the fundamental representation by the isomorphism in~\eqref{eq:dualFundamentalIsomorphism}.
Furthermore, we can reduce the rank-two tensor representation $\rho_{\rF}\otimes \rho_{\rF}$ into its symmetric and antisymmetric part
\begin{equation}
\label{eq:rhoF2Decomposition}
    \rho_{\rF}\otimes \rho_{\rF}=\mathrm{Sym}^2 \oplus \wedge^2 \,.
\end{equation}
The symmetric product representation was defined in section \ref{SubSec:Siegel}, while it is easy to see that the antisymmetric part here is simply the determinant representation $\wedge^2\simeq\det$. We thus find the decompositions
\begin{align}
\label{eq:blockRepresentations}
    \nonumber\rho_{11}&=\rho_{\rD}\otimes\rho_{\rF}\simeq \left(\mathrm{Sym}^2\otimes \mathrm{det}^{-1}\right)\oplus 1 \,, \\
    \rho_{12}&=\rho_{\rD}\otimes\rho_{\rD}\simeq \left(\mathrm{Sym}^2 \otimes \mathrm{det}^{-2}\right)\oplus \mathrm{det}^{-1} \,, \\
    \nonumber\rho_{21}&=\rho_{\rF}\otimes\rho_{\rF}\simeq \mathrm{Sym}^2 \oplus \det \,, \\
    \nonumber\rho_{22}&=\rho_{\rF}\otimes\rho_{\rD}\simeq \left(\mathrm{Sym}^2\otimes \mathrm{det}^{-1}\right)\oplus 1 \,.
\end{align}
On the level of the Siegel modular forms these decompositions can be implemented as follows. First one can insert a one in the form $\bepsilon^{-1} \bepsilon$, where the factor of $\bepsilon$ implements the isomorphism \eqref{eq:dualFundamentalIsomorphism}, see eq.\ \eqref{eq:eps_def}. Furthermore one can decompose matrices transforming under the product $\rho_{\rF}\otimes\rho_{\rF}$ into their symmetric and antisymmetric parts, implementing the decomposition \eqref{eq:rhoF2Decomposition}. For example, we can decompose
\begin{align}
\bcA^{-1}\bM_{11}\bcA&=\bepsilon^{-1}\bepsilon\bcA^{-1}\bM_{11}\bcA \\
    &=\bepsilon^{-1}\left[ \frac{1}{2}\left( \bepsilon\bcA^{-1}\bM_{11}\bcA+\bcA^T\bM_{11}^T\bcA^{-1T}\bepsilon^T\right)+\frac{1}{2}\left( \bepsilon\bcA^{-1}\bM_{11}\bcA-\bcA^T\bM_{11}^T\bcA^{-1T}\bepsilon^T\right)\right] \, , \nonumber
\end{align} 
abbreviating $\bM_{11}=\bbeta_1-\bbeta_2\bS$. The first term in the square brackets transforms in the symmetric representation up to an inverse factor of the determinant, while the second term transforms in the trivial representation, i.e., is modular invariant. Hence, we have decomposed the $2\times 2$ block appearing in the canonical differential equation matrix in terms of Siegel modular forms with irreducible weights (up to a constant factor of $\bepsilon^{-1}$). Similarly, we can decompose the other blocks according to the decomposition of the representations given above. In doing this
we find not only vector-valued Siegel modular forms but also classical Siegel modular forms of weights $-1,0$ or $1$ (and degree $2$).

The above observations show that the modular properties of the genus one canonical differential equation indeed generalize to Siegel modularity at higher genus. We can even observe a generalization of the weight grading in the above representations by considering the total degree of $\bM$ in the $\rho_{ij}$. Explicitly, we find total degree $(-2,0,2)$ along the anti-diagonal starting in the upper right, in agreement with the weights of the modular forms in the elliptic case (remembering that $\dd\tau$ has weight $-2$). Furthermore, we observe that the upper right block is precisely the differential of the normalized period matrix,
\begin{equation}
    \bcA^{-1}\bbeta_2\bcA^{-1T}=\dd\bOmega \,,
\end{equation}
nicely generalizing the analogous statement at genus one. We can then also consider the three independent entries of $\bOmega$ as the variables in the differential equation matrix $\bA(\blambda)$. The parameters $\blambda$ then become functions $\blambda(\bOmega)$, which can explicitly be given in terms of Riemann $\Theta$ functions \cite{enolskiRichterThomae,eilersRosenhain}, using Thomae's formula \cite{thomaeFormula},
\begin{equation}
    \lambda_1(\bOmega)=\left(\frac{\theta_6\theta_7}{\theta_8\theta_9}\right)^2 ,\qquad  \lambda_2(\bOmega)=\left(\frac{\theta_5\theta_7}{\theta_8\theta_{10}}\right)^2,\qquad \lambda_3(\bOmega)=\left(\frac{\theta_5\theta_6}{\theta_9\theta_{10}}\right)^2 \,,
\end{equation}
using the brief notation for the $\Theta$ constants introduced in appendix \ref{app_genus2thetas}. Note that these are Siegel modular invariant functions for $\Gamma_2(2)$. Also the period matrix $\bcA$ can be written explicitly as a function of the normalized period matrix \cite{enolskiRichterThomae,eilersRosenhain,eilersModularRepresentation} using the Rosenhain formula \cite{rosenhainFormula}
\begin{equation}
    \bcA(\bOmega)=\frac{2\theta_8\theta_9\theta_{10}}{\theta_1\theta_2\theta_3\theta_4\theta_5\theta_6\theta_7}
    \begin{pmatrix}
        \theta_8\theta_9\theta_{10}\partial_1\theta_{16} & \theta_8\theta_9\theta_{10}\partial_2\theta_{16}  \\
        \theta_5\theta_6\theta_{7}\partial_1\theta_{11} & \theta_5\theta_6\theta_{7}\partial_2\theta_{11}
    \end{pmatrix} \,.
\end{equation}

So far we have focused only on the first matrix in \eqref{eq:canonicalDELauricella3}, which does not contain the matrix $\bT_{\rA}$, i.e., the new function $a(\blambda)$. Indeed, we can show that the terms containing $a(\blambda)$ will spoil the modular properties, irrespective of the function $a(\blambda)$ having an expression solely in terms of periods or not. For details see appendix \ref{app_newFctConstraints}. Thus, in contrast to the elliptic case, we cannot interpret the entries of the full differential equation matrix as Siegel modular forms but only parts of the entries. The terms that spoil the modular properties contain the new function $a(\blambda)$. It is hence clear why this did not occur at genus one. We thus see that our canonical differential equation extends the modular properties encountered for genus one in a very direct and natural way.

%% file: Sec_OutConc.tex
In this paper we have taken first steps in studying the existence and structure of the canonical basis for Feynman integrals associated with higher-genus Riemann surfaces, in particular hyperelliptic curves. To this end we have studied three- and four-variable Lauricella hypergeometric functions which serve as prototypes for the maximal cuts of Feynman integrals associated to genus two hyperelliptic curves. In particular, this class of functions covers the maximal cut of the non-planar double box integral reviewed in section~\ref{subsec_Lauricella}. We have explicitly constructed a rotation into an $\eps$-factorized basis for these functions, generalizing the method of \cite{Gorges:2023zgv}, which was developed for elliptic curves and has also been applied to Calabi-Yau geometries \cite{Gorges:2023zgv,Klemm:2024wtd,Duhr:2024bzt,Driesse:2024feo,Forner:2024ojj}.  It is known that when performing this rotation one  generically  needs to introduce new functions defined as integrals over combinations of the (quasi-)periods of the geometry. In general, it is not an easy task to understand precisely how many such objects one needs. In the examples studied in this paper, we have shown how recent results on the structure of the intersection matrix in the canonical basis can be used to derive algebraic constraints on these functions allowing one to reduce the maximum number needed in an algorithmic way. Explicitly we were able to reduce the number of new functions from a naive counting leading to four and eight functions for the three- and four-variable Lauricella function, to only one and three functions, respectively.

While the differential equations constructed in this way are certainly $\eps$-factorized, it is not a priori clear that one has also found the canonical basis. While the fact that the intersection matrix is constant is already a strong hint that this is the case, we have provided further evidence by studying the modular properties of the $\eps$-factorized differential equation matrix, focusing on the three-variable Lauricella function. We were able to identify (parts of) the differential equation matrix as Siegel modular forms, nicely generalizing similar observations made in elliptic canonical forms \cite{Adams:2017ejb,Broedel:2018rwm,Abreu:2019fgk,Broedel:2021zij,Pogel:2022yat,Duhr:2024bzt}.
The modular properties are, however, more intricate as in the elliptic case, due to the appearance of the new function $a(\bs{\lambda})$, similar to what happens, e.g., in the case of the three-loop equal-mass banana integral~\cite{Pogel:2022yat}.

We only showcased genus two examples, but we expect the same method to work for higher genera with no new conceptual difficulties. Hence, our work provides strong evidence for the existence of the canonical basis for Feynman integrals associated with hyperelliptic curves, at least on the maximal cut, and points at some first structural properties. In particular, it shows that the algorithm of \cite{Gorges:2023zgv} generalizes in a straightforward way and continues to successfully construct the canonical form. 

There are however still some open questions which we leave for future investigation.
For example, we believe that the new function we found in the canonical basis of the three-variable Lauricella function is a genuine new function, in the sense that it can not simply be expressed in terms of the periods of the hyperelliptic curve. While the existence of such functions can be predicted in some cases \cite{Gorges:2023zgv}, these arguments do not seem to apply here. Rather, since we are working in a minimal setup there could be some purely geometric interpretation of this function, which would be very interesting to understand better in the future.

%% file: Sec_Intersection.tex
In this appendix we introduce some concepts related to twisted cohomology. We will keep it short and conceptual, whilst referring to \cite{aomotoBook, Mizera:2019gea, yoshida_hypergeometric_1997} for more in depth reviews and precise definitions. 

The central objects of study in the context of twisted cohomology are the twisted (co-)homology groups, whose periods are of the form 
\begin{align}
    \int_{\gamma} \Phi  \varphi \,,
\end{align} 
with $\Phi$ a multivalued function and $\varphi$ a single-valued differential forms whose only singularities are at branch points of $\Phi$. 
The twisted cohomology group $H^n_{\text{dR}}(X, \nabla_\Phi)$ is a cohomology group  with the connection $  \nabla_\Phi= \dd_{\mathrm{int}} + \dd_{\mathrm{int}} \log \Phi \wedge$
being sensitive to the multi-valued twist $\Phi$, i.e., the group of closed modulo exact differential forms with respect to $\nabla_\Phi$. Here $\dd_{\mathrm{int}}$ refers to the total differential with respect to the integration variables. Similarly, the corresponding twisted homology  group $H_n(X, \check{\mathcal{L}}_\Phi)$ with the local system $\check{\mathcal{L}}_\Phi$ contains contours that have a local branch of the twist assigned to them. We can choose bases $\{\varphi_j\}$ and $\{\gamma_i\}$ for each of these  and then define the period matrix to be the matrix of period pairings
\begin{align}
\label{eq:Periodmatrix}
    \bs{P}(\boldsymbol{x},\varepsilon)_{ij}=\int_{\gamma_j} \Phi\varphi_i\, .
\end{align}
Additionally, we define dual twisted cohomology and homology groups,   $H_{\text{dR}}^{n} (X,\check{\nabla}_\Phi)$ and $H_n({X},\mathcal{L}_\Phi)$ with bases $\{\check{\varphi}\}$ and $\{\check{\gamma}_j\}$ respectively. These have the twist $\Phi^{-1} $ and the dual twisted cohomology group is therefore defined with the connection $\check{\nabla}_\Phi = \dd_{\mathrm{int}} -\dd_{\mathrm{int}} \log \Phi\wedge$. Their period pairing is defined by 
\begin{align}
\bs{\check{P}}(\bs{x},\eps)_{ij}=\int_{\check{\gamma}_j}\Phi^{-1} \check{\varphi}_i \, .
\end{align}
 Information about the basis of differentials and the dual basis of differentials is captured by the intersection matrix, which is defined by
\begin{align}
\label{eq:CohomologyInt}
\bs{C}(\boldsymbol{x},\varepsilon)_{ij}= \frac{1}{(2\pi i)^n}\left(\int_X \varphi_i \wedge \check{\varphi}_j\right)_{ij} \, . 
\end{align}
The period matrix and its dual fulfill bilinear relations of the form 
\begin{align}
\label{generalriemannbil}
\frac{1}{(2\pi i)^n} \bs{P} \left(\bs{H}^{-1}\right)^T \bs{\check{P}}^T &\, =    \bs{C}\,. 
\end{align}
where $\bs{C}=\bs{C}(\boldsymbol{x},\varepsilon)$ is the cohomology intersection matrix defined in (\ref{eq:CohomologyInt}) and $\bs{H}=\bs{H}(\eps)$ is the intersection matrix for the homology group and its dual, defined by the topological intersections with the local branch choice of the twist taken into account (for more detail on the computation of homology intersection numbers, see also \cite{yoshida_hypergeometric_1997,Duhr:2023bku,kita_intersection_1994-2}). 

 As discussed in \cite{duhr2024twistedriemannbilinearrelations,Duhr:2024xsy}, the basis and the dual basis can always be chosen such that the period matrices are related by
\begin{align}
\label{sefldualityapp}
    \boldsymbol{\check{P}}(\boldsymbol{x},\varepsilon) = \left(\int_{\check{\gamma}_j}\Phi^{-1}\varphi_i\right)_{ij}= \boldsymbol{P}(\boldsymbol{x},-\varepsilon)\, , 
\end{align}
and this is in particular always true for the period matrices of maximal cuts. 
We interpret this relation as a notion of \textit{self-duality} for the maximal cut \cite{Duhr:2024xsy,Pogel:2024sdi}. It is reflected in the differential equation by the relation
\begin{align}
\label{eq:selfduality}
    \check{\bB}(\bs{x},\varepsilon)=\bB(\bs{x},-\varepsilon)\, . 
\end{align}
Since in this paper the twist always takes the form 
\begin{align}
    \Phi= \prod_{i=1}^{n+2} (x-\lambda_i)^{-\frac{1}{2}+ a_i\varepsilon} \, , 
\end{align}
(\ref{sefldualityapp}) is always fulfilled for some choice of basis $\{\varphi_i\}$ and the dual basis chosen as 
\begin{align}
    \check{\varphi}_i = \frac{\varphi_i}{\prod_{k=1}^{n+2} (x-\lambda_k)}\, . 
\end{align}
Whenever we compute an intersection matrix in the main text, this is how we choose the dual basis. For the purpose of this paper, the cohomology intersection matrix (\ref{eq:CohomologyInt}) is the most relevant object we get from twisted cohomology. 

\paragraph{Intersection numbers for one-forms.}

We want to briefly review how to compute this object practically. The Lauricella functions we consider here only have one integration variable, so we also restrict the review to the case of one-forms. For the calculation of intersection matrices of $n$-forms, see, e.g.,~\cite{Mizera:2019vvs,Brunello:2024tqf,Brunello:2023rpq,Chestnov:2022xsy,Chestnov:2022alh,Frellesvig:2020qot,Frellesvig:2019uqt}. 

We consider a twisted cohomology group with twist $\Phi$ and connection
\begin{align}
    \nabla_\Phi=\dd_{\text{int}} +  \dd_{\mathrm{int}} \log \Phi=\dd_{\text{int}} +\frac{\dd_{\mathrm{int}} \Phi}{\Phi}\, . 
\end{align}
We denote by $\mathcal{P}_{\Phi}$ the set of poles of $\dd_{\mathrm{int}} \log \Phi$. Let $\varphi_L\in H_{\text{dR}}^{1}(X,\nabla_\Phi)$ and $\check{\varphi}_R\in H^{1}_{\text{dR}}(X,\check{\nabla}_\Phi)$. We can compute the intersection number between these two differentials as a sum of residues 
\begin{align}
\label{deqloca}
  \langle \varphi_L |\check{\varphi}_R\rangle:=\frac{1}{2\pi i} \int_X \varphi_L\wedge \check{\varphi}_R =\sum_{p\in\mathcal{P}_\Phi} \text{Res}_{z=p} \left[\psi_p \check{\varphi}_R\right]\, , 
\end{align}
where $\psi_p$ is the local solution around $z=p$ of 
\begin{align}
\label{deqforpsi}
    \nabla_{\Phi} \psi_p =\varphi_L\,.
\end{align} The computation of the intersection numbers is
purely algebraic since we only need a small set of Laurent coefficients of $\Psi$ for extracting the residue, which can be achieved by making a power series ansatz and solving the resulting linear system. Note, that only a limited amount of terms need to be considered in this ansatz, more specifically, one makes an ansatz of the form 
\begin{align}
    \psi_p = \sum_{k=\text{min}}^{\text{max}} c_{p,k} x_p^k\, , 
\end{align}
with $x_p$ a local coordinate at $p$ and 
\begin{align}
     \text{min} = \text{ord}_p(\varphi_L) +1 \text{ and }      \text{max} = -\text{ord}_p(\check{\varphi}_R) -1\, ,
\end{align}
where $\text{ord}_p$ denotes the lowest non-vanishing order of the expansion in $p$. If $\text{max} < \text{min}$, there is no local solution for $\psi_p$ that contributes to (\ref{deqloca}).  We illustrate this with a pedagogical example for the reader not familiar with intersection theory. All other cases in this paper are also single-variable cases very similar to this example and can be obtained in the same way.  Additional instructive examples, also in the multi-variate case,  can  be found in \cite{Frellesvig:2020qot}.

\begin{example}[Hypergeometric ${}_2F_1$-function]

We consider as an example the hypergeometric ${}_2F_1$ function discussed in section \ref{SubSec:Elliptic}, i.e., we define the twisted cohomology $H_{\text{dR}}^1(X, \nabla_\Phi)$ with 
\begin{align}
    \Phi = x^{-\frac{1}{2}+a_1\varepsilon}(x-1)^{-\frac{1}{2}+a_2\varepsilon}(x-\lambda)^{-\frac{1}{2}+a_3\varepsilon} \,,
\end{align}
and $X=\mathbb{C}\mathbb{P} -\mathcal{P}_\Phi$, where 
\begin{align}
    \mathcal{P}_\Phi= \{0,1,\lambda,\infty\} \,,
\end{align} 
is the set of poles of $\dd \log \Phi$. For the basis and the dual basis we choose: 
\begin{align}
\label{basesf21twist}
    \varpi_1 & = \dd x\,, \qquad \check{\varpi}_1 = \frac{1}{x(x-1)(x-\lambda)} \dd x\,,\\
    \varpi_2 &= x\, \dd x\,, \qquad \check{\varpi}_2 = \frac{x}{x(x-1)(x-\lambda)} \dd x\,.
\end{align}
By expanding in local coordinates, we obtain the following lowest orders and corresponding minimal and maximal orders we need to compute of the $\psi_p$: 
\begin{table}[H]
\centering
\begin{tabular}{|c|c|c|c|c|}
\hline \text{Form} & $\text{ord}_0$&$\text{ord}_1$&$\text{ord}_\lambda$&$\text{ord}_\infty$  \\
\hline
    $\varpi_1$ & 0&0&0&-2\\
\hline
    $\varpi_2$ & 1&0&0&-3\\
\hline
    $\check{\varpi}_1$ &-1&-1&-1&1 \\
\hline
    $\check{\varpi}_2$ &0&-1&-1&0 \\ \hline 
\end{tabular}\, \, \, \, 
\begin{tabular}{|c|c|c|c|c|}\hline
$p$ &0&1&$\lambda$&$\infty$\\
\hline
$\text{min}_{1}$&1&1&1&-1\\
\hline
$\text{min}_{2}$&2&1&1&-2\\
\hline
$\text{max}_{1}$&0&0&0&-2\\
\hline
$\text{max}_{2}$&-1&0&0&-1\\
\hline 
\end{tabular}
\caption{Lowest orders in the local expansion of the bases (\ref{basesf21twist}) and the corresponding values for $\text{min}$ and $\text{max}$.}
\label{tab.orders}
\end{table}

\noindent $\langle \varpi_1 |\check{\varpi}_1\rangle$: Reading off from table \ref{tab.orders}, we find that for all poles $\text{min}_1> \text{max}_1$ and thus all residues in (\ref{deqloca}) vanish and 
\begin{align}
    \langle \varpi_1|\check{\varpi}_1\rangle = 0\, . 
\end{align}

\noindent $ \langle \varpi_1 |\check{\varpi}_2\rangle$: Again, we find from \ref{tab.orders}, that we do not get contributions from the poles $0,1,\lambda$ in (\ref{deqloca}), but we do get a contribution from the pole at $\infty$. We make the ansatz 
\begin{align}
    \psi_{\infty}^{12}= \sum_{k=\text{min}_1=-1}^{\text{max}_2=-1}c_{\infty,k}^{12} \, x_\infty^k \,,
\end{align}
and solve (\ref{deqforpsi}) for $c_{\infty,-1}$, finding 
\begin{align}
    \psi_\infty^{12} = \frac{2}{-1+ 2\eps(a_1+a_2+a_3)}\frac{1}{x_\infty}+ \mathcal{O}\left(x_{\infty}^{0}\right)\, . 
\end{align}
Consequently, we obtain: 
\begin{align}
    \langle \varpi_1 |\check{\varpi}_2\rangle= \text{Res}_{z=\infty} \left[\psi_\infty^{12}\cdot \left(-1 - (\lambda+1)x_\infty + \mathcal{O}(x_\infty^2) \right)\right]= -\frac{2}{-1+ 2\eps(a_1+a_2+a_3)}\, . 
\end{align}

\noindent $ \langle \varpi_2 |\check{\varpi}_1\rangle$ \textit{and} $ \langle \varpi_1 |\check{\varpi}_2\rangle$:  Similarly, we find 
\begin{align}
    \psi_{\infty}^{21} &= \frac{2}{1+2\eps(a_1+a_2+a_3)} \frac{1}{x_\infty^2} + \mathcal{O}(x_{\infty}^{-1}) \,,\\
    \psi_{\infty}^{22} &= \frac{2}{1+2\eps(a_1+a_2+a_3)} \frac{1}{x_\infty^2} + \frac{2-4\eps a_2+(2-4a_3\eps)\lambda}{(2\eps(a_1+a_2+a_3)-1)(2\eps(a_1+a_2+a_3)+1)}\frac{1}{x_\infty}+\mathcal{O}(x_{\infty}^{0}) \,,
\end{align}
and consequently
\begin{align}
    \langle \varpi_2 |\check{\varpi}_1\rangle&=-\frac{2}{1+2\eps(a_1+a_2+a_3)} \,, \\
      \langle \varpi_2 |\check{\varpi}_2\rangle&=-\frac{4\eps(a_1+a_3+\lambda(a_1+a_2))}{(2\eps(a_1+a_2+a_3)-1)(2\eps(a_1+a_2+a_3)+1)}\, . 
\end{align}
\end{example}

%% file: Sec_GenusTwoThetas.tex
There is a finite number of (half) characteristics for theta functions of given genus. In this appendix we record all theta constants (corresponding to even characteristics) and derivative theta constants (corresponding to odd characteristics) for genus two.

At genus two there are 10 even characteristics leading to non-trivial theta constants that we label as follows
\begin{align}
    &\theta_1=\RTheta{1\,1}{1\, 1}{\bOmega},\qquad \theta_2=\RTheta{0\,0}{1\, 1}{\bOmega},\qquad \theta_3=\RTheta{0\,0}{1\, 0}{\bOmega},\qquad \theta_4=\RTheta{0\,1}{1\, 0}{\bOmega}\,, \nonumber \\
    &\theta_5=\RTheta{0\,0}{0\, 1}{\bOmega},\qquad \theta_6=\RTheta{0\,0}{0\, 0}{\bOmega}, \qquad
    \theta_7=\RTheta{0\,1}{0\, 0}{\bOmega},\qquad
    \theta_8=\RTheta{1\,1}{0\, 0}{\bOmega} \,, \\
    &\theta_9=\RTheta{1\,0}{0\, 0}{\bOmega}, \qquad 
    \theta_{10}=\RTheta{1\,0}{0\, 1}{\bOmega}\,, \nonumber
\end{align}
where we suppress the $\bOmega$ dependence for brevity. Furthermore there are 6 odd characteristics at genus two, which lead to non-trivial derivative theta constants,
\begin{equation}
\begin{split}
    &\partial_i\theta_{11}=\partial_i\RTheta{0\,1}{0\, 1}{\bOmega},\qquad \partial_i\theta_{12}=\partial_i\RTheta{0\,1}{1\, 1}{\bOmega},\qquad 
    \partial_i\theta_{13}=\partial_i\RTheta{1\,0}{1\, 1}{\bOmega}\,, \\
    &\partial_i\theta_{14}=\partial_i\RTheta{1\,0}{1\, 0}{\bOmega},\qquad \partial_i\theta_{15}=\partial_i\RTheta{1\,1}{1\, 0}{\bOmega},\qquad \partial_i\theta_{16}=\partial_i\RTheta{1\,1}{0\, 1}{\bOmega} \,.
\end{split}
\end{equation}

%% file: ExpansionsDiff.tex
In this appendix, we give further technical details regarding the Abelian differentials. In section \ref{SubSec_expAroundInf} we give explicit expressions for the expansion of differentials of the form 
\begin{align}
    \frac{x^{j}\dd x}{y} 
\end{align}
around $\infty$ for even and odd hyperelliptic curves. These expansions allow us to construct combinations of these with a particular pole structure and in particular construct combinations with vanishing residue at infinity, i.e., second kind differentials. Using the expressions given here, it is easy to check that the differentials given in examples \ref{exdiff2} and \ref{exdiff4} are indeed of the kind claimed. Furthermore, in section \ref{SubSec_diffInThetas} we show how Abelian differentials can be written in terms of Riemann $\Theta$ functions using a global coordinate.

\subsection{Expansion Around $\infty$}
\label{SubSec_expAroundInf}

\paragraph{Odd Hyperelliptic Curve.}

Let us consider first the case of an {odd} hyperelliptic curve. 
In order to expand around $\infty$, we need to change coordinates via $u=\frac{1}{x^2}$. The square in the change of variables is necessary because $\infty$ is a branch-point of the square root in the differential. The expansion around $u=0$ of the differentials in question is then given by
\begin{equation}\label{expansionodd}
    \frac{x^{k-1}\dd x}{\sqrt{(x-\lambda_1)\dots(x-\lambda_{2g+1})}}=-2\sum_{N=g-k}^\infty \mathcal{S}_{N+k-g}u^{2N}\dd u \, ,
\end{equation}
 where
 \begin{equation}
    \mathcal{S}_N=(-1)^N\sum_{\sigma\subset\Sigma(N)}(-1)^{|\sigma|}\frac{(2|\sigma|)!}{(2^{|\sigma|} |\sigma|!)^2}s(\sigma) \,,
\end{equation}
and we define $\mathcal{S}_0=1$. Here the sum is taken over all ordered integer partitions of $N$, e.g.
\begin{equation}
    \Sigma(3)=\{\{3\},\{ 2,1\},\{ 1,2\},\{1,1,1 \}  \} \,.
\end{equation}
We defined $|\sigma|$ to be the cardinality of the ordered set $\sigma$ and
\begin{equation}
    s(\sigma)=\prod_{i=1}^{|\sigma|}s_{\sigma_i}(\Vec{\lambda})
\end{equation}
is a product of symmetric polynomials in the $\lambda_i$.\footnote{The undefined symmetric polynomials are considered to be zero.}

\paragraph{Even Hyperelliptic Curve.} 

Let us now consider  the case of an {even} hyperelliptic curve. Note that in this case, the differentials
\begin{equation}
    \frac{x^k \dd x}{y}\,,\quad k\geq g\,,
\end{equation}
have non-vanishing residues at $\infty$. To expand around $\infty$ we change variables with $x=1/u$. We find
\begin{equation}
\label{expansiondifferentialeven}
    \frac{x^{k-1}\dd x}{\sqrt{(x-\lambda_1)\dots(x-\lambda_{2g+2})}}=-\sum_{N=g-k}^\infty \mathcal{S}_{N+k-g}u^N\dd u \, ,
\end{equation}
This allows us to explicitly give a linear combinations of differentials with only a pole of order $k+1$. To this end, we  define the functions $\tilde{\Psi}_k$ by 
\begin{equation}
\label{noresiduesecond}
%\varpi_{2}^{(k)}
\tilde{\Psi}_k=\sum_{i=0}^{k}\sum_{\sigma\subset\Sigma(i)}(-1)^{|\sigma|}\mathcal{S}(\sigma)x^{g+k-i}  \text{ with } 
    \mathcal{S}(\sigma)=\prod_{i=1}^{|\sigma|}\mathcal{S}_{\sigma_i}.
\end{equation}
Then the differential
\begin{equation}
    \frac{\tilde{\Psi}_k\dd x}{y}=-\frac{1}{u^{k+1}}\dd u+\mathcal{O}(1).
\end{equation}
only has a pole of order $k+1$. Note, that of course we can add any linear combination of holomorphic differentials to this without changing this pole structure and we will do so in a manner that brings the bilinear relations between (quasi)-periods into a simple form, see examples \ref{exampleellipticlegendreeasy} and \ref{exampleoddlegendreeasy}. 

\subsection{Expressions in terms of Riemann $\Theta$-functions}
\label{SubSec_diffInThetas}
In this section we will show how the Abelian differentials on the hyperelliptic curve that we discussed in section \ref{Sec:background} can be written in a global coordinate in terms of first kind differentials and $\Theta$ functions. Such a coordinate could for example be provided by a Schottky parametrization \cite{schottky}, see \cite{Bobenko2011} for more details or \cite{Baune:2024biq} for a recent review in the physics literature. Numerically this coordinate change can e.g., be performed using the Myrberg algorithm \cite{myrbergAlgorithm}, see for example \cite{seppalaMyrberg1,seppalaMyrberg2,seppalaMyrberg3,li_myrbergs_nodate}.

\paragraph{Abelian Differentials of First Kind.} 

The Abelian differentials of the first kind are considered as elementary building blocks of the construction. We will choose a particularly normalized basis $\{\omega_1,\dots, \omega_g\}$ defined by the requirement
\begin{equation}
  \int_{\gamma_i}  \omega_{j}= \delta_{ij} \, . 
\end{equation}
That means, the basis $\varpi_i$ of holomorphic differentials on the hyperelliptic curve  defined in (\ref{eq_difffirst}) can be decomposed as 
\begin{align}
    \varpi_i = \sum_{j=1}^g \mathcal{A}_{ij}\omega_j\, . 
\end{align}
\paragraph{Prime Form and Bidifferential.}
The construction of the second and third kind differentials relies on the \emph{prime form} and the \emph{bidifferential}. The prime form is defined as (see e.g., \cite{dhokerStringReview,dhokerPrimeForm,fayThetaFunctions,tataTheta2})
\begin{equation}
    E(x,y| \bs{\Omega})=\frac{\Theta[\bepsilon](\mathfrak{u}(x,y), \bs{\Omega})}{\eta_{\bepsilon}(x)\eta_{\bepsilon}(y)} \,,
\end{equation}
where $\bepsilon=(\bepsilon_1,\bepsilon_2)$ is an odd 
%half-
(half-)characteristic and $\eta_{\bepsilon}(x)$ is the holomorphic function defined by
\begin{equation}
\eta_{\bepsilon}(x)^2=\sum_{i=1}^g\omega_i(x)\partial_i\Theta[\bepsilon](0, \bs{\Omega}) \,,
\end{equation}
and  
\begin{equation}
 \mathfrak{u}(x,y)=\left(\int_y^x\omega_{1},\dots,\int_y^x \omega_{g}\right)\, 
\end{equation}  
is referred to as \textit{Abel's map}. While not obvious, it turns out that the prime form is independent of the chosen characteristic. The fundamental bi-differential is defined by, see, e.g.,  \cite{bertolaLectures},
\begin{equation}
\label{eq_bidif}
    B(z,z')=\dd_z\dd_{z'}\log E(z,z')=\dd_z\dd_{z'}\log \Theta[\bepsilon](\mathfrak{u}(z,z'), \bs{\Omega}) \,.
\end{equation}
We can explicitly write this as 
\begin{equation}
    B(z,z')=(2\pi)^2\left[\frac{\Theta\langle z,z'\rangle(\mathfrak{u}(z,z')+c_{\bepsilon})}{\Theta(\mathfrak{u}(z,z')+c_{\bepsilon})} -\frac{\Theta\langle z\rangle(\mathfrak{u}(z,z')+c_{\bepsilon})\Theta\langle z'\rangle(\mathfrak{u}(z,z')+c_{\bepsilon})}{\Theta(\mathfrak{u}(z,z')+c_{\bepsilon})^2}\right]\dd z\wedge \dd z' \,,
\end{equation}
where we introduced the following notation for derivatives of $\Theta$ functions
\begin{equation}
\label{thetader}
    \Theta\langle z_1,\dots,z_k\rangle(x)=\sum_{n\in\mathcal{Z}^g}n^T\omega_1(z_1)\dots n^T\omega_1(z_k)\exp\left[ i\pi n^\mathrm{T}\bOmega n +2\pi in^\mathrm{T}x\right] \,,
\end{equation}
and 
\begin{equation}
    c_{\bepsilon}=%\bepsilon_2+\Omega\bepsilon_1=
    \frac{1}{2}\left(\bOmega{\bepsilon_1}+{\bepsilon_2}\right)\, . 
\end{equation}

\paragraph{Abelian Differentials of Third Kind.} 
Using the prime form we can give an explicit expression for the \emph{normalized Abelian differential of the third kind} with poles at $w_1,w_2$ \cite{fayThetaFunctions}:
\begin{equation}
    %\Omega_{w_1,w_2}
    \omega_{3|w_1,w_2}=\dd_z\log\frac{E(z,w_1)}{E(z,w_2)}=\dd z\left[\frac{\partial_z E(z,w_1)}{E(z,w_1)}-\frac{\partial_z E(z,w_2)}{E(z,w_2)}\right]\, . 
\end{equation}
The normalization here means that
\begin{align}
\label{third_norm}
    \int_{a_i}
    \omega_{3|w_1,w_2}=0\,,\qquad 
    \mathrm{res}_{z=w_1}
    \omega_{3|w_1,w_2}=1\,,\qquad
    \mathrm{res}_{z=w_2}
    \omega_{3|w_1,w_2}=-1\, . 
\end{align}
We can also write the differential directly in terms of $\Theta$ functions:
\begin{align}
    \omega_{3|w_1,w_2}&=\left[\frac{\omega_i(z)\partial_i\Theta(\mathfrak{u}(z,w_1),\bs{\Omega})}{\Theta(\mathfrak{u}(z,w_1),\bs{\Omega})}-\frac{\omega_i(z)\partial_i\Theta(\mathfrak{u}(z,w_2),\bs{\Omega})}{\Theta(\mathfrak{u}(z,w_2),\bs{\Omega})}\right]\dd z\, . 
\end{align}
We can now relate any third kind differential to the normalized differential defined above, by simply matching the pole structure and normalization. For the third kind differential with pole at $\infty$, which is part of the basis of Abelian differentials for any even hyperelliptic curve, this leads to
\begin{equation}
\label{thirdkindcurvesur}
\frac{x^g\dd x}{y}=-\omega_{3|z_\infty^+,z_\infty^-}+\sum_{i=1}^g \left( \oint_{a_i}\frac{x^g\dd x}{y}\right)\omega_{i}\, .
\end{equation}
Here $z_\infty^+$ and $z_\infty^-$ are the values of the global variable related to $\infty$ on the two sheets. It is easy to check that indeed both sides have the same pole structure which means that they can only differ by a linear combination of holomorphic differentials. This combination however has to be trivial which can be seen by integrating both sides over the $a$-cycles.

\paragraph{Abelian Differentials of Second Kind.}

The \emph{normalized differential of the second kind} with pole at $z=w$ of order $k+1$ is given by an appropriate residue of the fundamental bi-differential \cite{bertolaLectures},
\begin{equation}
\label{eq_standsec}
    \omega_{2|w}^{(k)}=-\frac{1}{k}\mathrm{Res}_{z'=w}\left[\frac{1}{(z'-w)^k}B(z,z') \right] \,,
\end{equation}
which satisfies
\begin{align}
\label{eq_normsec}
    \oint_{a_i} \omega_{2|w}^{(k)} = 0\,,\qquad
    \mathrm{res}_{z=w}\, \left[\omega_{2|w}^{(k)}\right]=0\, .
\end{align}
We can explicitly perform this residue and give a concrete formula for the differential, but we have to differentiate between the pole $w$ being finite or at $\infty$. If $w$ is finite, we find\footnote{Note that we drop the explicit dependence of the $\Theta$ functions on $\bs{\Omega}$ to keep the expressions shorter.}
\begin{align}
    \omega_{2|w}^{(k)}&
    =\frac{(2\pi)^2}{k!}\left.\frac{\partial^{k-1}}{\partial z'^{k-1}}\left[\frac{\Theta\langle z,z'\rangle(\mathfrak{u}(z,z')+c_{\bepsilon})}{\Theta(\mathfrak{u}(z,z')+c_{\bepsilon})} -\frac{\Theta\langle z\rangle(\mathfrak{u}(z,z')+c_{\bepsilon})\Theta\langle z'\rangle(\mathfrak{u}(z,z')+c_{\bepsilon})}{\Theta(\mathfrak{u}(z,z')+c_{\bepsilon})^2}\right]\right|_{z'=w}\dd z \,.
\end{align}
If the pole is at $\infty$ we need to take the residue in a chart containing $\infty$. Generally one can write (focussing on an even hyperelliptic curve for definiteness)
\begin{align}
    \omega_{2|\infty}^{(k)}&=\frac{(2\pi)^2}{k!}\frac{\partial^{k-1}}{\partial u'^{k-1}}\frac{1}{u'^2}\left[\frac{\Theta\langle z\rangle(\mathfrak{u}(z,1/u')+c_{\bepsilon})\Theta\langle 1/u'\rangle(\mathfrak{u}(z,1/u')+c_{\bepsilon})}{\Theta(\mathfrak{u}(z,1/u')+c_{\bepsilon})^2}\right.\nonumber\\
    &\qquad\qquad\qquad\qquad\qquad\left.\left. -\frac{\Theta\langle z,1/u'\rangle(\mathfrak{u}(z,1/u')+c_{\bepsilon})}{\Theta(\mathfrak{u}(z,1/u')+c_{\bepsilon})}\right]\right|_{u'=0}\dd z \,. 
\end{align}
We can now relate any second kind differential to a normalized one in the same way as detailed above for the third kind differentials. We match the pole structure on the two sides so that their difference is a linear combination of the holomorphic differentials which can be fixed using the normalization conditions of the various normalized differentials. Of course, this analysis depends on whether one considers an even or an odd hyperelliptic curve. Note also that the pole structure generally changes when changing coordinates and only the residue of a differential form is coordinate invariant.

%% file: Sec_NewFctConstraints.tex
In the main text we have seen that we need to introduce a new function $a(\blambda)$, naturally living in an antisymmetric matrix $\bT_{\bA}$, to find the canonical form of the three-variable Lauricella function. In this appendix we will provide some evidence that $a(\blambda)$ cannot be expressed solely in terms of (quasi-)periods and hence constitutes a genuine new function. As a side product this will allow us to show that the existence of this function will necessarily spoil the Siegel modularity of the canonical differential equation matrix.
The main tool we will use is a generalization of the \enquote{modular bootstrap} idea of \cite{Giroux:2022wav,Giroux:2024yxu}. To this end let us consider the differential equation
\begin{equation}
\label{eq:DiffEqTA}
    \dd \bT_{\rA}+\bP^{\rA}=0 \,,
\end{equation}
which defines the matrix $\bT_{\rA}$. 

First, let us assume that $\bT_{\rA}$ transforms in the same way as the symmetric part of $\bT$, given in \eqref{eq:TsymmPart}, i.e., as a Siegel modular form of weight $\rho_{21}$, see  \eqref{eq:blockRepresentations}. This is necessary for the Siegel modularity of the full differential equation matrix. Now apply a modular transformation to equation \eqref{eq:DiffEqTA}
\begin{equation}
    \begin{pmatrix}
        \bA & \bB \\ \bC & \bD
    \end{pmatrix} \in \Gamma_2(2) \,,
\end{equation}
using the known transformation behavior of the (quasi-)period matrices, see eqs.\ \eqref{eq:periodModularTransf} and \eqref{eq:quasiperiodModular}, and using the original differential equation \eqref{eq:DiffEqTA} to find
\begin{equation}
    (\dd \bg)\bT_{\rA} \bg^T+\bg \bT_{\rA} \dd\bg^T+8\pi i\left[\bg \bcA^T \bP_2^{\rA}\bcA^{-1T}\bC^T+ \bC\bcA^{-1}\bP_3^{\rA}\bcA\bg^T \right] =0\,,
\end{equation}
where we abbreviated $\bg=\bC\bOmega+\bD$. Using $\dd\bg=\bC\dd\bOmega$ and $\dd\bOmega=\bcA^{-1}\bbeta_2\bcA^{-1T}$ we find the two constraints
\begin{align}
    \bT_A\bcA^{-1}\bbeta_2+\bcA^T\bP_2^{\rA} &=0\,, \\
    \bbeta_2\bcA^{-1T}\bT_A +\bP_3^{\rA}\bcA &=0 \,.
\end{align}
These are equalities between matrix-valued differential forms. Considering different components of these equations (ensuring that the respective component of $\bbeta_2$ is invertible), we can derive various expressions for $\bT_{\rA}$ of the form
\begin{equation}
    \bT_A=\bcA^T\bM(\blambda)\bcA \,,
\end{equation}
for some matrices $\bM(\blambda)$. These equations, however, are not compatible with each other, leading to a contradiction. Hence, while we cannot prove that the antisymmetric matrix $\bT_{\rA}$ has no expression in terms of (quasi-)periods, we can see that in any case it has to break the Siegel modularity of the canonical differential equation. Since these terms all stem from the antisymmetric part of $\bT$ it is also clear why this did not happen at genus one. 

We can use a similar approach to exclude a more general ansatz of the form
\begin{equation}
    \bT_{\bA}=\bcA^T\bM_1(\blambda)\bcA+\bcA^T\bM_2(\blambda)\bcAt-\bcAt^T\bM_2^T(\blambda)\bcA+\bcAt^T\bM_3(\blambda)\bcAt \,,
\end{equation}
for some $2\times 2$ matrices $\bM_i(\blambda)$ where $\bM_1(\blambda),\bM_3(\blambda)$ are antisymmetric. Following the same strategy as above and using the bilinear relation \eqref{eq:quadraticARelation}, we find the two constraints
\begin{align}
    \br_1\bM_2-\br_1\bM_3&=0 \,, \\
    \bM_1\br_2^T-\bM_2 \br_1^T+\bP_2^{\bA}&=0 \,,
\end{align}
where $\br_1,\br_2$ are $2\times 2$ matrices of rational one-forms which can be read off from the differential equation 
\begin{equation}
    \dd\bcA=\br_1\bcA+\br_2\bcAt \,.
\end{equation}
Note that one needs to use the simple relation $\bbeta_2=8\pi i \br_2$, which is not difficult to show. By considering different components of the above constraint equations we can try to solve for entries of the $\bM_i$ and observe that there is no solution, allowing us to exclude that $\bT_{\bA}$ takes the form above. While this does not fully establish that the function $a(\blambda)$ can not be written be written as a rational combination of (quasi-)periods, this argument gives further evidence that $a(\blambda)$ is indeed a new function.

%% file: Sec_QuasiPeriodModular.tex
In this appendix we give some details on how to derive equation \eqref{eq:toProveInAppendix} necessary to find the modular transformation behavior of the $a$ quasi-period matrix $\bcAt$, given in the main text, see equation \eqref{eq:quasiperiodModular}.

Recall that we introduced the matrices $\bR_1,\bR_2$ to express $\bcAt$ in terms of the $a$ period matrix $\bcA$ and its derivative $\partial\bcA$ in \eqref{eq:defRMatrices}, or equivalently
\begin{equation}
    \partial\bcA=-\bR_2^{-1}\bR_1\bcA+\bR_2^{-1}\bcAt \,,
\end{equation}
with the same equation holding for the $b$ (quasi)-periods. Recall that $\partial$ is some first-order differential operator with respect to some parameters of the hyperelliptic curve and the matrices $\bR_1,\bR_2$ depend on this choice. Using $\bOmega=\bcA^{-1}\bcB$, we can now compute
\begin{align}
    \partial\bOmega&=-\bcA^{-1}\partial\bcA \bcA^{-1}\bcB+\bcA^{-1}\partial\bcB \nonumber\\
    &=-\bcA^{-1}\bR_2^{-1}\bcAt\bcA^{-1}\bcB+\bcA^{-1}\bR_2^{-1}\bcBt \\
    &=8\pi i\bcA^{-1}\bR_2^{-1}\bcA^{-1T} \nonumber\,,
\end{align}
where in the last equality we used
\begin{equation}
    \bcAt\bcA^{-1}\bcB=\bcAt\bcB^T\bcA^{-1T}=\bcBt-8\pi i\bcA^{-1T} \,,
\end{equation}
which follows from the quadratic relations \eqref{eq:genLegendre1} and \eqref{eq:genLegendre3}. It immediately follows that
\begin{equation}
    \bR_2\bcA\partial\bOmega=8\pi i\bcA^{-1 T}\,,
\end{equation}
as claimed.